\documentclass[pre,aps,showpacs,twocolumn]{revtex4}%
\usepackage{graphicx}
\usepackage{amsmath}
\usepackage{amsfonts}
\usepackage{amssymb}%
\providecommand{\U}[1]{\protect\rule{.1in}{.1in}}
\providecommand{\U}[1]{\protect\rule{.1in}{.1in}}
\providecommand{\U}[1]{\protect\rule{.1in}{.1in}}
\providecommand{\U}[1]{\protect\rule{.1in}{.1in}}
\begin{document}
\title{Theory of Coherent Van der Waals Matter}
\author{Igor M. Kuli\'{c}$^{1}$}
\email{kulic@unistra.fr}
\author{Miodrag L. Kuli\'{c}$^{2}$}
\email{kulic@th.physik.uni-frankfurt.de}
\date{\today}

\begin{abstract}
We explain in depth the previously proposed theory of the coherent Van der
Waals(cVdW) interaction - the counterpart of Van der Waals (VdW) force -
emerging in spatially coherently fluctuating electromagnetic fields. We show
that cVdW driven matter is dominated by many body interactions, which are
significantly stronger than those found in standard Van der Waals (VdW)
systems. Remarkably, the leading 2- and 3-body interactions are of the same
order with respect to the distance $(\propto R^{-6})$, in contrast to the
usually weak VdW 3-body effects ($\propto R^{-9}$). From a microscopic theory
we show that the anisotropic cVdW many body interactions drive the formation
of low-dimensional structures such as chains, membranes and vesicles with very
unusual, non-local properties. In particular, cVdW chains display a
logarithmically growing stiffness with the chain length, while cVdW membranes
have a bending modulus growing linearly with their size. We argue that the
cVdW anisotropic many body forces cause local cohesion but also a negative
effective "surface tension". We conclude by deriving the equation of state for
cVdW materials and propose new experiments to test the theory, in particular
the unusual 3-body nature of cVdW.

\end{abstract}
\affiliation{$^{1}$CNRS, Institute Charles Sadron, 23 rue du Loess BP 84047, 67034
Strasbourg, France}
\affiliation{$^{2}$ Institute for Theoretical Physics, Goethe-University D-60438 Frankfurt
am Main, Germany }

\maketitle

\section{Introduction}

\begin{figure}[pt]
\begin{center}
\includegraphics[
width=3.2in]{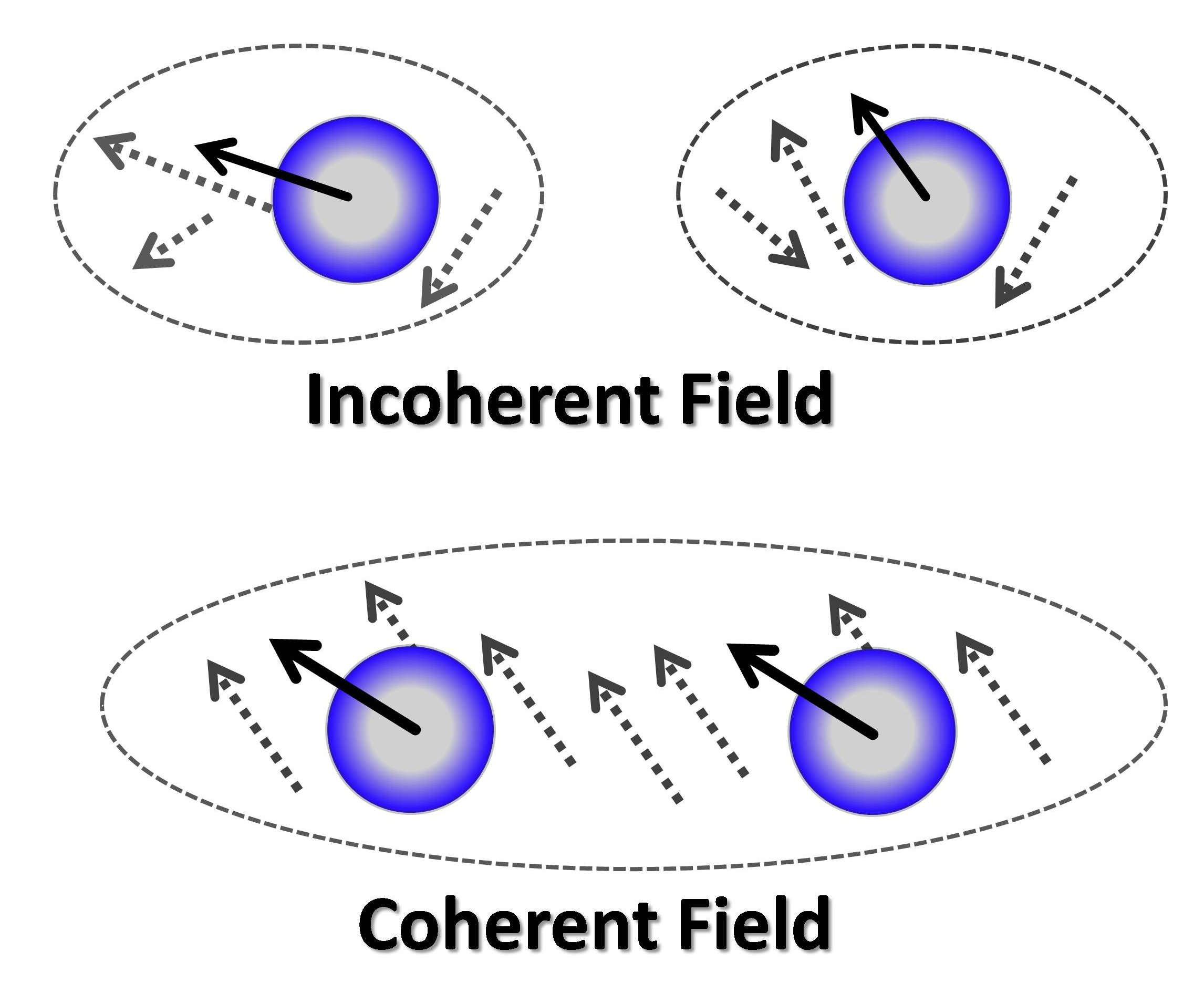}
\end{center}
\caption{(Top) The incoherent (Van der Waals-like) interaction VdW and
(Bottom) spatially coherent fluctuation interaction cVdW are both induced by
field fluctuations - but with different spatial correlations. }%
\end{figure}

The major goal of physics is the quest for understanding and controlling the
forces of Nature. In recent decades physicists and chemists have begun to
invent increasingly creative ways to combine the fundamental forces and to
generate new, effective interactions on microscopic and macroscopic scales.
Cold atoms could not be trapped and cooled \cite{ColdAtoms}, colloidal
suspensions would become unstable and flocculate \cite{Colloidal Suspensions},
and magnetic levitation would be impossible \cite{Levitron} if combined,
effective interactions were absent. As we know from condensed matter physics,
the interplay of attractive and repulsive forces of different origins can give
rise to highly complex structures. They range from gyroid phases in block
copolymers \cite{Block-Copolymers}, labyrinthine phases in ferrofluids
\cite{Labyrinthine} to nuclear pasta phases in neutron stars
\cite{Nuclear-Pasta} to name only a few. Not surprisingly, adding more
physical interactions naturally increases the structural complexity of the
resulting materials. Here we ask the opposite question: how much complexity
can emerge from a \textit{single}, simple to generate, effective interaction?

In the recent short paper \cite{KulicKulic-PRL} we studied, the probably
simplest \textit{effective interaction} able to generate surprisingly complex
structures. This effective interaction appears, for instance, between dipolar
magnetic (or dielectric) particles when a spatially uniform, isotropic but
time varying magnetic (or electric) field is externally applied (cf. Fig.1b).
The first instance of it was described in a series of important papers by
Martin et al. \cite{Martin1,Martin2} in a system of superparamagnetic colloids
in \textit{balanced} \textit{triaxial magnetic fields }(BTMF)\textit{\ }-
rotating magnetic fields spinning on a cone with the magic opening angle
$\theta_{m}\approx54,7^{\circ}$. The emerging effective interaction between
two colloids appeared to be, at the first glance, reminiscent of the
London-Van der Waals force \cite{Martin1,Martin2,Osterman}. Yet the structures
formed, including colloidal membranes and foams, were unexpectedly more
intricate and differing strongly from those expected in classical Van der
Waals (VdW) systems.

Inspired by the fascinating magnetic colloid superstructures generated
experimentally \cite{Martin1,Martin2,NatureCommentMartin} we have begun to
systematically investigate the physical ingredients and the consequences of
the induced interaction \cite{KulicKulic-PRL}. By starting out from the
analogy with the VdW interaction we have considered a generalization of
Martin's BTMF field structure \cite{Martin1,Martin2} and arrived at the
concept of the \textit{spatially coherent Van der Waals} (cVdW) interaction,
see Fig.1b \cite{NOTEcVdWName}.

In the previous short and rather dense paper \cite{KulicKulic-PRL}, we have
answered the important question, why cVdW generates complex structures like
chains, membranes and foams while its sister - the Van der Waals-like
\textit{incoherent fluctuation interaction} (VdW) -- see Fig.1a, merely forms
phase-separated lumps or 3D droplets of matter within a two phase system
\cite{Van Der Waals}. However, many details and numerous subtle questions were
omitted in \cite{KulicKulic-PRL} due to the lack of space. In the following we
close the gap and present a fairly complete theory of cVdW. \ 

As a new item going much beyond the previous Letter \cite{KulicKulic-PRL}, we
study the bulk and finite size effects in chains, rings, membranes, spherical
and cylindrical shells. In particular, we investigate the bending elasticity
of chains and membranes and show that there is a qualitative difference with
systems with short-range forces. As we will see, most structures formed by
cVdW, have properties which are inherently dictated by \textit{long range
anisotropic many body forces}. They exhibit collective (i.e. scale and shape
dependent) stiffness, surface tension and line tension.

\begin{figure}[pt]
\begin{center}
\includegraphics[
width=3.4in]{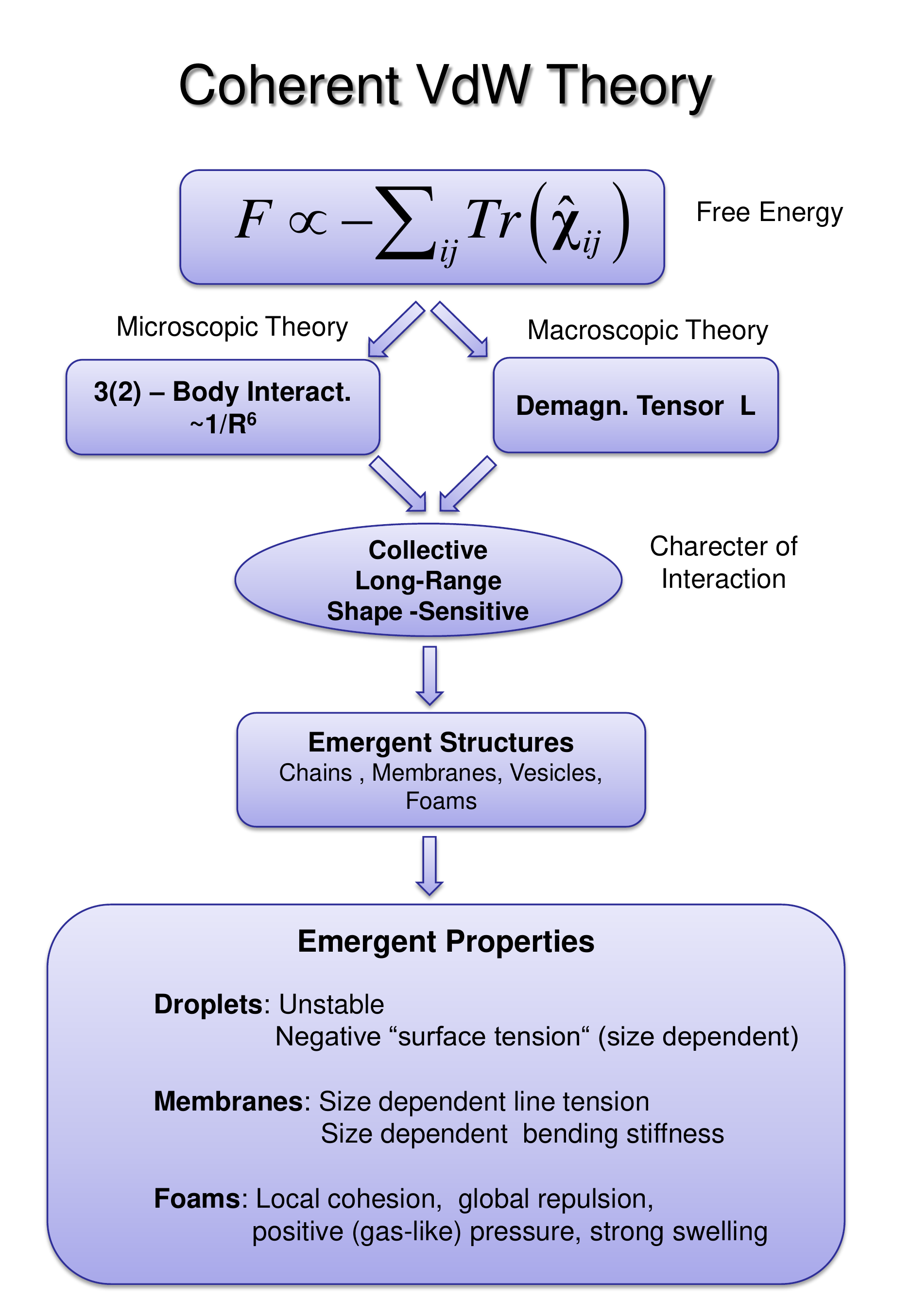}
\end{center}
\caption{ Theory of the cVdW interaction: Outline of results derived in this
paper. }%
\end{figure}

The physical content of the paper is schematically represented in the Diagram
in Fig.2, where the free-energy as a function of the \textit{effective
non-local susceptibility} $\hat{\chi}_{eff,ij}$ plays the central role in all
our studies of cVdW systems. In the microscopic theory the latter contains the
3-body (and higher order) interactions, while in the macroscopic theory it can
be expressed via an \textit{effective demagnetization tensor} $\hat{L}$. Both
approaches give rise to \textit{the collective, long range, shape-sensitive}
nature of the cVdW interaction.

The text is organized as follows. In $Section$ $II$ we briefly introduce the
reader into the basic properties of the magnetorheological (MR) colloids and
describe the first experimental realization of the cVdW interaction in such
systems by Martin et al\cite{Martin1,Martin2}. In $Section$ $III$ we develop
the basic physical and mathematical machinery to treat the cVdW interaction.
We then derive the first central result of this paper: The time-averaged
free-energy as the trace of the \textit{effective non-local susceptibility
tensor} $\hat{\chi}_{eff}$. The latter tensor describes an effective
interaction between two colloids mediated by all the other colloids - thus
containing all many body interactions. Based on the microscopic theory for the
free-energy in $Section$ $IV$ we discuss cVdW systems and self-assemblies of
colloids in various structures.

In $Section$ $V$ we study the formation of chains and membranes within the
framework of a macroscopic mean-field theory. The consistency of the latter
with the microscopic approach is discussed there as well. There we also show
that the cVdW systems can be considered as systems with an effective
\emph{negative} surface energy. In $Secton$ $VI$ we generalize the cVdW
interaction to anisotropic objects and study the interaction between multiple
elementary structures, including the bead-membrane and the two membrane
interaction. The interaction turns out to be very rich, anisotropic and
changes sign depending on the mutual orientation of the interacting objects.
Based on these preparatory results, in $Section$ $VII$ we study the formation
of the more complex emergent structures: the colloidal foams. We show that the
cVdW theory predicts a positive pressure of the foam and that it swells
against the gravitational field to notable heights.

The most notable quantitative results are summarized in two tables at the end
of $Section$ $X.$. We conclude by pointing out interesting experimental
effects and tests of the theory and by outlining some important open questions
concerning cVdW. More detailed derivations of some formulas are contained in
several Appendices for the interested reader. The mathematically less
interested reader is invited to browse through the figures, each of which
explains a new concept, and to run through them towards the Discussion.

\section{Preliminaries - Magnetic Colloids in Balanced Triaxial Fields}

\begin{figure}[pt]
\begin{center}
\includegraphics[
width=3.4in]{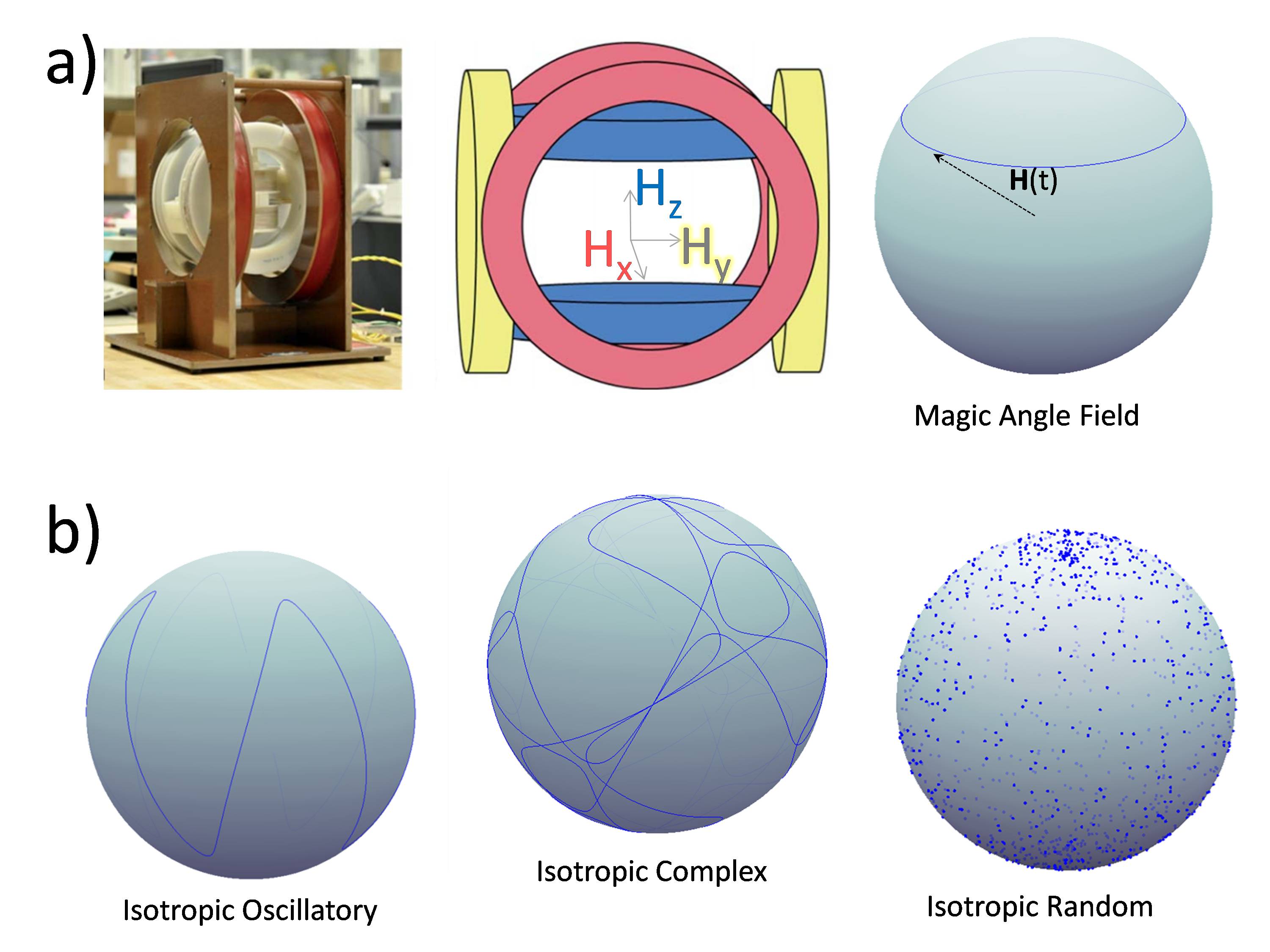}
\end{center}
\caption{ a) The first realization of the cVdW interaction with triaxial
magnetic fields by Martin et al. The field rotates on a cone with the "magic
angle" (cf. text). To avoid torques the field rotation is inverted every
cycle. b) Other possible realizations of the square-isotropic triaxial field.
If induced torques are negligible (no net rotation on average), all isotropic
excitations are equivalent and give rise to the same cVdW interaction.}%
\end{figure}

The study of responsive "smart materials" with remarkable properties has been
intensifying in the last decades. In that respect the magnetorheological (MR)
suspensions, made of magnetizable solid microparticles (colloids), dispersed
in nonmagnetic fluids and placed in magnetic fields, are of immense interest
due to the rapid, large and tunable transformations in their mechanical and
rheological properties. Having in mind the numerous applications
\cite{MR-applic}, a scientific challenge is to investigate which kind of
assembled structures are realized depending on combined static and oscillating
magnetic fields.


In a typical MR system consisting of superparamagnetic microbeads the induced
magnetic moment of a single bead $\mathbf{m}_{b}=\mathbf{M}_{b}V_{b}$, with
$V_{b}=(4\pi/3)d^{3}$ its volume and $\mathbf{M}_{b}$ its magnetization, is
proportional to the applied external magnetic field $\mathbf{H}_{0}$ i.e.
$\mathbf{m}_{b}=\chi_{b}V_{b}\mathbf{H}_{0}$. Here, $\chi_{b}(>0)$ stands for
the shape-dependent \textit{bead susceptibility} with respect to the external
field. This one should not be confused with the material susceptibility
$\chi_{b,m}(>\chi_{b})$
, which characterizes the physical properties of the material itself (not the
shape) out of which the bead is made. In the case when the bead is suspended
in a solvent with a material susceptibility $\chi_{s}$ in the magnetostatic
limit $\chi_{b}$ is given from $\chi_{b,m}$ by \cite{Landau,Jones}%

\begin{equation}
\chi_{b}=3\frac{\chi_{b,m}-\chi_{s}}{3+\chi_{b,m}+2\chi_{s}}. \label{hi-bead}%
\end{equation}

In the following we will study dipolar magnetic (dielectric) beads, which do
not carry permanent moments \cite{CommentNotation}. The beads, enumerated by
an index $i$, are assumed to be all identical, magnetically isotropic and
spherically shaped. They are placed in a spatially and temporally fluctuating
magnetic field $\mathbf{B}_{0,i}=\mu_{0}\mathbf{H}_{0,i}(t)$ with $\mu_{0}$
the vacuum permeability. We will focus here on the case when the field varies
on an intermediate timescale $\tau_{H}=2\pi/\omega$ fulfilling the condition
$\tau_{M}\ll\tau_{H}\ll\tau_{visc}$. Here $\tau_{M}$ is the typical magnetic
relaxation time of the paramagnetic bead, which is typically in the range of a
10th of second to microseconds. The other relevant characteristic time scale
is the aggregation time $\tau_{visc}\propto\eta$ which characterizes the
bead's motion in the surrounding viscous fluid over characteristic distances
comparable or larger than the bead diameter $d$. Under these conditions the
beads' magnetization $\mathbf{M}_{i}$ is equilibrated much faster than its
positional coordinate $\mathbf{R}_{i}(t)$, the beads move and aggregate slowly
and feel a net time-averaged force due to the dipole-dipole interaction. In
general , the susceptibility $\chi_{b}(\omega)$ can be a complex frequency
dependent function but in the following we will restrict ourselves to the case
when $\operatorname{Im}\chi_{b}(\omega)\ll\operatorname{Re}\chi_{b}(\omega)$.
In this case, the magnetic dissipation effects are in the first approximation
negligible compared to the magnetic free-energy effects.

In general, the dynamics of the $i-th$ bead is determined by the interplay of:
(1) the friction force $\mathbf{F}_{v}=-\xi\mathbf{v}_{b}$ with the friction
coefficient $\xi=6\pi d_{b}\eta$, (2) the average dipole-dipole force
$-\partial\mathcal{\bar{F}(}\mathbf{H}_{0},\mathbf{M}_{i},\{\mathbf{R}%
_{i}\})/\partial\mathbf{R}_{i}$ (with the dipole-dipole energy $\mathcal{\bar
{F}}(\mathbf{H}_{0},\{\mathbf{R}_{i}\})$) and (3) the fluctuating Brownian
force $\mathbf{F}_{B,i}$. However, in the following we shall consider the
(quasi-)equilibrium structures first, and postpone the bead dynamics to later works.

The first concrete instance of the cVdW interaction was realized
\cite{Martin1,Martin2,Osterman} a MR suspension that was placed in a magnetic
field rotating with a frequency $\omega$ on a cone with the opening angle
$\theta$, i.e. $\mathbf{H}_{0}(t)=\sqrt{3}H_{0}(\sin\theta\cos\omega
t,\sin\theta\sin\omega t,\cos\theta)$ - see Fig. 3a. More precisely, an ideal,
fully isptropic cVdW interaction is only realized when the cone opening angle
coincides with the magic angle $\theta=\arccos(1/\sqrt{3})\approx54,7^{\circ}$
. Such a \textit{balanced triaxial magnetic field} (BTMF) has very special
correlation properties: Its components have time averaged correlations,
denoted by $\overline{\left(  ...\right)  }$,
\begin{equation}
\overline{H_{i,0}^{\alpha}H_{j,0}^{\beta}}=\delta_{\alpha\beta}H_{0}^{2}.
\label{SqIsotropy}%
\end{equation}
that are formally (square) isotropic. Note, that this relation is true even
though the BTMF itself has a preferred orientation along the positive $z$
axis, see Fig. 3a. In the most general case we can consider any field
realization with such an isotropic correlation property and all particular
realizations (cf. Fig. 3(a-b)) will be considered equivalent within our
theory. In fact we will abstract away from any concrete representation of the
field and take the squared isotropy Eq.(\ref{SqIsotropy}) as the defining
property of the exciting field.

Note that in the most general case found in literature \cite{Martin1}%
,\cite{Martin2} the triaxial field is unbalanced and can have an arbitrary
in-plane $H_{\parallel}$ and perpendicular component $H_{\perp}:$
$\mathbf{H}_{0}=(H_{\parallel}\cos\omega t,H_{\parallel}\sin\omega t,H_{\perp
})$ with $2H_{\parallel}^{2}+H_{\perp}^{2}=H_{0}^{2}.$ One can show that in
this case the interaction can be linearly decomposed into a balanced triaxial
field (BTMF) magic angle interaction and a residual dipole-dipole interaction
along the orthogonal direction \cite{KulicKulic-PRL}. The latter is well
understood while the former is new and investigated here.


\section{The cVdW Free-Energy}

In this section we develop the mathematical formalism necessary to understand
the cVdW interaction. The cVdW interaction is a general phenomenon going
beyond the magnetic realm and the theory developed here is equally valid for
electrically polarizable colloids, i.e. the electrorheological materials. In
order to keep the continuity with Refs.
\cite{KulicKulic-PRL,Martin1,Martin2,Osterman}, we arbitrarily follow the
magnetic notation. The results for electrically polarizable colloids are
obtained by replacing the magnetic quantities with the corresponding electric ones.

Under the condition of quick variations of the external field $\mathbf{H}_{0}%
$, yet a much quicker equilibration of the magnetization $\mathbf{M}_{i}$ we
study equilibrium structures which minimize the effective (time
averaged)\ free-energy $\mathcal{\bar{F}}$, i.e.
\begin{equation}
-\frac{\partial\mathcal{\bar{F}(}\mathbf{H}_{i,0},\mathbf{M}_{i}%
,\{\mathbf{R}_{i}\})}{\partial\mathbf{M}_{i}}=0 \label{Min-cond}%
\end{equation}
for fixed particle positions $\mathbf{R}_{i}$.

In the following we will be dealing with purely athermal effects. This is
usually well justified: due to the large moments $\mathbf{m}_{i}$ of the beads
with diameters $D(=2d)>1$ $\mu m$, the energy per particle will be well in
excess of the thermal energies making contributions of a configurational
entropy negligible in practice. Therefore, all external field fluctuations
will be considered as extrinsically given.

The basic expression for the non-equilibrium free-energy $\mathcal{F}%
\{\mathbf{M}_{i},\mathbf{H}_{i,0}(t)\}$ of magnetic colloids (beads) with
respect to $\mathbf{M}_{i}$ in inhomogeneous external time-dependent field
$\mathbf{H}_{i,0}\left(  t\right)  $ can be written as \cite{Landau}%
\begin{equation}
\frac{\mathcal{F}}{\mu_{0}V_{b}}=\sum_{i=1}^{N}\left(  \frac{\mathbf{M}%
_{i}^{2}}{2\chi_{b}}-\mathbf{M}_{i}\mathbf{H}_{i,0}\right)  +\frac{1}{2}%
\sum_{i,j\mathbf{\neq}i}\mathbf{M}_{i}\hat{T}_{ij}\mathbf{M}_{j}, \label{S-1}%
\end{equation}
where $V_{b}$ is the bead-volume of one of the $N$ identical beads.
$\mathbf{M}_{i}=\mathbf{m}_{i}/V_{b}$ is the magnetization of the $i$-th bead
resulting from its magnetic moment $\mathbf{m}_{i}$ and $\chi_{b}$ is the
\textit{bead (sample) susceptibility} in the external field. The first term in
Eq.(\ref{S-1}) is the "self-energy" of the beads. It ensures that in absence
of external fields there is no magnetization. The second term represents the
$i-th$ bead's dipole-dipole interaction with all the other beads. It is
mediated by the dipole tensor $\hat{T}_{ij}$\
\begin{align}
\hat{T}_{ij}  &  =\varphi_{ij}\hat{t}(\mathbf{b}_{ij}),\text{ }\varphi
_{ij}=\frac{V_{b}}{4\pi\left\vert \mathbf{R}_{ij}\right\vert ^{3}%
}\label{T-ensor}\\
\hat{t}(\mathbf{b}_{ij})  &  =\hat{1}-3\left\vert \mathbf{b}_{ij}\right\rangle
\left\langle \mathbf{b}_{ij}\right\vert ,\text{ }\mathbf{b}_{ij}%
=\frac{\mathbf{R}_{ij}}{\left\vert \mathbf{R}_{ij}\right\vert },\nonumber
\end{align}
with $\mathbf{R}_{ij}\equiv\mathbf{R}_{i}-\mathbf{R}_{j}\neq0$ ( $i\neq j$ )
and $\mathbf{b}_{ij}$ the normalized bonding vector, i.e. the unit vector
pointing pointing from bead $j$ to $i$. Here, we decompose conveniently the
dipole tensor $\hat{T}_{ij}$ into a purely geometric $3\times3$ tensor
$\hat{t}(\mathbf{b}_{ij})$ - a linear combination of the unity matrix $\hat
{1}$ and the pure projector on the bonding vector $\left\vert \mathbf{b}%
_{ij}\right\rangle \left\langle \mathbf{b}_{ij}\right\vert $ (in the
"Bra-ket"\ notation). The second contribution in $\hat{T}_{ij}$ is the purely
distance dependent dipolar field-decay factor $\varphi_{ij}\left(
R_{ij}\right)  \propto R_{ij}^{-3}.$

Performing the minimization of $\mathcal{F}$ w.r.t. $\mathbf{M}_{i}$ and
reintroducing the result into Eq.(\ref{S-1}) we arrive at the
quasi-equilibrium free-energy for fixed coordinates $\{\mathbf{R}_{i}\}$
\cite{Landau}
\begin{equation}
\mathcal{F}\{\mathbf{H}_{i,0}\}=-\frac{1}{2}\mu_{0}V_{b}\sum_{i}\mathbf{M}%
_{i}\mathbf{H}_{i,0}, \label{S2}%
\end{equation}
where the quasi-equilibrium magnetization of the i-th bead $\mathbf{M}%
_{i}\left(  t\right)  =\chi_{b}\mathbf{H}_{i,loc}\left(  t\right)  $ is
determined by the \textit{total local} fields $\mathbf{H}_{i,loc}$ acting at
the position of the bead $i.$ These local fields are given implicitly as a
function of the applied external fields $\mathbf{H}_{i,0}(t)$ via%

\begin{equation}
\sum_{j}\left(  \delta_{ij}+\chi_{b}\hat{T}_{ij}\right)  \mathbf{H}%
_{j,loc}(t)=\mathbf{H}_{i,0}(t). \label{Hloc}%
\end{equation}
The last line is strictly valid if we adopt the practical convention that
$\hat{T}_{ii}=0$ for two identical bead indices, i.e. excluding a
self-interaction of beads. By inverting Eq.(\ref{Hloc}), the formal solution
for the local fields $\mathbf{H}_{i,loc}$ reads
\begin{equation}
\mathbf{H}_{i,loc}=\chi_{b}^{-1}\sum_{j}\hat{\chi}_{eff,ij}\mathbf{H}_{j,0}
\label{Hloc2}%
\end{equation}
Here we encounter a main player in the cVdW interaction - the
\textit{effective non-local microscopic susceptibility tensor} defined as:
\begin{equation}
\hat{\chi}_{eff}=\chi_{b}(\hat{1}+\chi_{b}\hat{T})^{-1}. \label{chi-eff}%
\end{equation}
This \textit{non-local microscopic susceptibility tensor }$\hat{\chi}_{eff}$
is crucial for understanding all the many body effects that will follow.
\ Lets take a few notes in order to understand some of its features.
\ Mathematically, it is a $3\times N$- dimensional matrix, relating the
external fields at all particle $j-th$ locations to the local fields at
particle $i-th$ position. For any fixed $i$ and $j,$ the single components
$\hat{\chi}_{eff,ij}$ are themselves $3$ dimensional second--rank tensors -
i.e. $3\times3$ matrices in $3$-dimensional space. In the following, dealing
with $\hat{\chi}_{eff}$ and its components will be conceptually easy with a
small caveat and the note of caution: the $i,j$ components $\hat{\chi
}_{eff,ij}$ are to be evaluated \textit{after} performing the operator
inversion in Eq. (\ref{chi-eff}). This is at the very origin of the many body
forces (cf. below).

With all theses issues and precautions about $\hat{\chi}_{eff}$ in mind we can
now insert Eq.(\ref{Hloc2}) into Eq.(\ref{S2} ) and average over the
fluctuating fields $\mathbf{H}_{i,0}(t)$ to obtain the averaged free-energy
$\mathcal{\bar{F}}$
\begin{align}
\mathcal{\bar{F}}\left(  \mathbf{H}_{0},\{\mathbf{R}_{i}\})\right)   &
=-\frac{\mu_{0}}{2}V_{b}\sum_{i,j}\overline{\mathbf{H}_{i,0}\hat{\chi
}_{eff,ij}\mathbf{H}_{j,0}}\label{S4}\\
&  =-\frac{\mu_{0}}{2}V_{b}\sum_{i,j,\alpha,\beta}C_{ij}^{\alpha\beta}%
\chi_{eff,ij}^{\alpha\beta},\nonumber
\end{align}
where $C_{ij}^{\alpha\beta}$ is the \textit{field-field correlation function}
defined by
\begin{equation}
C_{ij}^{\alpha\beta}=\overline{H_{i,0}^{\alpha}H_{j,0}^{\beta}}. \label{Cij}%
\end{equation}
Here, $\alpha,\beta=x,y,z$ stand for the 3-spatial directions and
$i,j=1,2....N$ are the particle indices. The Eq. (\ref{S4}) is completely
general and forms the backbone for all further analysis. It holds both for the
classic VdW interaction (with appropriately chosen $\mathbf{H}_{i,0}$) as well
as for cVdW and generally couples the field correlators $C_{ij}^{\alpha\beta}$
with the many-body interaction-encoding susceptibility $\chi_{eff,ij}%
^{\alpha\beta}$ functions. The former are given by the type of interaction
(VdW or cVdW) while the latter depend on the spatial configuration of all
particles in an interesting but (for now) very convoluted, little transparent
manner that we want to elucidate in the following.

\subsection{Incoherent vs. Coherent Fields}

The decomposition of the free energy in Eq. (\ref{S4}) into an "external
influence" term\ (the field correlator $C_{ij}^{\alpha\beta}$) and an
"internal response" function (susceptibility tensor components $\chi
_{eff,ij}^{\alpha\beta}$) is conceptually appealing. The remainder of the
paper we will spend on exploring the physical features associated with these
two terms.

In a first step, let us investigate how different types of driving field
correlations change the interactions. There are two important limiting cases
for the correlator $C_{ij}^{\alpha\beta}$ :

$(A)$ \textit{strong correlations} with perfect spatial coherence and

$(B)$ \textit{no correlations} in the driving field with perfect decoherence.

These are defined in the following way:

($A$) The \textit{spatially coherent fluctuation interaction} with the correlator%

\begin{equation}
C_{ij}^{\alpha\beta}=C^{\alpha\beta}=\delta_{\alpha\beta}H_{0}^{2}%
+h^{\alpha\beta} \label{C-coh}%
\end{equation}
The first term $C_{0}^{\alpha\beta}=\delta_{\alpha\beta}H_{0}^{2}$ describes
the \textit{isotropic} and spatially coherent (uniform) excitation, while the
second, $h^{\alpha\beta},$ describes a uniform constant field (anisotropic
contribution). In the following we restrict ourself only to the
\textit{isotropic coherent van der Waals interaction }(the cVdW one) with a
vanishing anisotropic component, $h^{\alpha\beta}=0$ i.e.
\begin{equation}
C_{ij}^{\alpha\beta}=C_{0}^{\alpha\beta}=\delta_{\alpha\beta}H_{0}^{2}.
\label{C0}%
\end{equation}
It is easy to see that the corresponding \textit{cVdW free-energy}, is then
given by
\begin{equation}
\mathcal{\bar{F}}_{cVdW}\left(  \mathbf{H}_{0},\{\mathbf{R}_{i}\}\right)
=-\frac{\mu_{0}}{2}H_{0}^{2}V_{b}\sum_{i,j}Tr\left(  \hat{\chi}_{eff,ij}%
\right)  , \label{S5}%
\end{equation}
where
\begin{equation}
Tr\left(  \hat{\chi}_{eff,ij}\right)  \equiv\chi_{eff,ij}^{xx}+\chi
_{eff,ij}^{yy}+\chi_{eff,ij}^{zz}%
\end{equation}
is the trace of the effective susceptibility. Note, that the coherent
isotropic correlation function $C_{\alpha\beta}=\delta_{\alpha\beta}H_{0}^{2}$
comprises also the case of the experimentally realized \textit{balanced
triaxial magnetic fields \cite{Martin1,Martin2,Osterman} }(BTMF) - see
$Section$ $II$ and Fig.3a. Even though the cone at which the field precesses
has a direction (opening) by itself, the square of the field is statistically
identical in all directions and mutually uncorrelated in all directions i.e.
$\overline{H_{i,0}^{\alpha}H_{j,0}^{\beta}}=\delta_{\alpha\beta}H_{0}^{2}$.
The BTMF is therefore only one instance of a general coherent isotropic field.
Any other realization, like e.g. one of those in Fig 3b satisfies the relation
(\ref{S4}) and consequently has the same energy Eq. (\ref{C0}).

($B$) \textit{The spatially\ incoherently excited fields} are realized for
$C_{ij}^{\alpha\beta}=\delta_{ij}C^{\alpha\beta}$. The latter may in principle
contain isotropic and anisotropic terms, too. Note, that the term proportional
to $\delta_{ij}$ means that the correlations of magnetic field fluctuations on
different particles vanish. In the completely \textit{isotropic }case one has
\begin{equation}
C_{ij}^{\alpha\beta}=\delta_{ij}\delta_{\alpha\beta}H_{0}^{2} \label{C-icFI}%
\end{equation}
and the fully incoherent VdW free-energy - of the \textit{VdW systems}, reads%
\begin{equation}
\mathcal{\bar{F}}_{VdW}\left(  \mathbf{H}_{0},\{\mathbf{R}_{i}\}\right)
=-\frac{\mu_{0}}{2}H_{0}^{2}V_{b}\sum_{i}Tr\left(  \hat{\chi}_{eff,ii}\right)
. \label{S6}%
\end{equation}

Note, that there is a \textit{significant difference} between the two
excitation cases ($A$) and ($B$), described by Eq.(\ref{S5}) and
Eq.(\ref{S6}), respectively. In the incoherent case ($B$) the free-energy
$\mathcal{\bar{F}}_{VdW}\left(  \mathbf{H}_{0},\{\mathbf{R}_{i}\}\right)  $
\textit{contains a summation over the index }$i$\textit{ only}, i.e. it
includes only the diagonal terms $i=j$ - sometimes called the self-energy
terms. This is equivalent to the usual Van der Waals (VdW) interaction and in
the following this part of the free-energy will be called the VdW one. However
in the novel case ($A$) $\mathcal{\bar{F}}_{cVdW}$ contains the more complex,
double summation over $i$ and $j$ which gives rise to unusual, non-local and
anisotropic many body effects in cVdW system - most of which are absent in a
standard VdW system (case $B$). In the coherent case, the terms $Tr(\hat{\chi
}_{eff,ij})$ with $i\neq j$, describe the effective coupling between the
$i$-th and $j$-th bead acting directly or indirectly via all other beads, thus
giving rise to very specific and anisotropic \textit{many body\ interactions}.
The latter will turn out to be a crucial effect for cVdW matter and will be
responsible for the formation of hierarchical assemblies of colloids. This is
in strong contrast with the standard VdW systems where 3D bulk structures such
as droplets and close-packed 3D crystal structures are favored and realized.

\section{Microscopic cVdW Theory - Many Body Interactions, the Formation of
Chains and Membranes}

In this Section we will explore how a \textit{ microscopic cVdW \ theory},
based on the effective energy Eq.(\ref{S5}) and the non-local many body
susceptibility Eq.(\ref{chi-eff}) works in practice. While the energy
Eq.(\ref{S5}) appears (deceptively) straightforward to evaluate, the many body
susceptibility operator Eq.(\ref{chi-eff}) is a sophisticated mathematical
object. To grasp physical insights about the latter, except for the simplest
case of two spherical beads, seems challenging.

After dealing with the elementary case of two particles, which can be treated
exactly, we will resort to the approximation of small bead susceptibility i.e.
$\chi_{b}\ll1$ - a limiting case that allows a controllable evaluation of the
many body interactions. In this spirit we will be making an energy expansion
up to lowest order in $\chi_{b}$. Notably, this lowest order expansion of the
energy, as we will see, comprises both the 2-body interaction of beads and the
non-local 3-body interactions at the same order.

As it will be shown both these interactions (2 and 3 body) scale identically
with distance for cVdW, i.e. $\mathcal{\bar{F}}_{2-body}\propto\mathcal{\bar
{F}}_{3-body}\propto R^{-6}$. The \emph{2+3 body inseparability} is the most
peculiar \emph{hallmark signature of the cVdW interaction}, and to our
knowledge stands out rather uniquely among other known many-body forces in Nature.

\subsection{Two-Body Interaction - Dimer Formation}

Let us start out elementary and consider a very dilute system. In such a case
the \textit{pairwise} bead-bead ($2$-body) interaction should dominate in the
free-energy $\mathcal{\bar{F}}_{cVdW}$. While this assumption of dominant
2-body forces turns out as too naive (see the next subsection) it is still
natural to consider only two interacting particles first. For two beads 1 and
2 the non-local susceptibility operator $\hat{\chi}_{eff}$ is easily
calculated by using Eq. (\ref{chi-eff}). The detailed derivation is given in
$Appendix$ $1A$ while the final result reads%

\begin{align}
Tr\hat{\chi}_{eff}  &  =C\cdot\left(
\begin{array}
[c]{cc}%
1-3\varphi_{12}^{2}\chi_{b}^{2} & 2\chi_{b}^{3}\varphi_{12}^{3}\\
2\chi_{b}^{3}\varphi_{12}^{3} & 1-3\varphi_{12}^{2}\chi_{b}^{2}%
\end{array}
\right) \label{TrChieff2Bead}\\
C  &  =\frac{\allowbreak3\chi_{b}}{\left(  1-4\varphi_{12}^{2}\chi_{b}%
^{2}\right)  \left(  1-\varphi_{12}^{2}\chi_{b}^{2}\right)  }%
\end{align}
Having the trace of $\hat{\chi}_{eff}$ , we can now evaluate the mean free
energy for the incoherent and the coherent case.

In the\textit{ incoherent (standard)\ VdW} case the energy $\bar{f}%
_{VdW}=\mathcal{\bar{F}}_{VdW}/\left(  \mu_{0}V_{b}H_{0}^{2}\right)  $ is the
sum of the $Tr\hat{\chi}_{eff}$ diagonals (note $\left(  Tr\hat{\chi}%
_{eff}\right)  _{11}=\left(  Tr\hat{\chi}_{eff}\right)  _{22}$) i.e.
\begin{align}
\bar{f}_{VdW}  &  =-\left(  Tr\hat{\chi}_{eff}\right)  _{11}\\
&  =-\frac{3\chi_{b}\left(  1-3\varphi_{12}^{2}\chi_{b}^{2}\right)  }{\left(
4\varphi_{12}^{2}\chi_{b}^{2}-1\right)  \left(  \varphi_{12}^{2}\chi_{b}%
^{2}-1\right)  }.\nonumber
\end{align}
$\bigskip$On the other hand, for the \textit{coherent cVdW} interaction we
have to sum all four elements of $Tr\hat{\chi}_{eff}$ \ (note $\left(
Tr\hat{\chi}_{eff}\right)  _{12}=\left(  Tr\hat{\chi}_{eff}\right)  _{21}$)
obtaining
\begin{align}
\bar{f}_{cVdW}  &  =-\left(  Tr\hat{\chi}_{eff}\right)  _{11}-\left(
Tr\hat{\chi}_{eff}\right)  _{12}\\
&  =-\frac{\allowbreak3\chi_{b}\left(  1-\chi_{b}\varphi_{12}\right)
}{\left(  1+\chi_{b}\varphi_{12}\right)  \left(  1-2\chi_{b}\varphi
_{12}\right)  }.\nonumber
\end{align}

Interestingly, the coherent and the incoherent 2 bead interaction energy look
very similar but are \textit{not} identical. After expanding the energies in
powers of $\varphi_{12}$ we see that $\bar{f}_{VdW}=\allowbreak-3\chi
_{b}-6\varphi_{12}^{2}\chi_{b}^{3}-18\varphi_{12}^{4}\chi_{b}^{5}+...%
\acute{}%
$\ and $\bar{f}_{cVdW}\approx-3\chi_{b}-6\varphi_{12}^{2}\chi_{b}^{3}%
-6\varphi_{12}^{3}\chi_{b}^{4}+...$ or in terms of the bead-bead distance
$R_{12}$:%

\begin{align}
\bar{f}_{VdW}  &  \approx-3\chi_{b}-\frac{3V_{b}^{2}\chi_{b}^{3}}{8\pi^{2}%
}R_{12}^{-6}-\frac{9\chi_{b}^{5}V_{b}^{4}}{128\pi^{4}}R_{12}^{-12}%
\label{Bead-Bead}\\
\bar{f}_{cVdW}  &  \approx-3\chi_{b}-\frac{3V_{b}^{2}\chi_{b}^{3}}{8\pi^{2}%
}R_{12}^{-6}-\frac{3V_{b}^{3}\chi_{b}^{4}}{32\pi^{3}}R_{12}^{-9}\nonumber
\end{align}
We observe that the first interaction terms $\propto R_{12}^{-6}$ exactly
coincide. This interesting $2$-body result was first obtained by Martin and
coworkers \cite{Martin1,Martin2} and confirmed experimentally by Osterman et
al. \cite{Osterman}. However, the higher order terms in $\bar{f}_{VdW}$ and
$\bar{f}_{cVdW}$ scale quite differently, and they are $\propto R_{12}^{-12}$
and $\propto R_{12}^{-9}$, respectively. This makes the cVdW interaction
slightly stronger (more attractive) than the usual incoherent VdW.

Now, if it was only for this slight difference between the two, investigating
the cVdW would hardly be very interesting. But we will see soon that the
3-body forces are a real game changer, giving the cVdW\ interaction its unique
character and flavor.

\subsection{Many Body Interactions}

For the standard (incoherent) VdW interaction the $2$-body interaction is
$\propto\left\vert \mathbf{R}_{12}\right\vert ^{-6}$ in leading order and
since the 3-body (and higher order) interactions are shorter ranged (cf.
below) and much smaller in magnitude than the 2--body ones, VdW favors the
formation of close packed droplet-like or 3D bulk (crystalline) structures
with high symmetries \cite{Van Der Waals}. If the 2-body interaction - given
by Eq.(\ref{Bead-Bead}) - would dominate the behavior of cVdW as well, one
would also expect the formation of bulk droplets. However this is in sharp
contrast to experimental evidence \cite{Martin1,Martin2,Osterman} which shows
a clear \textit{tendency for chain and membrane formation}, i.e. for
low-dimensional structures under cVdW. What is the microscopic origin of these
complex and low dimensional (anisotropic) structures in cVdW systems?

In the following we will explore how this remarkable difference of the two
forces emerges once the three body forces are considered.

\begin{figure}[pt]
\begin{center}
\includegraphics[
width=3.2in]{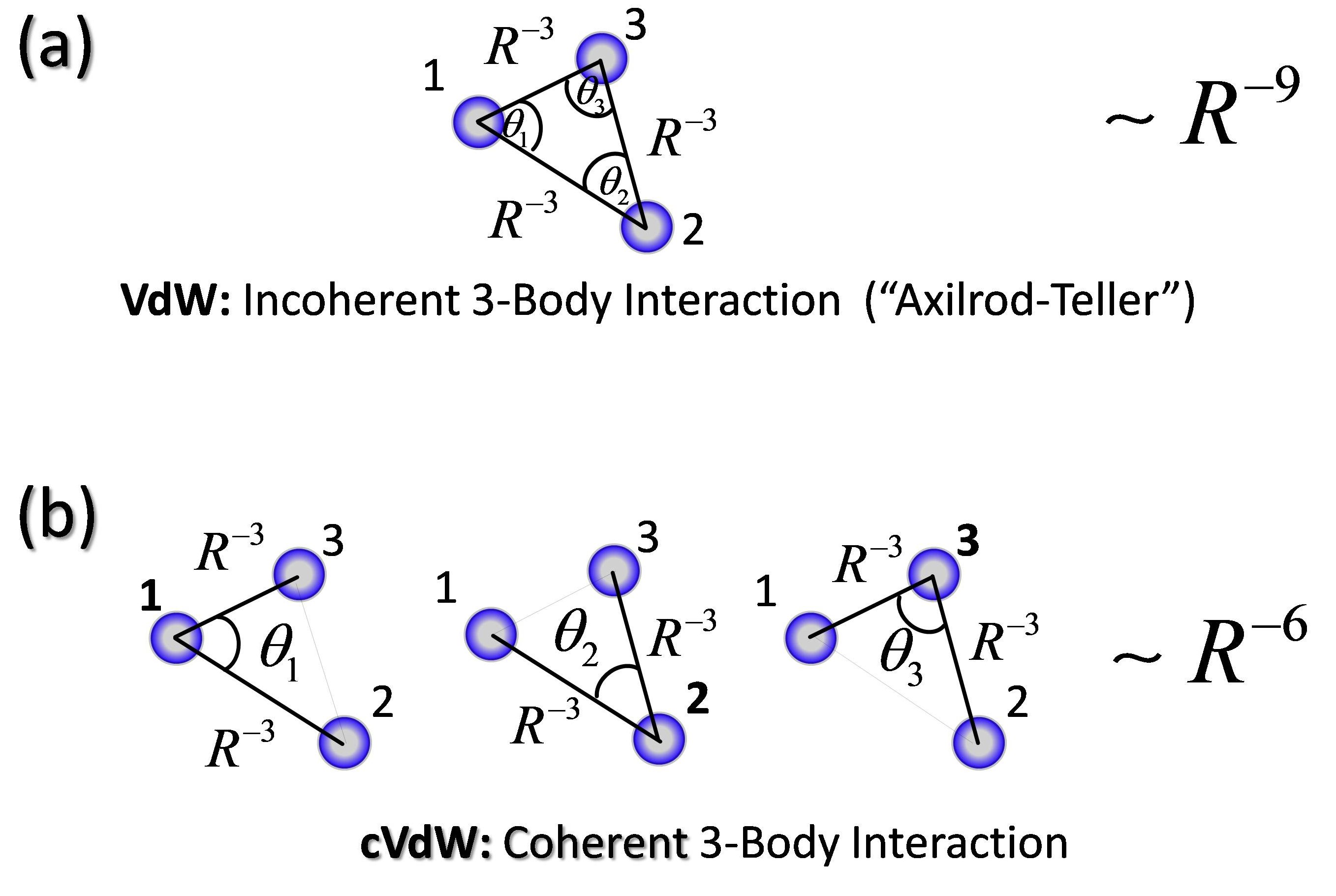}
\end{center}
\caption{The 3-body forces of (a) the incoherent (VdW) and (b) the coherent
(cVdW) interaction have different angular character and are longer ranged for
cVdW, Eq.(\ref{3-body}). }%
\end{figure}

\begin{figure}[pt]
\begin{center}
\includegraphics[
width=3.2in]{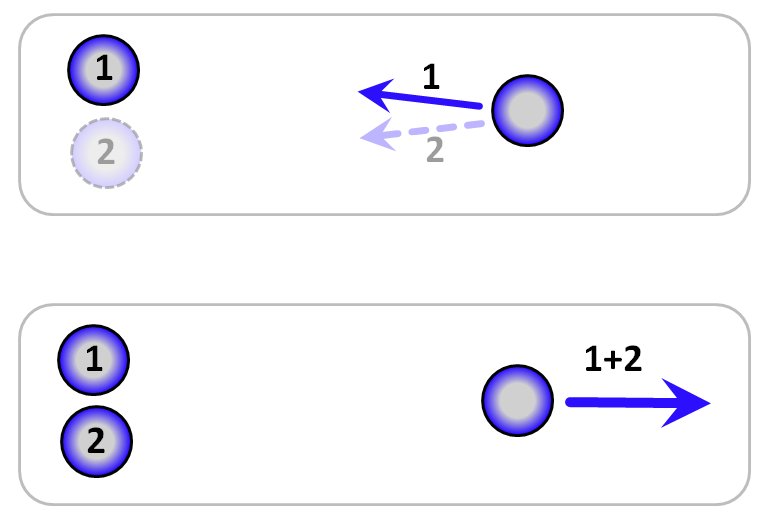}
\end{center}
\caption{Top - The most surprising consequence of the cVdW 3-body nature: Bead
1 and bead 2 individually (if alone) attract the bead 3 on the right. Bottom -
However when placed together beads 1+2 repel the bead 3 due to a strong 3-body
force. }%
\end{figure}

To this end we consider the free energy Eq.(\ref{S5}) with Eq.(\ref{chi-eff})
in the general case of $N\geq3$ particles. We expand the non-local
susceptibility tensor $\chi_{b}(\hat{1}+\chi_{b}\hat{T})^{-1}$ for small bead
susceptibility and large distances $\chi_{b}\varphi_{ij}\ll1$ (i.e. $\chi
_{b}\hat{T}\ll1$) into a Taylor series and take its trace over spacial
directions:
\begin{equation}
Tr\hat{\chi}_{eff}=Tr\left(  \chi_{b}\hat{1}-\chi_{b}^{2}\hat{T}+\chi_{b}%
^{3}\hat{T}^{2}-\chi_{b}^{4}\hat{T}^{3}+...\right)
\label{ChiEffTaylorExpansion}%
\end{equation}

The resulting (scaled) free energy in the coherent case is then $\bar
{f}_{cVdW}\equiv\mathcal{\bar{F}}_{cVdW}/(\mu_{0}V_{b}H_{0}^{2})=-\frac{1}%
{2}\sum_{i,j}Tr\left(  \hat{\chi}_{eff,ij}\right)  $ can be split in terms of
ascending order in $\chi_{b}$ :%
\begin{equation}
\ \bar{f}_{cVdW}=\bar{f}_{cVdW}^{(1)}+\bar{f}_{cVdW}^{(2)}+\bar{f}%
_{cVdW}^{(3)}+O\left(  \chi_{b}^{4}\hat{T}^{3}\right)  .
\end{equation}
with $\bar{f}_{cVdW}^{(k)}\propto\chi_{b}^{k}\sum_{i,j}Tr\left(  \hat{T}%
^{k-1}\right)  _{ij}.$ Similarly we can expand the incoherent VdW energy in
terms like $\bar{f}_{VdW}\propto\chi_{b}^{k}\sum_{i}Tr\left(  \hat{T}%
^{k-1}\right)  _{ii}$ as $\ $%
\begin{equation}
\bar{f}_{VdW}=\bar{f}_{VdW}^{(1)}+\bar{f}_{VdW}^{(2)}+\bar{f}_{VdW}^{(3)}%
+\bar{f}_{VdW}^{(4)}+... \label{f-VdW}%
\end{equation}
Note that for VdW systems, it is necessary to include the term $\bar{f}%
_{VdW}^{(4)}$ since this term contains the leading 3-body interaction
(Axilrod-Teller interaction - see below) in the Van der Waals case.

To shed light on the difference between cVdW and VdW energies, let us have a
closer look at the terms of the expansion Eq.(\ref{ChiEffTaylorExpansion}).
The first term $Tr\left(  \chi_{b}\hat{1}\right)  =3\chi_{b}$ is particle
distance independent and describes noninteracting beads, while the second term
of Eq.(\ref{ChiEffTaylorExpansion}) trivially vanishes as the tensors $\hat
{T}_{ij}$ are traceless: $Tr\hat{T}_{ij}=\varphi_{ij}Tr\left(  \hat{1}%
-3\hat{N}_{ij}\right)  =0.$ Here again \ $\hat{N}_{ij}=\left\vert
\mathbf{b}_{ij}\right\rangle \left\langle \mathbf{b}_{ij}\right\vert $ is the
bond vector projector with $Tr\left(  \hat{N}_{ij}\right)  =1$. It is only the
third term of Eq.(\ref{ChiEffTaylorExpansion}) $\propto Tr\left(  \hat{T}%
^{2}\right)  _{ij}=\sum_{k=1}^{N}\varphi_{ik}\varphi_{kj}Tr\left[  \left(
\hat{1}-3\hat{N}_{ik}\right)  \left(  \hat{1}-3\hat{N}_{kj}\right)  \right]  $
that gives rise to the first non-trivial interaction contribution. Using
$Tr\left[  \hat{N}_{ik}\hat{N}_{kj}\right]  =$ $Tr\left[  \left\vert
\mathbf{b}_{ik}\right\rangle \left\langle \mathbf{b}_{ik}\right\vert
\left\vert \mathbf{b}_{jk}\right\rangle \left\langle \mathbf{b}_{jk}%
\right\vert \right]  $ $=\left(  \left\langle \mathbf{b}_{ik}\right\vert
\left\vert \mathbf{b}_{kj}\right\rangle \right)  ^{2}:=\cos^{2}\theta_{k,ij}$
which involves the angle $\theta_{k,ij}$ between the bond vectors
$\mathbf{b}_{ik}$ and $\mathbf{b}_{jk}$ (at the particle $k$) we obtain:\
\begin{equation}
\left[  Tr\left(  \hat{T}^{2}\right)  \right]  _{ij}=3\sum_{k=1}^{N}%
\varphi_{ik}\varphi_{kj}\left(  3\cos^{2}\theta_{k,ij}-1\right)
\label{TrTSquare}%
\end{equation}
Similarly, using the relation $Tr\left[  \hat{N}_{kl}\hat{N}_{ik}\hat{N}%
_{lj}\right]  =\cos\theta_{kl,ik}\cos\theta_{ik,lj}\cos\theta_{kl,lj}$ \ with
the angles between the bond vectors defined by $\cos\theta_{ik,jl}%
=\left\langle \mathbf{b}_{lj}\right\vert \left\vert \mathbf{b}_{ik}%
\right\rangle $ we can expand also the 4-th term $\propto\hat{T}^{3}$ of
Eq.(\ref{ChiEffTaylorExpansion}) :%

\begin{align}
\left[  Tr\left(  \hat{T}^{3}\right)  \right]  _{ij}  &  =\sum_{k=1}^{N}%
\sum_{l=1}^{N}\varphi_{ik}\varphi_{kl}\varphi_{lj}C_{iklj}\text{
\ with}\label{TrTQube}\\
C_{iklj}  &  =9\left(  \cos^{2}\theta_{ik,jl}+\cos^{2}\theta_{ik,kl}+\cos
^{2}\theta_{jl,lk}\right) \nonumber\\
&  -27\cos\theta_{kl,ik}\cos\theta_{ik,lj}\cos\theta_{kl,lj}-6\nonumber
\end{align}

With these results in our hands we are now well equipped to analyze the 3-body
terms of the two interactions, cVdW and VdW, and understand how they differ.

\subsubsection{3-Body Energy for the Incoherent VdW}

From the $Tr\left(  \hat{T}^{2}\right)  $ term given by Eq.(\ref{TrTSquare})
we obtain the incoherent 2-body interaction $\bar{f}_{VdW}^{(3)}=-3\chi
_{b}^{3}\sum_{i,k}\varphi_{ki}^{2}$. To calculate the lowest order VdW 3-body
term $\bar{f}_{VdW}^{(4)}$ we define in the triangle $(ikl)$ the angles
$\theta_{i},\theta_{k},\theta_{l}$ - cf. Fig.4a, with the properties
$\cos\theta_{i,kl}=-\cos\theta_{i}$, etc. By using the geometrical rule
$\cos^{2}\theta_{i}+\cos^{2}\theta_{k}+\cos^{2}\theta_{l}=1-2\cos\theta
_{i}\cos\theta_{k}\cos\theta_{l}$ for a triangle we obtain%

\begin{align}
\bar{f}_{VdW}^{(4)}  &  =\frac{1}{2}\sum_{i,k,l}\varphi_{ik}\varphi
_{kl}\varphi_{li}C_{ikl}\label{f-VdW-3}\\
C_{ikl}  &  =3(3\cos\theta_{i}\cos\theta_{k}\cos\theta_{l}+1).\nonumber
\end{align}
This term coincides exactly with the \textit{Axilrod-Teller }$3$\textit{-body
potential} \cite{Teller} for the Van der Waals interaction. Due to its weaker
($\sim R^{-9}$) scaling than the 2-body force, it is typically \textit{small
and overridden} by the $2$-body VdW interaction $\sim R^{-6}$
\cite{NOTEAxilrodTeller} giving rise to close packed structures and droplets
in VdW systems.

\subsubsection{The cVdW 3-Body Energy}

In the case of the cVdW interaction the effective free-energy is given by
Eq.(\ref{S5}) where the double summation over $i,j$ must be performed by
including $i=j$, as well . The presence of non-local terms with $i\neq j$
gives rise to qualitatively new many-body effects in cVdW matter with respect
to the VdW one.
As seen from Eq.(\ref{TrTSquare}), the leading 3-body cVdW interaction term
arises from $\bar{f}_{cVdW}^{(3)}$ and is thus proportional to $\hat{T}^{2}$,
so it scales as $\propto\chi_{b}^{3}R^{-6}$. This means that the leading order
3-body cVdW interaction term is of the same order as the 2-body cVdW one
\cite{Note4Body}. The \textit{3-body free-energy} $\bar{f}_{cVdW}^{(3)}$ is
given by%

\begin{equation}
\bar{f}_{cVdW}^{(3)}=-\bar{\beta}\sum\nolimits_{i,j,k}^{\prime}\frac{3\cos
^{2}\theta_{k,ij}-1}{\left\vert \mathbf{R}_{ik}\right\vert ^{3}\left\vert
\mathbf{R}_{kj}\right\vert ^{3}}, \label{3-body}%
\end{equation}
where $\bar{\beta}=(3/32\pi^{2})\chi_{b}^{3}V_{b}^{2}$ and the sum running
over all triplets $\left(  i,j,k\right)  $ with $k\neq i,j$ (for angles
$\theta_{k,ij}$ cf. Fig.4b).

Since the $2$-body interaction in cVdW systems is \textit{contained }in
$\bar{f}_{cVdW}^{(3)}$ (for $k\neq i=j$) it is
physically\textit{\ inseparable} \textit{from the }$3$\textit{-body one}.
Therefore the $3$-body interaction must be treated on the same footing as the
$2$-body one. This fact shows us the pitfall in the dimer formation section
which considered the $2$-body interaction alone. Interestingly, $\bar
{f}_{cVdW}^{(3)}$ in Eq.(\ref{3-body}) is very anisotropic and has a specific
angular dependence. This angular dependence intuitively hints towards the
explanation of the tendency of cVdW interaction not to form 3D bulk
structures, but to drive the formation of anisotropic 1D and 2D structures.
For instance, the $-\cos^{2}\theta_{k,ij}$ term in $\bar{f}_{cVdW}^{(3)}$
favors either $\theta_{k,ij}=0$ or $\pi\ $- i.e. a colloidal chains, membranes
are preferred by the many body interactions.

\subsection{The Principle of "Anisotropic Lumping": An Emergent 2-Body
Interaction from the 3-Body One}

In this section we investigate how the 3-body cVdW interaction works in a
simple physical limit. Let us consider only 3 beads and place two of them, say
$1$ and $2$, very close to each other at distance $R_{12}=r$, while the bead
number $3$ is at large distance from $1$ and $2$, i.e. we assume
$R_{13}\approx R_{23}\approx R\gg r$. The free-energy Eq.(\ref{3-body}) is
calculated in $Appendix$ $2A$ and the expression up to the the lowest order in
the distance is
\begin{equation}
\frac{\mathcal{\bar{F}}_{cVdW}^{(3)}}{\beta}=-\frac{4}{r^{6}}-\frac{4\left(
3\cos^{2}\theta_{3,12}-1\right)  }{r^{3}R^{3}}+O\left(  R^{-6}\right)  ,
\label{F3-R}%
\end{equation}
where the first term is the 2-body interaction between particles $1$ and $2$.
The peculiarity of the second term in Eq.(\ref{F3-R}), i.e. the three-body
interaction is readily seen, when two particles are very near and the third
one is far away. From Eq.(\ref{F3-R}) it comes out that for the bead
arrangement shown in Fig.5, although the two body interactions $1-3$ and $2-3$
are attractive, the third bead is repelled from the first two as a consequence
of the anisotropic 3-body interactions. This is due to the condition $R\gg r$
making $\cos^{2}\theta_{3,12}\ll1$ for $\theta_{3,12}\approx\pi/2$ in the
configuration of Fig. 5 .

The second interesting observation that we can make from Eq.(\ref{F3-R}) is
that for a fixed dimer size $r$, the third particle interacts with the
point-like two particle complex via a "long-range" force $\propto R^{-3}$. The
$1+2$ dimer, instead of the two single monomers becomes now the emerging
elementary unit governed by different laws than single particles.

This example is quite instructive and tells us a lot about the very nature of
cVdW. On the one hand particles display many-body effects which override their
individual pairwise interaction in most general configurations. However there
is another idea emerging from the same example, which simplifies dealing with
the cVdW quite a bit. When two beads come together - i.e. "lump" together by
attractive forces - and find themselves much closer then to the rest of the
beads, they can be considered as a new combined entity- the dimer. Now, the
dimer itself interacts with the rest of the world by an anisotropic, angle
dependent, longer ranged interaction $\propto R^{-3}$ (instead of $R^{-6}$),
and this interaction is now 2-body, pairwise and is attractive for $3\cos
^{2}\theta_{3,12}-1>0$.

The concept of "lumping" works generally, also for more than 3 particles which
are forming anisotropic lumps out of closely packed aggregates of particles.
The usefulness of the lumping idea\ will become more clear in sections that
will follow. In particular, we will derive how beads lumped together inside of
chains or membranes interact with other beads and other lumps in the far
field. The consideration of the pairwise interactions of such lumps, instead
of all many-body interactions of all particles, will be an enormous simplification.

\subsection{Anomalous Elasticity of cVdW Chains and Membranes}

Here we study consequences of the many body effects in cVdW chains and
membranes with finite number ($N$) of particles, - called finite N-effects.
Additionally, we study their unusual elastic properties and compare with more
classical systems with short range forces. In order to grasp the physics of
the problem in the simplest form, we calculate the free-energy for small bead
susceptibility, $\chi_{b}\ll1$ - see Eq.(\ref{3-body}), which includes first
non-trivial and leading $3$-body effects.

\subsubsection{3-Body Effects in cVdW Chain - Free-Energy and Elasticity}

($i$) \textit{Finite linear chain - }The free-energy per particle of the
finite chain $\bar{f}_{cVdW,ch}^{(3),N}(\equiv\mathcal{\bar{F}}_{cVdW,lch}%
^{(3)}/N)$ is calculated by direct summation of 3-body forces in $Appendix$
$2B$ and the result is
\begin{equation}
\frac{\bar{f}_{cVdW,ch}^{(3),N}}{\beta}\approx-\frac{8\zeta^{2}\left(
3\right)  }{D^{6}}+\allowbreak\frac{3.4}{ND^{6}}, \label{lin-chain}%
\end{equation}
with $\beta=(3/32\pi^{2})\mu_{0}H_{0}^{2}\chi_{b}^{3}V_{b}^{3}$ and the zeta
function $\zeta\left(  3\right)  \simeq1.2$. The first $O(1)$ term is the
free-energy per particle of an infinite chain, while the leading finite $N$
term is of order $O\left(  1/N\right)  $. The latter is the price the last few
edge particles (at both free ends) pay for being at the end of the chain. Note
that, this edge energy being positive indicates that the chain would like to
be closed eventually (i.e. elliminate free ends), provided that the bending
energy for doing so is less than the gained edge energy.

($ii$)\textit{ Young's modulus for linear chain} - If one generates a slight
increase of the distance $D$ between bead centers, i.e $D\rightarrow
D(1+\varepsilon)$ then the Young's modulus of the stretched chain can be
formally defined by the second derivative of the free energy like
$Y(\varepsilon=0)=-(1/DS_{b})(\partial^{2}\mathcal{\bar{F}}_{cVdW,ch}%
^{(3)}/\partial\varepsilon^{2})$, where the bead cross-section surface is
$S_{b}\simeq\pi(D/2)^{2}$. From Eq.(\ref{lin-chain}) one obtains
\[
Y(\varepsilon=0)\approx0.84\cdot\mu_{0}H_{0}^{2}\chi_{b}^{3}\text{,}%
\]
i.e. the effective Young's modulus scales quadratically with external field
(for small $\chi_{b}$) and is independent of the bead diameter $D$. For
$\chi_{b}>0$ (paramagnetic beads) it is of course $Y(\varepsilon=0)>0$, which
guarantees stability of the chain. In case of $\chi_{b}<0$ (effective
diamagnetic beads - when the medium susceptibility $\chi_{m}>\chi_{b}$) one
has $Y(\varepsilon=0)<0$ and the chain is unstable due to the repulsive forces
of effectively diamagnetic beads.

($iii$) \textit{Bending energy of ring} - In another situation, instead of
stretching the chain, we can bend its center line and close it into a ring.
The free-energy per particle in this case is calculated in $Appendix$ $2C$ and
reads%
\begin{equation}
\frac{\mathcal{\bar{F}}_{cVdW,ring}^{(3)}}{N\beta}\approx-\frac{8\zeta
^{2}\left(  3\right)  }{D^{6}}+\frac{16\pi^{2}\zeta\left(  3\right)  }{D^{6}%
}\frac{\ln N}{N^{2}}. \label{ring}%
\end{equation}
The first (leading) term is the same as for the infinite linear chain, while
the second term in Eq.(\ref{ring}) can be related to the bending elasticity
energy per particle of the coherent VdW chain. Usually, the bending modulus
(stiffness) $K$ for the chain with short range forces is defined as the
prefactor in the bending energy%
\begin{equation}
\mathcal{\bar{F}}_{bend,ring}-\mathcal{\bar{F}}_{ch}=\frac{K}{2}%
{\displaystyle\int\limits_{0}^{2\pi R}}
ds\left(  \frac{\partial\mathbf{t}}{\partial s}\right)  ^{2}=K\frac{\pi}{R}.
\label{Fbend}%
\end{equation}
where the latter is true for a ring. Here, $\mathbf{t}$ is the unit tangent
vector $\mathbf{t}=(-\cos(s/R),\sin(s/R),0)$ and $R=ND/2\pi$.

By interpreting the cVdW ring in this elasticity framework a first surprise
comes out. From Eq.(\ref{ring}) and Eq.(\ref{Fbend}) we see an anomalous
behavior of the effective bending modulus
\begin{equation}
K_{cVdW}\simeq\frac{\zeta\left(  3\right)  D}{48}(\mu_{0}H_{0}^{2}\chi_{b}%
^{3}V_{b})\ln N, \label{K-cVdW}%
\end{equation}
i.e. $K_{cVdW}\propto\ln N\propto\ln R/D$ grows logarithmically with the chain
size. This behavior is qualitatively very different from the case of the chain
with short range interaction, where $K\ $is always a N-independent constant.
The logarithmic stiffness of chains is caused by the long-range many body
nature of cVdW.

It is interesting to note that even though the bending stiffness is growing,
the chain closure cost (per bead)\ $\propto N^{-2}\ln N$ is becoming quickly
smaller with large N. A comparison of the ring energy Eq.(\ref{ring}) with the
straight chain result Eq.(\ref{lin-chain}) tells us that the rings will indeed
become more preferable for long enough chains with $N\gtrsim320.$

\subsubsection{cVdW Membranes - Free-Energy and Elastic Properties}

As in the case of linear chains we can consider the elastic and finite size
properties of flat and curved membranes. Most of the results in this section
can be derived from the phenomenological macroscopic (demagnetization tensor)
approach for fat cylinders, hollow spheres etc. presented in the forthcoming
sections, while others are obtained by discrete summations. Here we only
present the main physical results and point out the unusual size dependent
scaling of various material properties.

($i$) \textit{The free-energy of the finite flat membrane} - The discrete
summations of the free-energy $\mathcal{\bar{F}}_{cVdW,fl-me}^{(3),N}$
\textit{ }for large $N$ and the radius $R\propto\sqrt{N}$ are difficult due to
absence of a convenient symmetry for all particles (like present for the
ring). However, if we combine the results for the tubular membrane in the
limit $N\rightarrow\infty$ - see Eq.(\ref{F-tube}) and $Appendix$ $2E$, with
the finite $N$ corrections in the cylindrical scheme 2 of the macroscopic
approach - see Eq.(\ref{f2-mac-me}). In this approach we can estimate
$\mathcal{\bar{F}}_{cVdW,fl-me}^{(3),N}$
\begin{equation}
\frac{\mathcal{\bar{F}}_{cVdW,fl-me}^{(3),N}}{N\beta}\approx-\frac{8\pi^{4}%
}{27D^{6}}\allowbreak+\frac{B}{D^{6}}\frac{\ln N}{\sqrt{N}}, \label{fl-mem-N}%
\end{equation}
with $B=192\rho_{pack}^{2}$ . Here, $\rho_{pack}<1$ is the \textit{packing
}(\textit{volume})\textit{ fraction} of the beads in membrane. The second term
in Eq.(\ref{fl-mem-N}) is due to the line tension of the membrane. This line
tension, similarly as the bending stiffness of chains, scales logarithmically
with the system size.

($ii$) \textit{Bending energy of the spherical membrane (vesicle) - }For a
fluid, spherical membrane of radius $R$ with classical elasticity (due to a
finite range interaction) the energy density would be proportional to
$(1/R^{2})$ (=curvature$^{2}$). In that case classic "Helfrich-like" membrane
case the total energy coming solely from the bending is then $\sim\frac
{1}{R^{2}}R^{2}\sim1$, i.e. it is constant for all vesicle sizes. What happens
in the case of a cVdW vesicle? To answer that, we can calculate the 3-body
part of the free-energy per particle with the diameter $D$ of the spherical
shell - see $Appendix$ $2D$, which gives%
\begin{equation}
\frac{\mathcal{\bar{F}}_{cVdW,sph-me}^{(3),N}}{N\beta}\approx-\frac{50\pi
\rho_{sph}^{2}}{D^{6}}\allowbreak+\frac{50\sqrt{\pi}\rho_{sph}^{9/4}}%
{D^{6}\sqrt{N}}, \label{sph-mem-N}%
\end{equation}
where $\rho_{sph}\approx0.4$ is the surface packing factor of the spherical
beads on the sphere, obtained by comparing Eq.(\ref{sph-mem-N}) and
Eq.(\ref{fl-mem-N}). The relation between $\rho_{sph}$, $N$ and the radius of
the shell $R$ is given by $N=4\pi R^{2}/\rho_{sph}^{-1}\pi\left(  D/2\right)
^{2}(=\allowbreak16\rho_{sph}R^{2}/D^{2})$. The free-energy expressed in terms
of $R(\propto\sqrt{N})$ is given by
\begin{equation}
\frac{\mathcal{\bar{F}}_{cVdW,sph-me}^{(3),N}}{\beta}\approx
-const.N\allowbreak+\frac{204\sqrt{\pi}\rho_{sph}^{11/4}}{D^{7}}R,
\label{sph-mem-R}%
\end{equation}
where the first term is the constant energy per particle. The second term is
remarkable as the bending energy grows with radius $\propto R(\propto\sqrt
{N})$. This means that the effective bending energy density is $\propto
R^{-1}(\propto N^{-1/2})$ which is much larger than in classical membranes
where $\propto R^{-2}(\propto N^{-1})$. To put it differently, the bending
stiffness of the membrane $K_{me}$ becomes size dependent with $K_{me}%
\sim\sqrt{N}$. This behavior is also confirmed in the macroscopic theory
(studied below) for the continuous hollow sphere.

After the logarithmic stiffness of the chain (with $K_{ch}\propto\ln N$), the
non-locality of the bending stiffness of membrane is another signature of the
anisotropic and long range 3-body nature of the coherent interaction in cVdW
systems. This is somehow reminiscent, of the physics of classic elastic
cross-linked (i.e. non-fluid) membranes where the non-locality is due to the
coupling of in-plane strains with the flexural deformation \cite{Nelson}.
These effects of the elastic in plane coupling are neglected in our case, as
the curved cVdW membranes can be considered to be a well shaken, i.e. behaving
like a fluid and without in-plane stresses. The size dependent stiffness
effects emerge in our case entirely from the many-body, long-range nature of
the cVdW interaction.

By comparing the free-energies for the flat- and spherical-membrane in
Eq.(\ref{fl-mem-N}) and Eq.(\ref{sph-mem-N}) one sees that for finite, but
large, $N$ the spherical membrane has lower energy than the flat one, i.e.
$\mathcal{\bar{F}}_{cVdW,sph-me}^{(3),N}<\mathcal{\bar{F}}_{cVdW,fl-me}%
^{(3),N}$. This result is also confirmed in the macroscopic approach - see
below. At the first glance this result is not conform with experiments where
only flat membranes were observed \cite{Martin1,Martin2,Osterman}. This can be
explained by the fact that, in order to form a spherical membrane large energy
barriers, far beyond thermal energies ($\mathcal{\bar{F}}_{cVdW,sp-me}%
^{(3),N}\gg kT$), have to be overcome. This might prevent the spherical
membranes from being observed experimentally, so far.

($iii$) \textit{Bending energy of the tubular membrane} - We consider the case
when the thickness of the tube is the bead radius $D$, i.e. $R_{2}-R_{1}=D$.
The approximate free-energy is calculated in $Appendix$ $2E$ . The obtained
free-energy in the leading order is given by
\begin{equation}
\frac{\mathcal{\bar{F}}_{cVdW,tub}^{(3)}}{\beta N}\simeq-\frac{8\pi^{2}}%
{27}\frac{1}{D^{6}}(\pi^{2}-5\frac{D}{R_{\perp}}), \label{F-tube}%
\end{equation}
where $N=N_{1}N_{2}$, $R_{\perp}\sim N_{1}$ is the external radius of the tube
and the limit $(R_{\perp}/D)\gg1$ is assumed - see Fig.11. The first
nontrivial term is proportional to $1/R_{\perp}$ which means that the tubular
membrane\ (cylindrical shell) has a similar type of anomalous and non-local
elasticity (at least in scaling) as the spherical membrane.
This result is also confirmed within the macroscopic approach - see the next Subsection.

\section{Macroscopic Approach to cVdW}

By studying the microscopic $3$-body interaction in Eq.(\ref{3-body}) we have
understood, intuitively and qualitatively, why chains form initially. However,
in order to capture quantitatively their transition to membranes for arbitrary
values of $\chi_{b}<3$, higher $O\left(  \chi_{b}^{4}\varphi_{ij}^{4}\right)
$ terms beyond the 3-body interactions (in Eq.(\ref{3-body})) are necessary.
This appears as a difficult task at present. In order to study the assembly of
magnetic colloids, especially in dense systems, hard and physically much less
transparent numerics would be necessary. Therefore, it is conceptually
instructive to take a more macroscopic, continuous mean-field approach
\cite{Landau}, where \textit{dense chains/membranes} are modelled as a
continuum medium. In this approach the dipolar tensor $\hat{T}_{ij}$ is
replaced by its macroscopic analogue - the demagnetization tensor $\hat{L}$,
while the microscopic effective susceptibility $\hat{\chi}_{eff,ij}$ in
Eq.(\ref{chi-eff}) is replaced in the continuum limit by the corresponding
macroscopic, (shape-dependent) tensor-susceptibility $\hat{\chi}^{(L)}$ given
by \cite{Landau}
\begin{equation}
\hat{\chi}^{(L)}=\chi(1+\hat{L}\chi)^{-1}. \label{chi-L}%
\end{equation}
Here, $\chi$ is the \textit{material susceptibility }(with respect to an
internal field $\mathbf{H}_{int}$) which is due to local field effects in an
aggregate of beads and depends in a nonlinear way on the bead susceptibility
$\chi_{b}$. For $\chi_{b}>0$ one has $\chi>\chi_{b}$. The demagnetization
tensor $\hat{L}$ depends on the shape of the sample, which is chosen in such a
way to mimic the composite structure like e.g. a chain and membrane. The
time-averaged cVdW free-energy $\mathcal{\bar{F}}$ in the macroscopic approach
is generally given by \cite{KulicKulic-PRL}%
\begin{equation}
\mathcal{\bar{F}}^{mac}(\mathbf{H}_{0},\hat{L})=-\frac{1}{2}\mu_{0}H_{0}%
^{2}VTr\left\{  \hat{\chi}^{(L)}\right\}  , \label{F-macro1}%
\end{equation}
together with Eq.(\ref{chi-L}), where the demagnetization coefficients $L_{x}%
$, $L_{y}$,$L_{z}$ - the eigenvalues of $\hat{L}$ - depend on the shape and
aspect ratio of the sample.

As already mentioned the main structures that form on the intermediate scales
are initially chains and then membranes. In the macroscopic approach we can
model them within the framework of two different schemes:

$(1)$ the \textit{spheroid scheme}, where the structure is replaced with a
spheroidal shape with the semi-axes $a=b\neq c$ and the volume $V=(4\pi
/3)a^{2}c$, where we have $c\gg a$ for \textit{prolates (chains)} and $c\ll a$
\ for \textit{oblates (membranes)}, or

($2$) the \textit{cylinder scheme}, where the sample is modelled by a cylinder
with height $h$ and radius $R$. For a \textit{long cylinder} we have $h\gg R$
and $h\ll R$ for a \textit{thin} (flat)\ one.

It will be shown below that for infinite systems both approaches (1) and (2)
give the same results, while for a finite number of particles (the
$N$-effects) they differ. We study also the energetics of the
\textit{spherical membrane (spherical shell - coated sphere)} which in the
large $N$ limit mimics a membrane and its bending stiffness. At the end of
this chapter, we shall also compare the microscopic and macroscopic approach
for various shapes and ask for consistency of the two approaches.

\subsection{Infinite Chains and Membranes}

Here we consider the case $N\rightarrow\infty$ and obtain asymptotic results
for large (infinite) membranes and chains.

($i$) \textit{Chain} - The two perpendicular demagnetization factors of
infininte chains are given by $L_{a,\infty}=L_{b,\infty}=1/2$, while for the
one along the long chain axis we have $L_{c,\infty}=0$ in both schemes
(cylinder and ellipsoid). In that case Eq.(\ref{F-macro1}) gives the
macroscopic approximation of the free-energy $\mathcal{\bar{F}}_{ch,\infty
}^{mac}$ of chain
\begin{equation}
\mathcal{\bar{F}}_{ch,\infty}^{mac}(\mathbf{H}_{0})=-\frac{1}{2}\mu_{0}%
H_{0}^{2}V\chi(1+\frac{2}{1+\chi/2}) \label{Fchain}%
\end{equation}

($ii$) \textit{Membrane} - In both schemes in the limit $N\rightarrow\infty$
one has $L_{a,\infty}=L_{b,\infty}=0$, $L_{c,\infty}=1$ and the asymptotic
\textit{free-energy of the membrane} is
\begin{equation}
\mathcal{\bar{F}}_{me,\infty}^{mac}(\mathbf{H}_{0})=-\frac{1}{2}\mu_{0}%
H_{0}^{2}V\chi(2+\frac{1}{1+\chi}) \label{Fmem}%
\end{equation}

Note, that for both chains and membranes the macroscopic free-energy is
dominated by the smallest demagnetization factors $L^{m/ch}\rightarrow0$. For
membranes two demagnetization factors vanish, while for chains only one
vanishes. Then, for $N\rightarrow\infty$ and by assuming the same $\chi$ for
chains and membranes, the free-energy of the membrane is always smaller than
that of the chain for any $\chi$, i.e. $\mathcal{\bar{F}}_{me,\infty}%
^{mac}<\mathcal{\bar{F}}_{ch,\infty}^{mac}$. \ 

The macroscopic approach is well in agreement with the experiments
\cite{Martin1,Martin2,Osterman} which show that the formation of 3D spherical
droplets is unfavorable. Namely, from Eq.(\ref{Fmac-exp}) (further below) it
is seen that for a \textit{spherical droplet,} where $L_{x}=L_{y}=L_{z}=1/3$,
the free-energy is given by $\mathcal{\bar{F}}_{drop}^{mac}(\mathbf{H}%
_{0},\hat{L})\approx-(3/2)\mu_{0}H_{0}^{2}V\chi/(1+\chi/3)$. This means that
$\mathcal{\bar{F}}_{me,\infty}^{mac}<\mathcal{\bar{F}}_{ch,\infty}%
^{mac}<\mathcal{\bar{F}}_{drop}^{mac}$, i.e. also in the macroscopic approach
the formation of chains and membranes (once they become large) is more
favorable than the creation of spherical droplets.

\subsection{Finite Chains and Membranes}

For finite $N$ the demagnetization factors $L_{i}$ depend on the aspect ratio
$a/c$, i.e. on $N$. The chain is characterized by the long semi-axis $c\sim
ND$ ($D$ the bead size), and the short semi-axis$\ a\sim D\ll c$ , with the
corresponding demagnetization factors $L_{a},L_{b},L_{c}$ ($L_{a}+L_{b}%
+L_{c}=1$) \cite{Landau}, \cite{1.Beleggia, 4.Beleggia, 2.Beleggia}. The
\textit{chain-membrane transition} in the macroscopic approach is reached on
the critical line $N_{c}(\chi)$, where $\mathcal{\bar{F}}_{me}^{mac}%
(N_{c})=\mathcal{\bar{F}}_{ch}^{mac}(N_{c})$. To demonstrate the existence of
such a transition the critical line was calculated for the spheroid scheme 1
by assuming, for simplicity, that the chain and membrane material
susceptibilities are the same \cite{KulicKulic-PRL}. It was found that the
critical cluster size $N_{c}$ grows with the material susceptibility $\chi$.
For $\chi\approx1-3$, it was estimated $N_{c}\approx10-20$. Such a tendency is
also observed in experiments \cite{Osterman}, where for $N_{c}\approx10$
initial signs of the chain-membrane transition are found. In these
experiments, chains are formed for small $N\sim10$ in the very dilute limit,
while a further addition of colloids results in a branching of chains (via the
so called $Y$-junctions), further followed by a network of inter-connections
and finally dense membrane patches are formed. The conclusion is that the
chain-membrane transition in the macroscopic approach is qualitatively in
accord with experiments.

Let us calculate the finite $N$-effects in the macroscopic approach and
compare it with the microscopic theory given in Section IV. To remind the
reader, in the \textit{microscopic approach} the free-energy in Section IV is
calculated for small $\chi_{b}$ by expanding it up to $\chi_{b}^{3}$, i.e. it
is given by $\mathcal{\bar{F}}_{cVdW}^{micro}\simeq\mathcal{\bar{F}}%
_{cVdW}^{(1)}+\mathcal{\bar{F}}_{cVdW}^{(3)}$ (since $\mathcal{\bar{F}}%
_{cVdW}^{(2)}=0$) where $\mathcal{\bar{F}}_{cVdW}^{(1)}\propto N\chi_{b}$
describes $N$ non-interacting beads, while the first nontrivial term due to
the dipole-dipole interaction in Eq.(\ref{3-body}) is $\mathcal{\bar{F}%
}_{cVdW}^{(3)}\propto\chi_{b}^{3}$ . As already discussed, the term
$\mathcal{\bar{F}}_{cVdW}^{(3)}$ is due to both, 2-body and 3-body interactions.

In order to compare these two approaches we need to expand the free-energy in
the macroscopic approach as a function of $\chi_{b}$ up to $\chi_{b}^{3}$, as
well. In the macroscopic approach, described by Eq.(\ref{chi-L}%
)-Eq.(\ref{Fmem}), the free-energy depends on the material susceptibility
$\chi$ which is related to the bead susceptibility $\chi_{b}$ and in systems
with small bead susceptibility ($\chi_{b}$) $\chi$ is also small. Therefore,
in order to make the expansion of the macroscopic free-energy
Eq.(\ref{F-macro1}) with respect to $\chi_{b}$ a relation between $\chi$ and
$\chi_{b}$ is necessary. In fact $\chi$ is related to the averaged bead
susceptibility $\tilde{\chi}_{b}=\rho_{pack}\chi_{b}$, where $\rho_{pack}(<1)$
is the \textit{packing }(\textit{volume})\textit{ fraction} of the beads in a
given assembly. (Note that the total volume of the bead $V_{tot}=NV_{b}$ is
related to the macroscopic volume $V$ by $V_{tot}=\rho_{pack}V$.) By assuming
high local symmetry around each bead within the assembled structures it is
easy to show that in that case the Lorenz-Lorenz relation $\chi=\tilde{\chi
}_{b}(1-(\tilde{\chi}_{b}/3))^{-1}$ holds. By making an expansion of $\chi$ up
to $\tilde{\chi}_{b}^{3}$ and using that $L_{x}=L_{y}$ and $L_{x}=(1-L_{z})/2$
the \textit{macroscopic free-energy} in Eq.(\ref{F-macro1}) reads
\begin{equation}
\frac{\mathcal{\bar{F}}^{mac}(\mathbf{H}_{0},\hat{L})}{(\mu_{0}H_{0}^{2}%
V/2)}\approx-3[\tilde{\chi}_{b}+\frac{\tilde{\chi}_{b}^{3}}{2}(L_{z}-\frac
{1}{3})^{2}]. \label{Fmac-exp}%
\end{equation}
Note, that in the case of the chain one has $L_{z}\rightarrow0$ for
$N\rightarrow\infty$, while for the membrane one has $L_{z}\rightarrow1$ for
$N\rightarrow\infty$ - see below.

\subsubsection{Linear Chain}

As mentioned before in the case of finite $N$ the demagnetization factor
$L_{z}(N)$ is different in the spheroid scheme 1 and the cylindrical scheme 2.
Let us have a look at the difference \cite{Landau, 1.Beleggia, 4.Beleggia,
2.Beleggia}:

(i) the \textit{spheroid scheme 1} - In this case the chain corresponds to a
prolate spheroid and the exact demagnetization factor $L_{z}$ of a prolate
spheroid is given in $Appendix$ $3A$ Eq.(\ref{chain-prol}) with $\tau_{a}=N$
one has $L_{z}\approx\left(  \ln N\right)  /N^{2}$ and the dimensionless
free-energy $f_{ch,N}^{mac,1}$($\equiv\mathcal{\bar{F}}_{ch,N}^{mac,1}%
(\mathbf{H}_{0},\hat{L})/(\mu_{0}H_{0}^{2}NV_{b}/2)$) given by:
\begin{equation}
f_{ch,N}^{mac,1}\approx f_{ch,\infty}^{mac}+\rho_{pack}^{2}\chi_{b}^{3}%
\frac{\ln N}{N^{2}}, \label{f1-ch-N}%
\end{equation}
where the energy per bead of the infinite chain is
\begin{equation}
f_{ch,\infty}^{mac}\approx-(3\chi_{b}+\frac{\rho_{pack}^{2}\chi_{b}^{3}}{6}).
\label{f-ch-inf}%
\end{equation}

(ii) the \textit{cylinder} \textit{scheme} 2 - In the cylinder scheme 2 a
chain corresponds to a long cylinder with the aspect ratio $\tau
\equiv(h/2R)\approx2N/3\gg1$, $h$ is the hight (along the c-axis) and $R$ the
radius of cylinder, one has $L_{z}\approx2N/\pi$ - see in $Appendix$ $3A$
Eq.(\ref{A3-Lz-cyl-l}) and the corresponding free-energy $f_{ch,N}^{mac,2}%
$($\equiv\mathcal{\bar{F}}_{ch,N}^{mac,2}(\mathbf{H}_{0},\hat{L})/(\mu
_{0}H_{0}^{2}NV_{b}/2)$) reads%
\begin{equation}
f_{ch,N}^{mac,2}\approx f_{ch,\infty}^{mac}+\frac{2\rho_{pack}^{2}\chi_{b}%
^{3}}{\pi N}. \label{f2-ch-N}%
\end{equation}
By comparing Eq.(\ref{f1-ch-N}) and Eq.(\ref{f2-ch-N}) of the macroscopic
approach with the corresponding microscopic free-energy for the finite linear
chain in Eq.(\ref{lin-chain}), it turns out that the linear chain is slightly
better described by the long cylinder in the cylinder scheme 2.

\subsubsection{Flat Membrane}

(i) the\textit{ spheroid scheme }1 - In that case the flat membrane
corresponds to extreme oblate spheroids with $L_{z}\approx1-\pi/2\sqrt{N}$ -
see Eq.(\ref{A3-Lz-sp-0}) in $Appendix$ $3A$ and the dimensionless free-energy
$f_{me,N}^{mac,1}$($\equiv\mathcal{\bar{F}}_{me,N}^{mac,1}(\mathbf{H}_{0}%
,\hat{L})/(\mu_{0}H_{0}^{2}NV_{b}/2)$) is given by%
\begin{equation}
f_{me,N}^{mac,1}\approx f_{me,\infty}^{mac}+\frac{\pi\rho_{pack}^{2}\chi
_{b}^{3}}{\sqrt{N}}, \label{f1-mac-me}%
\end{equation}
where the free-energy per bead of the infinite flat membrane is%
\begin{equation}
f_{me,\infty}^{mac}\approx-(3\chi_{b}+\frac{2\rho_{pack}^{2}\chi_{b}^{3}}{3}).
\label{f-mac-me}%
\end{equation}

(ii) \textit{the cylinder scheme }2 \textit{- }In this case a thin cylinder
with aspect ratio $\tau\approx1/\sqrt{N}\ll1$, one has $L_{z}\approx1-\ln
N/\pi\sqrt{N}$ which gives the dimensionless free-energy $f_{me,N}^{mac,2}%
$($\equiv\mathcal{\bar{F}}_{me,N}^{mac,2}(\mathbf{H}_{0},\hat{L})/(\mu
_{0}H_{0}^{2}NV_{b}/2)$ )%

\begin{equation}
f_{me,N}^{mac,2}\approx f_{me,\infty}^{mac}+\rho_{pack}^{2}\chi_{b}^{3}%
\frac{\ln N}{\sqrt{N}} \label{f2-mac-me}%
\end{equation}
By comparison with the microscopic energy in Eq.(35) one expects that the
cylinder scheme 2 mimics the membrane better than the spheroid scheme 1.

\subsection{Spherical Shell Membrane}

In the microscopic approach we calculated the energy of a closed, monolayered
spherical membrane (spherical shell) under the action of cVdW interaction. It
is interesting to study its energetics in the macroscopic approach where the
outer and inner radius of the spherical shell are $R_{o}$, $R_{i}$,
respectively, while the relative magnetic permeability in the shell is
$\mu_{shell}=1+\chi$. Outside this shell we assume $\mu_{out}=1$. In this case
the symmetry implies that Eq.(\ref{F-macro1}) is simplified to
\begin{equation}
\mathcal{\bar{F}}_{cVdW,shell}^{mac}(\mathbf{H}_{0},\hat{L})=-\frac{3}{2}%
\mu_{0}H_{0}^{2}\alpha_{shell}, \label{Fshell}%
\end{equation}
where $\alpha_{shell}\equiv V\chi_{shell}$ can be calculated in the
magneto-static limit by using standard boundary conditions \cite{Jones}%
\begin{equation}
\alpha_{shell}=4\pi R_{o}^{3}\frac{\chi(3+2\chi)(1-\phi)}{(\chi+3)(3+2\chi
)-2\phi\chi^{2}}. \label{alpha-shell}%
\end{equation}
with $\phi=(R_{i}/R_{o})^{3}$ where $R_{i}$ is the inner and $R_{o}$ \ is the
outer radius of the spherical shell. For a small shell thickness $D\ll
R_{o},R_{i}$ one has $R_{o}^{3}-R_{i}^{3}\approx N(D/2)^{3}$. For the outer
surface with $4\pi R_{o}^{2}\approx\pi N(D/2)^{2}$ one has $\phi
=1-N(D/2R_{o})^{3}=1-8/\sqrt{N}$. For a small bead susceptibility $\chi_{b}<1$
the material susceptibility is given by $\chi\approx\tilde{\chi}_{b}%
(1+(\tilde{\chi}_{b}^{2}/3)+(\tilde{\chi}_{b}^{3}/9))$. By making an expansion
for large $N$ it is straightforward to obtain $\alpha_{shell}$ and
$\mathcal{\bar{F}}_{shell}^{mac}$ in Eq.(\ref{Fshell})%
\begin{equation}
f_{shell,\infty}^{mac}=f_{me,\infty}^{mac}+\frac{16}{3}\frac{\varrho
_{pack}^{2}\chi_{b}^{3}}{\sqrt{N}}. \label{Fshell-N}%
\end{equation}
The first term characterizes the infinite flat membrane, while the second one
$\propto(16\varrho_{pack}^{2}\chi_{b}^{3}/3)N^{-1/2}$ is asymptotically
smaller than that for the flat membrane $\propto(\varrho_{pack}^{2}\chi
_{b}^{3})N^{-1/2}\ln N$ in Eq.(\ref{Fmac-exp}. This means that the free-energy
of the spherical membrane (with some large but finite $N$) becomes overall
smaller than the microscopic free-energy of the finite flat membrane. This
confirms our previous analysis that a large spherical membrane is slightly
more favorable than the flat one.

\subsection{Consistency of Microscopic and Macroscopic cVdW Theory}

Let us check if the macro- and microscopic theory agree in concrete cases. For
that purpose we compare the corresponding interacting parts of the
free-energy, i.e. $\Delta f_{\infty}^{mac}$($=f_{\infty}^{mac}-\frac{3}{2}%
\chi_{b}$) and $\Delta f_{\infty}^{mic}$, for chains and membranes. In the
case of the chain one has $V=\rho_{pack}V_{0}$ and $\Delta f_{ch,\infty}%
^{mac}\approx-\chi_{b}^{3}\rho_{pack}^{2}/6$, while the first term of the
microscopic free-energy in Eq.(\ref{lin-chain}) is $\Delta f_{ch,\infty}%
^{mic}\approx-\varsigma^{2}(3)\chi_{b}^{3}H_{0}^{2}/24$. The equality $\Delta
f_{ch,\infty}^{mac}=\Delta f_{ch,\infty}^{mic}$ gives the consistency
condition for the bead packing fraction $\rho_{pack}=0.5\varsigma
(3)\approx0.6$ - indeed a rather plausible and realistic value for the packing
density. A similar situation holds for membranes, where Eq.(\ref{Fmac-exp})
gives $\Delta f_{me,\infty}^{mic}\approx-2\chi_{b}^{3}\rho_{pack}^{2}/3$. The
microscopic free-energy of the flat membrane can be obtained, for instance,
from Eq.(\ref{F-tube}) for the tubular membrane in the limit $R_{\perp
}\rightarrow\infty$, i.e. $\Delta f_{me,\infty}^{mic}=-(\pi^{4}/162)\chi
_{b}^{3}$. The condition $\Delta f_{me,\infty}^{mac}=\Delta f_{me,\infty
}^{mic}$ gives the packing fraction $\rho_{pack}\approx0.9$ which is a
reasonable value.

The good news overall is, that for several types of assemblies we confirm a
qualitative and a satisfactory quantitative agreement of the macroscopic
approach with the microscopic one in cases considered. This justifies the more
coarse grained but simpler macroscopic approach in studying several cVdW structures.

\subsection{Instability of The Spherical Droplet}

We have seen that the many body interactions in cVdW systems favor the
formation of large membrane structures. The latter behavior could be
interpreted as coming from an effective negative "surface tension". It appears
that cVdW tends to "flatten out" every aggregate into a thin monolayered sheet
down to the smallest cut-off length (the constituent bead size)- as if a
negative effective surface tension was at work.

In this $Section$ we show on a concrete example that the cVdW assemblies
indeed behave as systems with negative effective surface tension
$\sigma_{cVdW}<0$. Although this quasi-surface tension is in reality a rather
complex and anisotropic \textit{bulk} term (and depends on the shape of
assemblies) it generates effects which are very reminiscent of a real surface tension.

\begin{figure}[pt]
\begin{center}
\includegraphics[
width=3.2in]{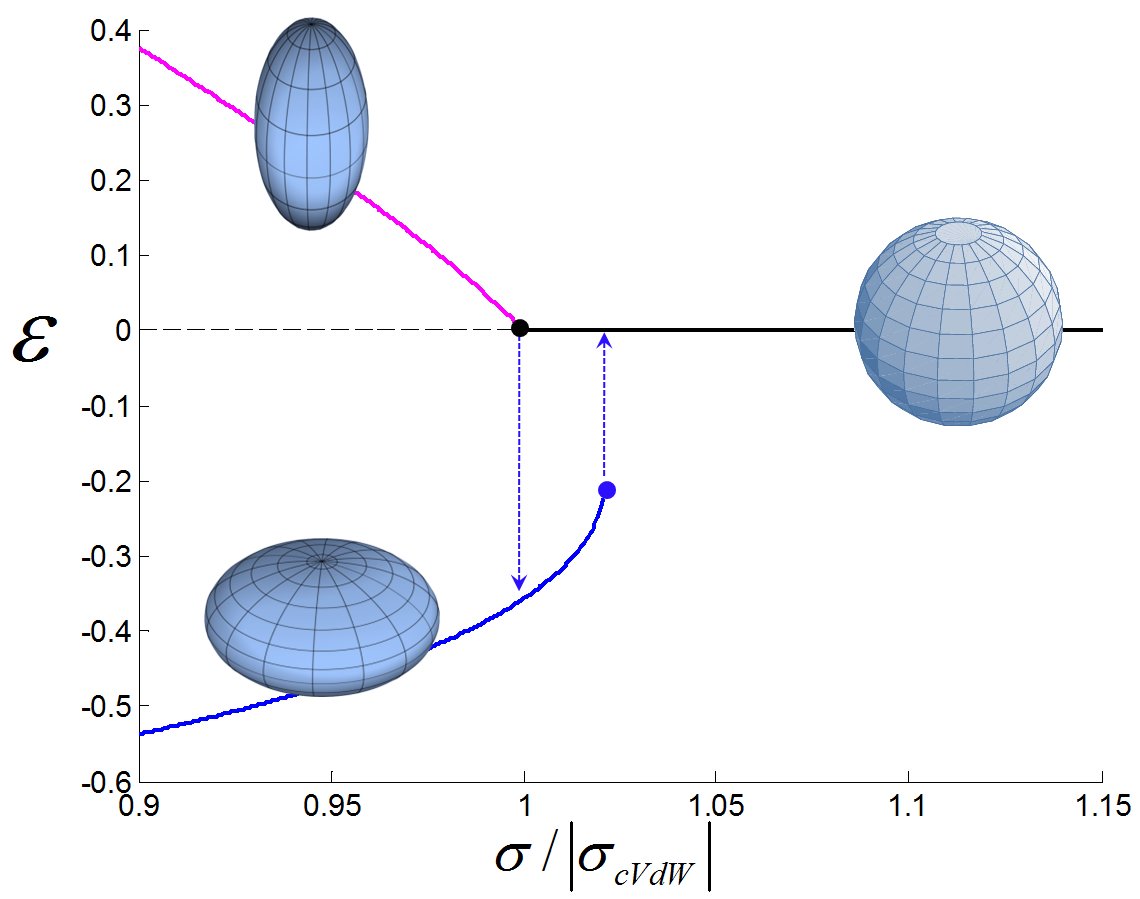}
\end{center}
\caption{ cVdW induces buckling of a fluid droplet as function of the scaled
surface tension. $\varepsilon$ is the eccentricity of the ellipsoid with
$\varepsilon<0$ - oblate ellipsoid; $\varepsilon>0$ - prolate ellipsoid}%
\end{figure}

To crystallize out the physics, in the following we will let the cVdW
interaction directly compete with an additional real (positive) fluid surface
tension. Concretely, we consider a cVdW system made of paramagnetic beads
which are embedded in a spherical droplet of the solvent$,$ say oil or water,
with the surface tension $\sigma$. Now, the question is : What is the critical
surface tension below which the spherical droplet becomes unstable and starts
forming an oblate ($\tau_{s}<1$) or prolate ($\tau_{s}>1$) spheroid? Here
$\tau_{s}=c/a$ is the aspect ratio of the spheroid.

For shapes very close to the sphere, one can expand the total free-energy in
terms of the small $z$-axial stretch $\varepsilon=\tau_{s}-1\ll1$, where
$\varepsilon<0$ represents an oblate ellipsoid and $\varepsilon>0$ means a
prolate one. By using Eq.(\ref{A3-Lz-sp}) given in $Appendix$ $3B$ one obtains
the $z$-axis demagnetization factor $L_{z}$. The results for $L_{z}$ holds in
both cases the oblate spheroid and prolate one, and we furthermore have
$L_{x}=L_{y}=(1-L_{z})/2$. By using the latter property the dimensionless
free-energy in Eq.(\ref{Fmac-exp}) takes the form%

\begin{equation}
\frac{\mathcal{\bar{F}}_{cVdW}-\mathcal{\bar{F}}_{0}}{E_{0}}=-\left(  \frac
{8}{75}\varepsilon^{2}-\frac{24}{175}\varepsilon^{3}+\frac{1382}%
{11\,025}\varepsilon^{4}\right)  +O\left(  \varepsilon^{5}\right)  ,\nonumber
\end{equation}
with $E_{0}=\mu_{0}\tilde{\chi}_{b}^{3}VH_{0}^{2}/2$ the convenient energy
scale and $\mathcal{\bar{F}}_{0}$ the non-interacting (self-) free-energy of
all beads.

Introducing also the surface energy term $\sigma A$ to the total energy we
have then%
\begin{equation}
\mathcal{\bar{F}}_{tot}=\mathcal{\bar{F}}_{cVdW}+\sigma A
\end{equation}
where the surface area of spheroid $A$ depends on the aspect ratio $\tau_{s}$
and is given in $Appendix$ $3B$. A short calculation then gives%
\begin{align}
\frac{\mathcal{\bar{F}}_{tot}}{E_{0}}  &  =-\left(  \frac{8}{75}%
\varepsilon^{2}-\frac{24}{175}\varepsilon^{3}+\frac{1382}{11\,025}%
\varepsilon^{4}\right) \label{Ftot}\\
&  +\frac{\sigma A_{0}}{E_{0}}\left(  \frac{8}{45}\varepsilon^{2}-\frac
{584}{2835}\varepsilon^{3}+\frac{118}{567}\varepsilon^{4}\right)  .\nonumber
\end{align}
We see that the sphere with radius R is only stable when the $\varepsilon^{2}$
term is positive which implies%

\begin{align}
\frac{\sigma A_{0}}{E_{0}}  &  >\allowbreak\frac{3}{5}\equiv\frac{\sigma
_{cr}A_{0}}{E_{0}}\label{sigma-cr}\\
\sigma_{crit}  &  =\frac{1}{5}\mu_{0}\tilde{\chi}_{b}^{3}H^{2}R,\text{
\ }\nonumber
\end{align}
where $\mu_{0}\tilde{\chi}_{b}^{3}H^{2}R/5$ may be considered as an effective
cVdW surface tension $\sigma_{cVdW},$ which has however a negative sign and
counteracts (reduces) the actual surface tension of the surrounding liquid.

The phase diagram for the cVdW spherical droplet is shown in Fig.6. Notably
for a subcritical fluid surface tension $\sigma<\sigma_{crit}$ the droplet can
be either prolate or oblate. Which branch is actually chosen might subtly
depend on the dynamics and history of the shape. However, from energetic point
of view, the absolute energy minimum in the subcritical regime is reached for
the flatter i.e. oblate shape (lower $\varepsilon$ stable branch). The
elongated prolate ellipsoid forms only a shallow local minimum and is
therefore thermodynamically only metastable.

The effective surface tension $\sigma_{cVdW}\left(  R\right)  $ is in reality
a many-body bulk term and thus size dependent. Let us estimate on which scale
it becomes relevant, for instance, in a drop of water, with surface tension
$\sigma_{H_{2}O}=\allowbreak0.073\,J/m^{2}$ (at room temperature). For
$\chi_{b}\sim1$ and $B=\mu_{0}H_{0}=0.01T$, the cVdW surface tension is
$\sigma_{crit}\approx-B^{2}R/5\mu_{0}=\allowbreak-15.\,9(J/m^{3})R$. For this
to be of the same order as $\sigma_{H_{2}O}$ we need the radius of the droplet
to be $R_{c}=5\sigma_{H_{2}O}/(\mu_{0}\chi_{b}^{3}H^{2})=4.\,\allowbreak
6\times10^{-3}m=4mm$. For larger droplets made of paramagnetic beads with
$R>R_{c}$ the spherical shape becomes unstable.

In the dielectric analogue of this phenomenon in cVdW systems we would have
$\sigma_{crit}=\varepsilon_{0}\kappa_{el}^{3}E^{2}R/5$. Let us assume that the
dielectric droplet has much smaller dielectric susceptibility than of the
electric beads ($\kappa_{el,m}\ll\kappa_{el}$) and that the surface tension is
of the order as that of water. Then for a feasible electric field $E=10V/mm$,
$\ \varepsilon_{0}=$ $8.8\ast10^{-12}J/mV^{2}$ , and for $\kappa_{el}^{3}%
\sim1$ we have $R_{el}=5\sigma/(\varepsilon_{0}\kappa_{el}^{3}E^{2}%
)\approx\allowbreak500m$ ! This means that this effect is less favorable in
the electric case. However, for magnetic colloids (such as ferrofluids) under
magnetically induced cVdW, the "negative surface tension" instability effect
should be easily observable.

\section{Formation of Superstructures in cVdW Systems}

In magnetically driven cVdW systems chains and membranes are the predominant
structures formed on intermediate length and timescales
\cite{Martin1,Martin2,Osterman}. However, the experiments also show that in
more dense colloid systems which are placed in containers of finite volume,
more complex structures like foams are formed on larger scales. The existence
of these foam structures is also confirmed in numerical simulations
\cite{Martin1,Martin2}. The basic motif underlying such foams, is a complex
network of interconnected membrane patches, which apart from touching along
their edges do not stack and aggregate. Instead the membranes seem, at least
by visual inspection, to repel each other and the whole foam structure appears
to swell against gravity. What is the origin of such large scale cVdW foam structures?

In this Section we give a plausible physical explanation by combining both the
microscopic and macroscopic approaches. First, we will study the cVdW
interaction free-energy of two flat membranes. We show below that this
interaction switches from an attraction to a repulsion, depending on the
mutual orientation.

While in general, the interactions between two membranes can have both signs,
it turns out that in the majority of possible configurations the interaction
is in fact repulsive on the average. Then, we calculate the free-energy of the
foam structure by modelling it by a cubic shelf structure ansatz. We show that
such a structure indeed tends to swell and at the end we derive something that
reassembles an equation of state of a cVdW foam, i.e. a
pressure-concentration-field relation. We show that the magnitude of this
pressure is quite notable and can indeed lead to a rise of the foam to
measurable heights.

\subsection{cVdW Interaction Generalization to Anisotropic Objects}

In the previous sections we were concerned with interactions of isotropic
spherical particles whose susceptibility tensors were merely diagonal
$(\hat{\chi}_{b})_{\alpha\beta}=\chi_{b}\delta_{\alpha\beta}$ \ i.e. simply a
number. Here we generalize the interaction to any two arbitrary shaped bodies.
In general this is a complicated problem when the bodies are very close.
However, when they are far enough, say much further than their typical body
extensions, the field of any object can be replaced with a corresponding
effective ellipsoid field. In this sense it is sufficient to consider the
interaction of two ellipsoids, with orientation dependent and non-trivial
susceptibility tensors $\hat{\chi}_{1},\hat{\chi}_{2}$. The free-energy in
this case can be rewritten in the form
\begin{equation}
\mathcal{\bar{F}}_{cVdW}\left(  \mathbf{H}_{0},\{\mathbf{R}_{i}\}\right)
=-\frac{\mu_{0}H_{0}^{2}}{2}Tr(V_{1}\hat{\chi}_{1,eff}+V_{2}\hat{\chi}%
_{2,eff}), \label{M1}%
\end{equation}
where $\hat{\chi}_{1,eff}$ and $\hat{\chi}_{2,eff}$ are now effective
susceptibility tensors of the two bodies (ellipsoids) with respective volumes
$V_{1/2}$. The effective susceptibility tensors are now given by the
expression
\begin{equation}
\hat{\chi}_{1,eff}=(1-\varphi_{12}^{2}\hat{\chi}_{1}\hat{t}_{12}\hat{\chi}%
_{2}\hat{t}_{12})^{-1}(\hat{\chi}_{1}-\varphi_{12}\hat{\chi}_{1}\hat{t}%
_{12}\hat{\chi}_{2}) \label{M2}%
\end{equation}
and same for the second $\hat{\chi}_{2,eff}$ which is obtained by replacing
$1\rightarrow2$. The slightly more intricate form of $\hat{\chi}_{1/2,eff},$
which is obviously a generalization of the corresponding isotropic expression
Eq.(\ref{chi-eff}), \ comes now from the fact that the operator $\hat{t}%
_{12}=\hat{1}-3\left\vert \mathbf{b}_{12}\right\rangle \left\langle
\mathbf{b}_{12}\right\vert $ and the susceptibilities $\hat{\chi}_{i}$ are
operators with spacial orientations which don't commute now any more in general.

The expressions in Eq.(\ref{M1} -\ref{M2}) are general and contain the
distance dependence through the scalar factor $\varphi_{12}\propto1/\left\vert
\mathbf{R}_{12}\right\vert ^{3}$ in a slightly scrambled form that hides the
leading order scaling. Thus, it is interesting to expand $\hat{\chi}%
_{1/2,eff}$ \ to the first order w.r.t. $\varphi_{12}\ $and obtain the leading
order interaction part%

\begin{equation}
\mathcal{\bar{F}}_{cVdW,inter}(1,2)=\frac{\mu_{0}}{2}H_{0}^{2}V\varphi
_{12}Tr\{(\hat{\chi}_{1}\hat{\chi}_{2}+\hat{\chi}_{2}\hat{\chi}_{1})\hat
{t}_{12}\}. \label{M3}%
\end{equation}
Here we omit trivial self-energies, i.e. consider the interaction energy term
only, and assume the two bodies to have the same volume $V$. Note that, when
the objects are isotropic (e.g. spheres, point like) then the $\hat{\chi
}_{1/2}$ turn simply into numbers. Reminding ourselves that $\hat{t}_{12}$ is
traceless we see that the whole term $\propto\varphi_{12}$ vanishes in the
isotropic case. This is why the cVdW interaction for two spheres only starts
out with a higher order leading $\propto\varphi_{12}^{2}\propto1/\left\vert
\mathbf{R}_{12}\right\vert ^{6}$ interaction term. However in general, for
anisotropic objects the trace in Eq.(\ref{M3}) is non-zero, giving rise to a
strong and long range interaction $\propto$ $1/R^{3}$. That is, anisotropic
objects interact much stronger than isotropic ones under cVdW. Once growing
aggregates become shape anisotropic and notably they always tend to do so
(forming chains and membranes), they interact in a long range manner. This can
be seen as another manifestation of the cooperative many-body nature of cVdW.

\subsection{Interaction of Membrane and Single Spherical Bead}

\begin{figure}[pt]
\begin{center}
\includegraphics[
width=3.2in]{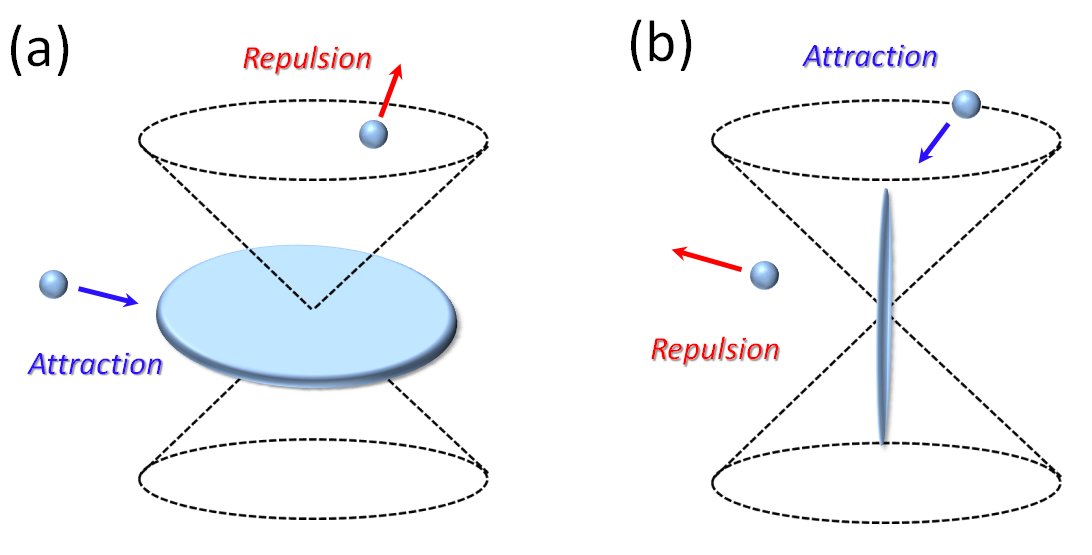}
\end{center}
\caption{The cVdW interaction for a bead interacting with (a) an oblate
membrane (cf. Eq.(\ref{Wmb})) and (b) a prolate chain.}%
\end{figure}

To understand the content of the anisotropic cVdW interaction from
Eq.(\ref{M3}), let us have a look at how a flat membrane interacts with a
single spherical bead. The effects of beads approaching chains and membranes
were previously studied numerically by Osterman et al. \cite{Osterman}.
Interestingly, it turns out, that even in this simplest case the interaction
can vary in sign. It is attractive when the membrane is approached by the bead
from the edge side. When however approaching from the top, in the direction of
the membrane normal,\ the bead is repelled in the far field. If the bead
approaches the membrane even further (against the repulsive force) and comes
closer in this normal direction, the interaction switches again to a short
range attraction. Obviously there is a barrier to cross in this normal
direction. The largest barrier for joining of the bead to the membrane is when
the former is above the center of the membrane and we study this case first -
see Fig.(7a). By applying Eq.(\ref{S5}) on two particles, where the first one
is big and anisotropic - it mimics a membrane, while the second one is small
and isotropic - it mimics a bead, one obtains the free-energy in the form%
\begin{align}
W^{m-b}  &  =-\frac{1}{2}\mu_{0}H_{0}^{2}V_{0}(q_{m}Tr\hat{\chi}_{m,eff}%
+q_{b}Tr\hat{\chi}_{b,eff})\label{Wmb}\\
&  =W^{m}+W^{b}%
\end{align}
Here, $V_{0}=V_{m}+V_{b}$, $q_{m,b}=V_{m,b}/V_{0}$ and $V_{m}$, $V_{b}$ is the
volume of the membrane and bead, respectively, while $\hat{\chi}_{m,eff}$ and
$\hat{\chi}_{b,eff}$ the effective susceptibilities of membrane and bead
respectively with%

\begin{equation}
\hat{\chi}_{m,eff}=\frac{1}{(1-q_{m}q_{b}\varphi_{0}^{2}\hat{\chi}_{m}\hat
{t}\hat{\chi}_{b}\hat{t})}\hat{\chi}_{m}(1-q_{m}\varphi_{0}\hat{t}\hat{\chi
}_{b}). \label{chi-m}%
\end{equation}
and $\hat{\chi}_{b,eff}$ is obtained by replacing $m\leftrightarrow b$ in
Eq.(\ref{chi-m}) and $\varphi_{0}=V_{0}/4\pi R_{mb}^{3}$. In this approach the
membrane is replaced by an oblate spheroid with the susceptibility $\hat{\chi
}_{m}=\chi_{\min}\left\vert \mathbf{n}_{m}\right\rangle \left\langle
\mathbf{n}_{m}\right\vert +\chi_{\max}(1-\left\vert \mathbf{n}_{m}%
\right\rangle \left\langle \mathbf{n}_{m}\right\vert )$, while for the
spherical bead we have $\hat{\chi}_{b}=\chi_{b}\hat{1}$; $\mathbf{n}_{m}$ is
the unit vector normal to the membrane. The tensor $\hat{t}$ in
Eq.(\ref{chi-m}) is $\hat{t}(\equiv\hat{t}_{m-b})=\hat{1}-3\left\vert
\mathbf{b}_{m-b}\right\rangle \left\langle \mathbf{b}_{m-b}\right\vert $.
Since we assume that the membrane's dimensions are much larger than the bead
size $D$ and the bead is above the membrane one has $\mathbf{n}_{m}\parallel$
$\mathbf{b}_{m-b}$, i.e. $\hat{t}_{m-b}=$ $1-3\left\vert \mathbf{n}%
_{m}\right\rangle \left\langle \mathbf{n}_{m}\right\vert $.

The final expression for the dimensionless free-energy $w^{m-b}=W^{m-b}%
/(\frac{1}{2}\mu_{0}H_{0}^{2}V_{0})=w^{m}+w^{b}$ is given by (cf. $Appendix$
$4A$ )
\begin{equation}
w^{m}=-\frac{q_{m}\chi_{\max}}{1-b}\{2(1-\alpha_{b})+\frac{\chi_{\min}}%
{\chi_{\max}}\frac{1+2\alpha_{b}}{1+c}\} \label{wm}%
\end{equation}%
\begin{equation}
w^{b}=-\frac{q_{b}\chi_{\max}}{1-\alpha_{m}\alpha_{b}}\{\frac{3+2c}%
{1+c}+2\frac{\chi_{\min}}{\chi_{\max}}\frac{\alpha_{m}}{1+c}-2\alpha_{m}\},
\label{wb}%
\end{equation}
where $\alpha_{b}=q_{b}\chi_{b}\psi_{0}$, $\alpha_{m}=q_{m}\chi_{\max}\psi
_{0}$; $c=(\alpha_{m}\alpha_{b}(\chi_{\max}-4\chi_{\min}))/\chi_{\max
}(1-\alpha_{m}\alpha_{b}))$. Let us discuss the energy of the membrane-bead
complex as a function of their distance $r=R_{m-b}/D$ by assuming (for
simplicity) that $\chi_{\min}\ll\chi_{\max}$ i.e. a very flat membrane. Since
$q_{b}\ll q_{m}\simeq1$ one has $\alpha_{m}\alpha_{b}\ll1$ and the expression
simplifies to
\begin{equation}
\frac{w^{m-b}}{2\chi_{\max}}\simeq-1+\frac{\chi_{b}}{24}(\frac{2}{r^{3}}%
-\frac{\chi_{\max}}{24q_{b}}\frac{1}{r^{6}}). \label{w-mb}%
\end{equation}
From Eq.(\ref{w-mb}) we see that there is an energy barrier for the bead, i.e.
for $R_{m-b}>R_{c}=D(\chi_{\max}/24q_{b})^{1/3}$ the membrane repels the bead
since $F_{m-b}=-(\partial W^{m-b}/\partial R_{m-b})>0$, while for
$R_{m-b}<R_{c}$ the force is attractive ($F_{m-b}<0$). As an example we take
$\chi_{\max}\sim10$ and $q_{b}>10^{-3}$ which gives us the barrier at a
notable distance $R_{c}>10D$ - larger then the bead size. Based on the same
formalism it is straightforward to show that when the bead is placed in the
plane containing the membrane, i.e. when $\mathbf{n}_{m}\perp$ $\mathbf{b}%
_{m-b}$ holds, it is always attracted to the membrane, i.e. $F_{m-b}<0$.

To conclude, the above considerations show that the most favorable and fastest
membrane growth pathway is addition of beads along the membrane edges in
membrane's plane. Those beads found above the membrane must move parallel to
the membrane and finally descend toward the ends of membranes. This is
schematically shown in Fig.7a for a bead interacting with a membrane and in a
Fig.7b for a bead interacting with a chain (prolate ellipsoid). In the latter
case calculations go along the same lines as for membranes, with the only
difference being in flipping the signs of interaction. The beads are repelled
laterally and attracted along the symmetry axis of the chains. The easy
calculation being very similar as for membranes is omitted here.

\subsection{Interaction of Two Membranes}

Once they emerge, what is the fate of the membranes as they continue growing?
At some point the membranes will run out of free beads in the solution and
start interacting only with the remaining aggregates which turn into membranes
once they are large enough. To understand, how two membranes mutually order,
we need the\textit{\ 2-membrane interaction} for arbitrary membrane
orientations $\mathbf{n}_{1,2}$ and anisotropic susceptibilities $\hat{\chi
}_{i}^{(L)}=\chi(1+\hat{L}_{i}\chi)^{-1}$. For identical membranes and at
large distances ($\varphi_{12}\ll1$, i.e. for $\left\vert \mathbf{R}%
_{12}\right\vert \gg V_{m}^{1/3}$) one expands $\hat{\chi}_{eff,12}\approx
\hat{\chi}_{1}^{(L)}(1-\varphi_{12}$ $\hat{t}(\mathbf{b}_{12})\hat{\chi
}_{\mathbf{2}}^{(L)})$ and the long range interaction energy in
Eq.(\ref{chi-eff}) after a short calculation reads (see details in $Appendix$
$4B$),%

\begin{equation}
\frac{\mathcal{\bar{F}}_{int}}{\alpha}=\frac{C_{1}^{2}+C_{2}^{2}%
+\frac{1-\gamma}{3}C_{3}^{2}-(1-\gamma)C_{1}C_{2}C_{3}-\frac{2}{3}}{\left\vert
\mathbf{R}_{12}\right\vert ^{3}} \label{me-me}%
\end{equation}
with $\alpha=3(1-\gamma)\chi_{\max}^{2}\mu_{0}H_{0}^{2}V_{m}^{2}/16\pi,$ and
$\gamma=\chi_{\min}/\chi_{\max}$ the ratio of the minimal/maximal eigenvalue
of the membrane susceptibility tensor $\hat{\chi}^{(L)}$.

The dimensionless factors $C_{1}=\mathbf{n}_{1}\cdot\mathbf{b}_{12}$,
$C_{2}=\mathbf{n}_{2}\cdot\mathbf{b}_{12}$, $C_{3}=\mathbf{n}_{1}%
\cdot\mathbf{n}_{2}$ reveal all the geometrical beauty of cVdW: the 2-membrane
interaction is angle dependent and repulsive in many configurations - see
Fig.8. Notably, for a fixed $\left\vert \mathbf{R}_{12}\right\vert $,
$\mathcal{\bar{F}}_{int}$ becomes minimal for the orthogonally
\textit{twisted} membrane orientation with $\mathbf{n}_{1}\perp\mathbf{n}_{2}%
$, $\mathbf{n}_{1}\perp\mathbf{b}_{12}$ and $\mathbf{n}_{2}\perp
\mathbf{b}_{12}$ ($C_{1/2/3}=0$). The twisted membranes attract each other
since $\mathcal{\bar{F}}_{int}^{\left(  tw\right)  }<0$ (up to the point of
mutual contact), as in the \textit{coplanar} case, yet the \textit{twisted}
configuration has lower energy. This interesting result should affect the
kinetics of membrane formation: If two distant membranes start growing within
a large distance they will rotate to a $90^{\circ}$ position before touching.
Therefore, some type of glassy state in their orientation may be kinetically
favored. In other relevant configurations, such as the \textit{top}, with two
out of plane parallel membranes ($C_{1/2/3}=1$) or the \textit{generic} one
(cf. Fig.8), the interaction is repulsive with $0<\mathcal{\bar{F}}%
_{int}^{(gen)}<\mathcal{\bar{F}}_{int}^{(top)}$.

\begin{figure}[pt]
\begin{center}
\includegraphics[
width=3.2in]{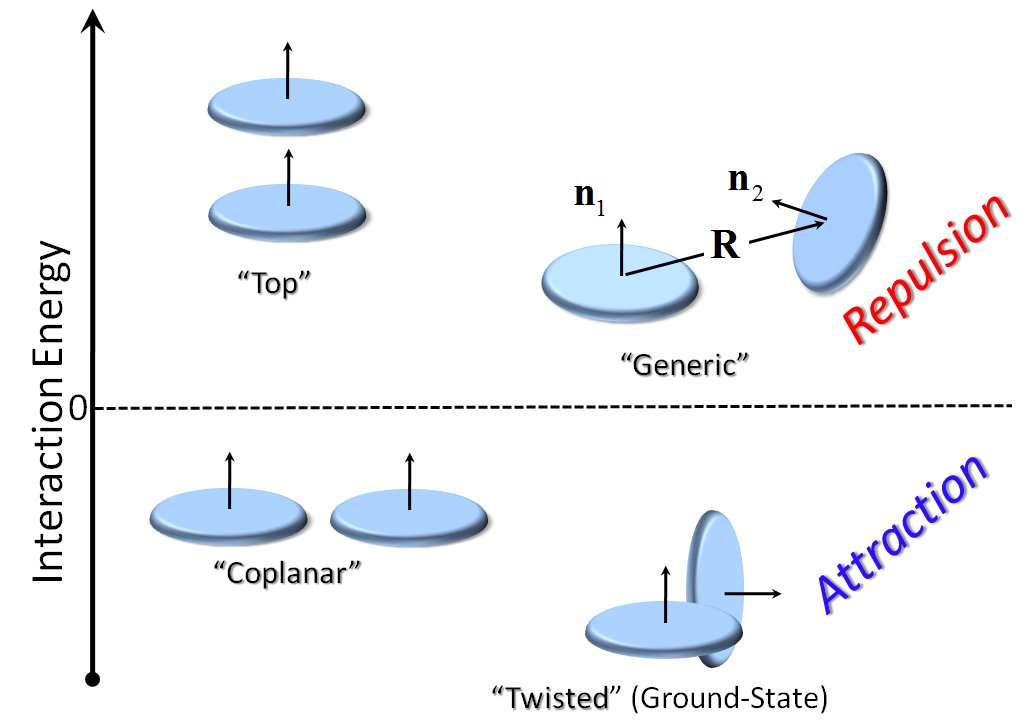}
\end{center}
\caption{The cVdW for two interacting membranes is unexpectedly complex. The
2-membrane interaction is attractive or repulsive depending on orientation,
cf. Eq.\ref{me-me} with a ground state in the "twisted" configuration.}%
\end{figure}

\subsection{Emergence of Foams}

Simulations and experiments \cite{Martin1,Martin2} provide some empirical
evidence for the existence of a hollow foam-like superstructure forming on
large scales (cf. Fig 9a). What is the physical mechanism driving such a cVdW
foam formation?

We have seen above that large aggregates prefer to form membranes, and that
these membranes mutually interact. Specifically, when two distant membranes
are stacked over each other they repel each other ($\mathcal{\bar{F}}%
_{int}^{(top)}>0$). In the opposite limit - in close contact distance- a
simple estimate implies their preference to split as well. Namely, when a
thick membrane, with the thickness $2D$, radius $R$ and volume $2V_{m}$, is
cut into two parallel membranes, with the thickness $D$ and radius $R$ each
and separated to infinite distance there is a gain in the energy
$\Delta\mathcal{F}=2\mathcal{F}_{1m}-\mathcal{F}_{2m}\approx-V_{2m}L\chi
^{2}(1-(1+\chi)^{-2})<0$ for $L\chi\ll1$, where $L\propto D^{3/2}V_{2m}%
^{-1/2}$. Physically this means that the second membrane lying above the first
one is repelled to increase the local fields with respect to the thicker
membrane case.

It is this remarkable reluctance of membranes to mutually stack that in fact
sets the microscopic structure of the foam: It is formed out of the thinnest
possible membrane patches, whose thickness is collapsed onto the smallest
available physical scale - the bead size $D$. The characteristic lateral size
$a_{M}$ of these membrane patches, on the other hand, is set by the bead
volume fraction in the container $f_{V}=V_{b}^{tot}/V\ll1$ (with $V_{b}%
^{tot}=NV_{b}$ the\ total volume of all beads and $V$ the container's volume).
By assuming a cubic shelf structure as an ansatz, cf. Fig 9b, one obtains a
patch size $a_{M}\approx3D/f_{V}$.

In order to calculate the pressure in such a foam structure we need the total
interaction free-energy of all membranes in the system. It turns out that the
interaction part of the free-energy ($\mathcal{\bar{F}}_{int}$) of the assumed
cubic shelf structure is positive ($\mathcal{\bar{F}}_{int}>0$) due to global
average repulsion of membranes - see $Appendix$ $4C$.

\section{Equation of State of cVdW Foams}

In the previous Section we have argued qualitatively that in the cVdW foam
structure the positive interaction free-energy should favor an effective
repulsion between constituent membranes forming this structure, i.e. that the
foam should exert a pressure on walls of the container and in fact swell. In
this Section we calculate this pressure as a function of the volume fraction
of magnetic beads, i.e. we derive the equation of state for a cVdW material.

\begin{figure}[pt]
\begin{center}
\includegraphics[
width=3.2in]{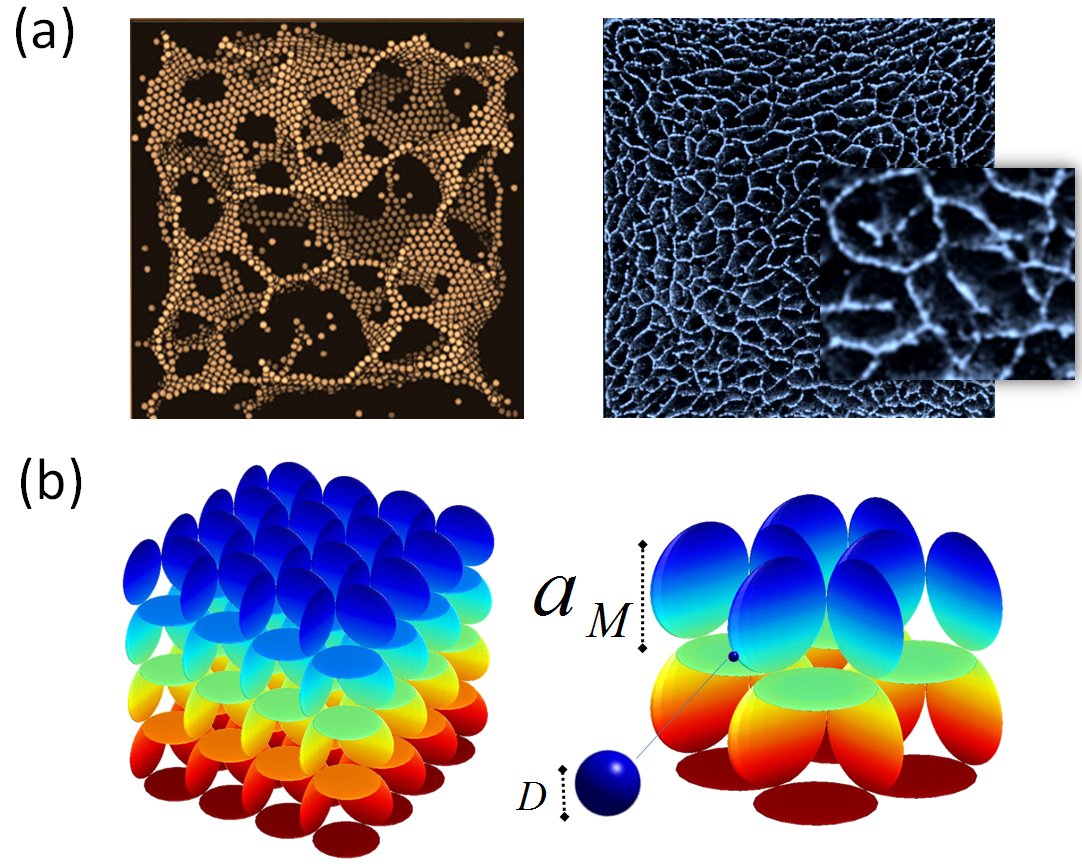}
\end{center}
\caption{(a) The large scale structure of a dipolar cVdW foam (from
\cite{Martin1,Martin2}) in experiment (right) and in simulation (left). (b)
The theoretical 3D shelf-model for cVdW foam's structure.}%
\end{figure}

As above we define the volume fraction of all beads in the container
$f_{V}=(N_{m}V_{m}/V)\approx V_{b}^{tot}/V$ where $V_{m}(\approx Da_{M}^{2})$
is the volume of the single membrane and $N_{m}$ is the total number of
(equal) membranes in the container volume $V\approx Na_{M}^{3}$, and
$V_{b}^{tot}$ is the total volume of the beads. Here, $D$ is the bead diameter
and $a_{M}$ is the size of the single membrane - see Fig.9b. It follows that
$f_{V}\approx3D/a_{M}$.

In the following we fix the total volume of all $N_{m}$ membranes $V_{m}%
^{tot}=N_{m}V_{m}$ and vary the size of the container $V$. The pressure is
then defined by $p=-\partial\mathcal{\bar{F}}^{tot}/\partial V$ where
$\mathcal{\bar{F}}^{tot}\mathcal{=\bar{F}}_{self}^{tot}\mathcal{+\bar{F}%
}_{int}^{tot}$ is the total energy of the membranes. $\mathcal{\bar{F}}%
_{self}^{tot}$ is the self-energy of (non-interacting) membranes and
$\mathcal{\bar{F}}_{int}^{tot}$ is the interaction energy of membranes. From
Eq.(\ref{F-macro1}) the free-energy of the $N_{m}$ single membranes (the
self-energy) with the total volume $V_{m}^{tot}=N_{m}V_{m}$ is given by
\begin{equation}
\mathcal{\bar{F}}_{self}^{tot}=-(2\chi_{\max}+\chi_{\min})V_{m}^{tot}%
\frac{B_{0}^{2}}{2\mu_{0}}, \label{p1}%
\end{equation}%
\begin{equation}
\chi_{\max}=\frac{\chi}{1+L_{m}\chi},\text{ }\chi_{\min}=\frac{\chi
}{1+(1-2L_{m})\chi}. \label{p2}%
\end{equation}
For simplicity, we study here only the case with large material susceptibility
$\chi>1$ (note that, the material susceptibility fulfills $\chi>\chi_{b}\leq
3$, where $\chi_{b}$ is the bead susceptibility with respect to the applied
(external) field) and at the same time $L_{m}\chi\ll1$ ($L_{m}\ll1$).

In the following we approximate, for simplicity, the membranes by oblate
spheroids. The demagnetization factor of the flat membranes in the plane
direction can be related with the membrane aspect ratio, which itself is set
by the volume fraction $L_{m}\approx f_{V}/4$ (for $f_{V}\ll1$). After a
straightforward expansion with respect to small $L_{m}$ one obtains%
\begin{equation}
\mathcal{\bar{F}}_{self}^{tot}(V)\approx-(Const-\frac{1}{2}f_{V}\chi^{2}%
)V_{m}^{tot}\frac{B_{0}^{2}}{2\mu_{0}}, \label{p3}%
\end{equation}
where $Const$ is independent of the volume fraction $f_{V}$ .

The \textit{total interaction energy} of membranes $\mathcal{\bar{F}}%
_{int}^{tot}$ is on the other hand%
\begin{equation}
\mathcal{\bar{F}}_{int}^{tot}(V)=\frac{1}{2}\sum_{i,j}\mathcal{\bar{F}}%
_{int}(i,j), \label{p4}%
\end{equation}
where the pair-interaction energy $\mathcal{\bar{F}}_{int}(i,j)$ is given by
Eq.(\ref{me-me}) where the summation goes over all membranes in the container.
Note, that for the nearest neighbor membranes with $\left\vert \mathbf{R}%
_{12}\right\vert \approx a_{M}$ the far field approximation Eq.(\ref{me-me})
holds qualitatively only, while for the next-nearest neighbors it holds
already quantitatively. For $\chi\gg1$ and $L_{m}\chi\ll1$ Eq.(\ref{p4})
gives
\begin{equation}
\mathcal{\bar{F}}_{int}^{tot}(V)\approx(\frac{1}{8\pi}f_{V}\chi^{2}%
S)V_{m}^{tot}\frac{B_{0}^{2}}{2\mu_{0}}, \label{p5}%
\end{equation}
where the explicit expression for the sum $S(\equiv a_{M}^{3}\mathcal{\bar{F}%
}_{int}/\alpha)\approx10$ - a numeric dimensionless constant - is calculated
by explicitly summing over all the pairwise membrane-membrane interactions
(given by Eq.(\ref{me-me})) in the cubic shelf lattice , for details cf.
$Appendix$ $4C$.

Finally, combining both contributions to the free energy (self-energy and
total interactions), the total pressure of the foam in a container with the
volume $V$ is given by
\begin{equation}
p=-\frac{\partial\mathcal{\bar{F}}^{tot}}{\partial V}\approx(\frac{1}{2}%
+\frac{S}{8\pi})\chi^{2}f_{V}^{2}(\frac{B_{0}^{2}}{2\mu_{0}}). \label{p7}%
\end{equation}
As a result the foam's pressure is given by the approximate expression
\begin{equation}
p\approx\frac{1}{2}\mu_{0}\chi^{2}f_{V}^{2}H_{0}^{2}. \label{pressure-1}%
\end{equation}
Interestingly, this pressure can assume notable magnitudes in practice. For
moderate volume fractions, reasonable fields and susceptibilities
($f_{V}\approx5\cdot10^{-2},$ $\mu_{0}H_{0}\approx20mT,$ and $\chi\approx10$
in densely packed $Ni$-beads membranes) we obtain $p\approx40$ $Pa$. Since
$p\propto H_{0}^{2}$, the pressure is very sensitive to the strength of the
excitation (field) $H_{0}$ and can lead to strong swelling of the foam against
gravity. The latter effect is also observed experimentally \cite{Jim-swelling}
and can be used to practically test the equation of state Eq.(\ref{pressure-1}).

\subsection{Gravitational Pressure of the Foam}

Since a real foam is formed in the gravitation field, the gravity can limit
its swelling. As we see from Eq.(\ref{pressure-1}) the foam's pressure $p$ is
proportional to $f_{V}^{2}$ and in the gravitational field both are dependent
on the vertical height position $h$ along the gravity direction. If one
assumes that at $h=0$ the volume fraction takes the value $f_{V,0}$ and the
pressure $p_{0}$ then (in case of constant $f_{V}$ and $p$) the foam would
grow up to the maximal hight $h_{\max}^{0}=p_{0}/\Delta\rho gf_{V,0}$, where
$g\simeq10m/s^{2},$ is the gravitational acceleration and $\Delta\rho
=\rho_{bead}-\rho_{s}$ is the difference in densities of magnetic beads and
solvent. For instance, for water immersed $Ni$-beads as in Refs.
\cite{Martin1,Martin2} one has $\Delta\rho\simeq8\cdot10^{3}kg/m^{3}$. For
$f_{V,0}\approx5\cdot10^{-2},$ $\mu_{0}H_{0}\approx20mT,$ and $\chi\approx10$
in densely packed $Ni$-beads membranes one obtains the pressure $p^{(1)}%
\approx40$ $Pa$ and the equilibrium foam height $h$ is reached once the
internal and the gravitational pressure balance, i.e. $p\approx\Delta\rho
gf_{V,0}h_{\max}^{0}$ and the foam will swell strongly up to $h_{\max}^{0}%
\sim1$ $cm$.

The variation of pressure $p\left(  h\right)  $ and the volume fraction
$f_{V}\left(  h\right)  $ with the hight in the gravitational field changes
this approximative analysis slightly. In the gravitational field one has
\begin{equation}
\frac{dp}{dh}=-\Delta\rho gf_{V}. \label{dp-dh}%
\end{equation}
By using the equation of state in Eq.(\ref{pressure-1}) - with $f_{V}%
=C\cdot\sqrt{p}$, the solution of Eq.(\ref{dp-dh}) reads
\begin{align}
p\left(  h\right)   &  =p_{0}(1-\frac{h}{2h_{\max}})\label{p1-f-h}\\
f_{V}\left(  h\right)   &  =f_{V,0}(1-\frac{h}{2h_{\max}}).\nonumber
\end{align}
The the maximal height is reached when $p=0$, i.e. when $h_{\max}%
^{(1)}=2h_{\max}^{0}$. For the above parameters one obtains $h_{\max}%
^{(1)}\sim2$ $cm$.

Therefore, the strong swelling behavior of magnetic foams can be used as a
sensitive test of the theory.

\section{Summary and Discussion}

We have studied the formation of hierarchical superstructures in systems
driven by the spatially coherent Van der Waals (cVdW) interaction. We have
developed a fairly general formalism involving the effective susceptibility
tensor which allowed us to walk through all the important aspects of the cVdW
interaction. Within this setting, in a bottom up approach we investigated
numerous phenomena, from dimer formation, over 3 body forces, then collective
elasticity of intermediate structures (chains and membranes) up to the
presumably highest scale of pattern formation, i.e. to the cVdW foams.

In the theory we took a bird's view approach, and we have shown that the cVdW
interaction can be equivalently created in many types of excitation fields,
generalizing the triaxial balanced fields used in the past. It turned out that
the consideration of a general square isotropic uniform field (rather then any
particular realization of it), brings the cVdW and its classical incoherent
VdW "sister-"interaction onto a common footing. This parallel consideration of
cVdW- and VdW-matter allowed us also to crystallize out the common behavior,
but more importantly the central differences between the two types of forces
behind them.

The most remarkable difference is found in the 3-body interactions. For the
standard VdW matter the 3-body forces are recovered in the fully incoherent
limit of our formalism and they agree with the classic result of Axilrod and
Teller \cite{Teller}. These VdW 3-body forces are much weaker and shorter
ranged than the corresponding 2-body forces, i.e. one could say they are
\textit{subdominant }and give only higher order corrections. In sharp
contrast, in the cVdW-matter the 3-body forces derived here are as strong and
often even stronger than the pairwise 2-body ones. Thus, the 3-body effects
under the cVdW interaction can be considered as \textit{essential} and
\textit{dominant} forces in the system. To our knowledge, this "many body
dominance" makes the cVdW force stand out among other known interactions and
gives it a very unique, interesting character. We have studied the physical
origin of these dominant cVdW many body forces and we found them originating
from the fact that the direct (induced) dipole-dipole interactions between
isotropic objects vanish (are averaged out) and only the many- body mutual
polarization effects survive the statistical averaging over the external
excitation fields.

The pronounced \textit{anisotropy} of the many-body interactions in the
cVdW-matter systems also gives rise to a number of phenomena that are
qualitatively different from standard VdW-matter, in particular the growth of
anisotropic, low-dimensional assemblies - chains, then membranes once a
critical size is reached. In a container of finite size, smaller membrane
patches are formed, which tend to repel on the average, thus giving rise to
dipolar foam structures. The foam exerts a positive pressure onto the walls of
the sample container due to the tendency of membranes to increase their
surface areas as well as their mutual repulsion. The dipolar foam represents a
new and intriguing state of colloidal matter, formed by a delicate interplay
of an attractive local interaction and a net repulsive longer range force.
Remarkably, both types of forces are born out of a single, conceptually simple
cVdW interaction - given by Eq.(\ref{S5}).

The interactions driving the hierarchy of the assembly processes, from dimers
to foams are summarized in Figs.10 and 11, where the 2-body and anisotropic
3-body interactions are responsible for the formation of chains, membranes and
vesicles, while the membrane-membrane interaction is responsible for the
formation of foams in a container with finite volume.

\begin{figure}[pt]
\begin{center}
\includegraphics[
width=3.2in]{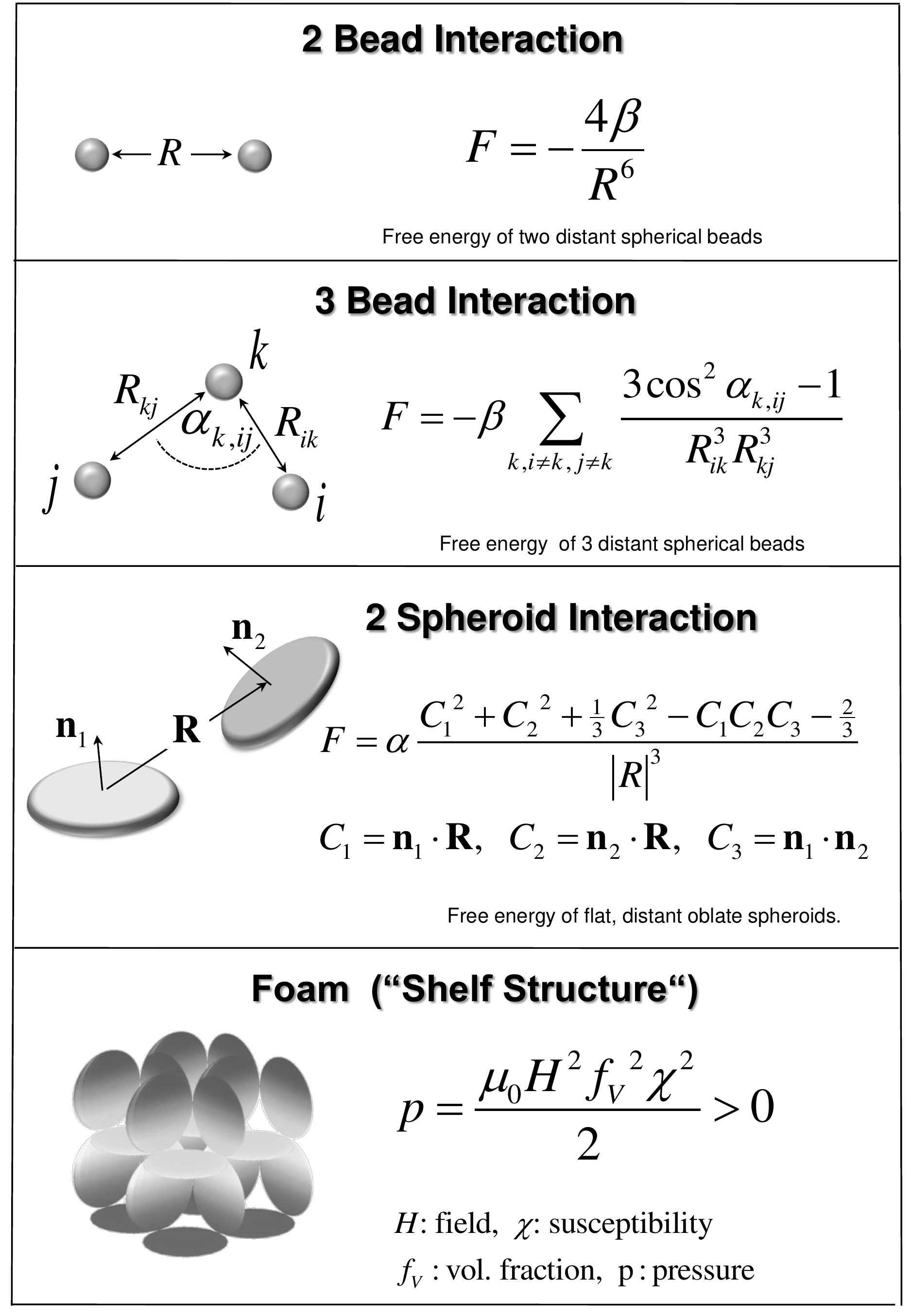}
\end{center}
\caption{ Summary of main results : Interactions induced by the cVdW
interaction. }%
\end{figure}

\begin{figure}[pt]
\begin{center}
\includegraphics[
width=3.2in]{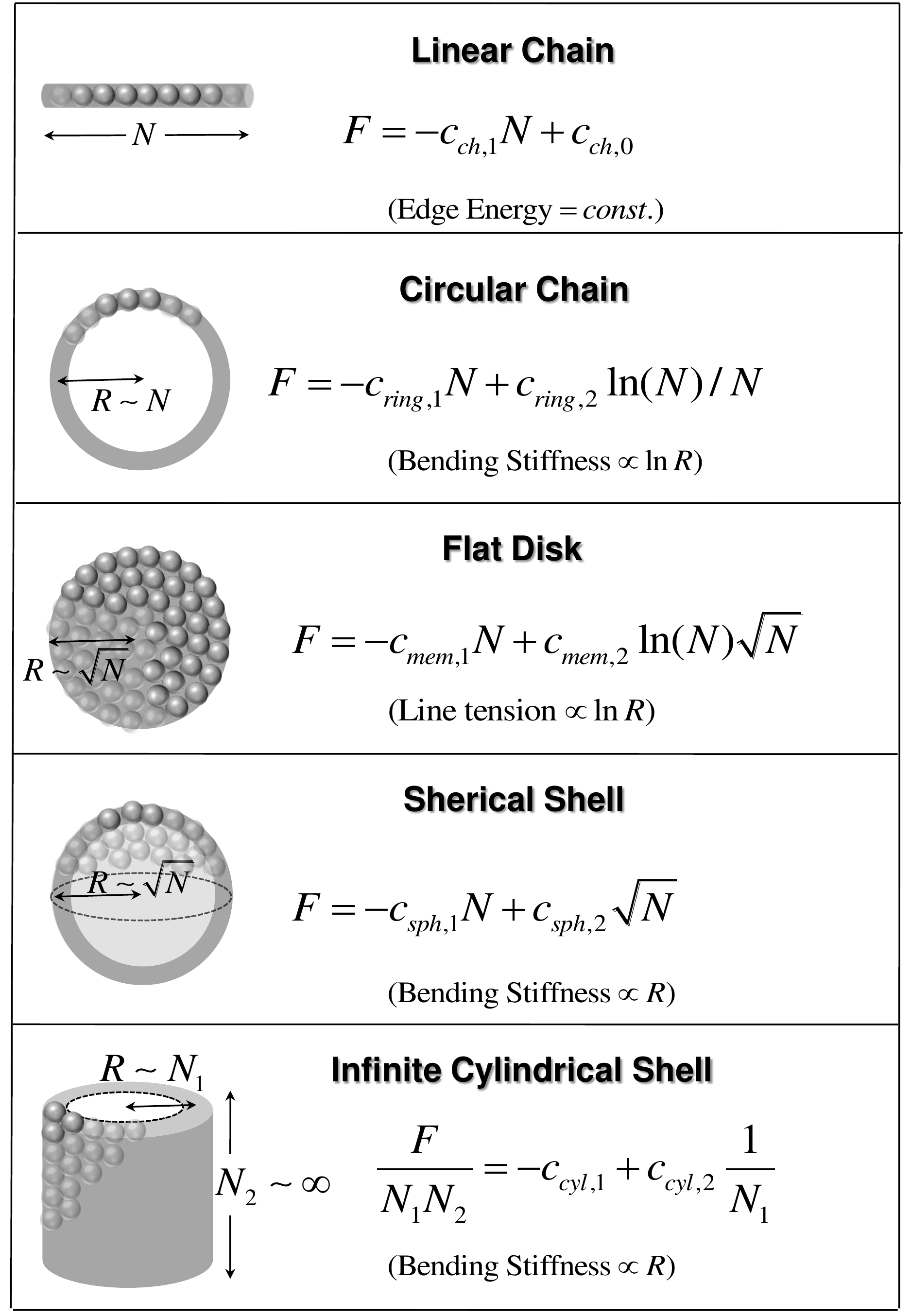}
\end{center}
\caption{ Summary of results : Finite size and elastic properties of various
structures. The finite size effects ($N$-effects) in chains, vesicles and
membranes are giving rise to anomalous elasticity effects. The free-energies
of these respective structures are written in terms of two leading order terms
with respect to the particle numbers. The corresponding pre-factors are found
in the corresponding sections of the main text.}%
\end{figure}

We have also argued that the finite size (finite particle number $N$) effects
in cVdW-matter are very different from other common interactions like VdW or
e.g. for classical magnetic beads with permanent moments. The many-body forces
are also found to play a crucial role in the \textit{anomalous elastic
properties} of chains and membranes. For, instance the bending stiffness of a
cVdW ring and the cVdW spherical membrane's stiffness are proportional to $\ln
N$ and $\sqrt{N}$, respectively, which is a direct consequence of the
specifically induced long-range many body effects in these systems.

The theory suggests a number of interesting and feasible \emph{experiments}
that can be performed to test the theoretical predictions about the
interactions in the cVdW-matter:

$1$. It would be very interesting to experimentally probe the dynamics of
exactly $3$ beads and the behavior of 3-body forces, cf. for instance the
surprising attraction/repulsion effects in Fig. 5. The experiment can be
performed for microscopic or macroscopic beads (the effect is scale
independent). Since 3 beads always span a common plain, the most general
dynamics can be observed directly in a single focal plain, e.g. on the
microscopy glass-slide on which the beads naturally settle down by gravity.

$2$. The predicted negative effective surface tension and the instability of
millimetric ferrofluid droplets, like in Fig. 6, would be a rather simple
experimental test of the theory. Also the shape bistability, i.e. the
coexistence of prolate and oblate shapes of the droplets, would be an
interesting qualitative outcome to be tested.

$3$. Bending cVdW chains and flat membrane patches, either by active forces or
passively by their own weight and observing their deflections should
experimentally reveal the presence of the predicted anomalous, size dependent stiffness.

$4$. The most telling and fundamental experiment would be to directly probe
the equation of state for a foam material. Measuring actively the forces on
the container walls or passively observing the rising height of the foam
against gravity would be two simple possibilities to test the predicted
internal pressure equation for the cVdW foams.

Finally, the central theoretical and experimental question, in our opinion,
remains\ if and how the cVdW can be generally realized in Nature. In
particular, we might ask if it can be induced in a truly \textit{equilibrium}
system. The previous realizations, in field driven colloidal systems were all
non-equilibrium. However there is no principal aspect of the theory that is
specific and restricted to a non equilibrium system only. While the driving
field amplitudes in our case are externally set, in an equilibrium system they
would satisfy a fluctuation dissipation condition which would relate them to
the temperature and the susceptibilities of the particles in the system. We
might speculate that in some long-range correlated fluctuating media, like
those considered in \cite{Kardar} (see also \cite{ReviewCasimir}) the cVdW can
indeed be realized even in equilibrium. If the fluctuations of the medium are
sufficiently longer ranged than the typical sizes of the formed structures,
the assembly will be driven by the cVdW interactions instead of the VdW ones
on these scales.

It is important to note that a simple tweak in the way how the interaction is
induced (by switching from incoherent to coherent\ excitation), enormously
increases the "morphogenic capacity" of the interaction i.e. its ability to
form complex structures. If we are interested in the self-assembly of anything
more complex than a spherical droplet (for which the standard VdW-matter is
good enough), cVdW-matter would be a better candidate than the simple VdW one.
The exploration and utilization of novel non-equilibrium (field-driven) or
equilibrium realizations of cVdW interactions is an interesting future
challenge. It could open the doors to deeper many-body studies of complex
self-assembled materials, and more importantly to technological applications
of the potentially very versatile and powerful cVdW-matter.

\section{\textit{Acknowledgements}}

We thank Jim Martin, A.Johner, H.Mohrbach, for discussions and comments.

\section{Appendix 1}

\subsection{Calculation of The cVdW Dimer Free-Energy}

In the two-particle problem the dipole operator $\hat{T}$ in Eq.
(\ref{chi-eff}) has only one non-vanishing component ($\hat{T}_{12}=\hat
{T}_{21}\neq0$) and can be written as:%

\[
\hat{T}=\varphi_{12}\cdot\left(
\begin{array}
[c]{cc}%
0 & \hat{1}-3\hat{N}\\
\hat{1}-3\hat{N} & 0
\end{array}
\right)
\]
where $\hat{N}=\left\vert \mathbf{b}_{12}\right\rangle \left\langle
\mathbf{b}_{12}\right\vert $ is the projector on the bond vector of the two
particles and the scalar factor $\varphi_{12}=V_{b}/4\pi\left\vert
\mathbf{R}_{12}\right\vert ^{3}$ as introduced before. \ Using the relations
$\hat{N}^{2}=\hat{N}$ , $(\hat{1}-a\hat{N})^{-1}=\hat{1}+a(1-a)^{-1}\hat{N}$
and the fact that $\hat{N}$ and $\hat{1}$ (or any scalar function like
$\varphi_{12}$) commute the operator inversion in $\hat{\chi}_{eff}=\chi
_{b}(\hat{1}+\chi_{b}\hat{T})^{-1}$ is quickly evaluated%

\begin{equation}
\hat{\chi}_{eff}=\left(
\begin{array}
[c]{cc}%
A_{1}\hat{1}+A_{2}\hat{N} & B_{1}\hat{1}+B_{2}\hat{N}\\
B_{1}\hat{1}+B_{2}\hat{N} & A_{1}\hat{1}+A_{2}\hat{N}%
\end{array}
\right)  \label{Chi2Bead}%
\end{equation}
with the bead distance dependent (scalar) coefficients%
\begin{align}
A_{1}  &  =\frac{\chi_{b}}{1-\chi_{b}^{2}\varphi_{12}^{2}}\text{ , }%
A_{2}=\frac{3\chi_{b}^{3}\varphi_{12}^{2}}{\left(  1-\chi_{b}^{2}\varphi
_{12}^{2}\right)  \left(  1-4\chi_{b}^{2}\varphi_{12}^{2}\right)  }\nonumber\\
B_{1}  &  =-\frac{\chi_{b}^{2}\varphi_{12}}{1-\varphi_{12}^{2}\chi_{b}^{2}%
}\text{ , }B_{2}=\frac{3\chi_{b}^{2}\varphi_{12}\left(  1-2\varphi_{12}%
^{2}\chi_{b}^{2}\right)  }{\left(  1-\varphi_{12}^{2}\chi_{b}^{2}\right)
\left(  1-4\varphi_{12}^{2}\chi_{b}^{2}\right)  }\nonumber
\end{align}
The relevant quantity for the coherent and the incoherent VdW - the trace of
$\hat{\chi}_{eff}$ \ over the 3 spacial directions- is directly obtained by
taking into account that $Tr\hat{N}=1$ and $Tr\hat{1}=3$, which gives
Eq.(\ref{TrChieff2Bead}) in the main text.

\section{Appendix 2}

\subsection{The Free-Energy for The Three-Body Problem in cVdW Systems}

If we consider only $3$ beads the interaction energy is given by
$\mathcal{\bar{F}}_{cVdW}^{(3)}=-\beta\sum\nolimits_{k=1,2,3}\sum
\nolimits_{i\neq k,j\neq k}w_{k,ij}$ $\ $with $w_{k,ij}=(3\cos^{2}%
\theta_{k,ij}-1)/\left\vert \mathbf{R}_{ki}\right\vert ^{3}\left\vert
\mathbf{R}_{kj}\right\vert ^{3}$. We put two beads $1$ and $2$ very close to
each other at distance $R_{12}=d$ and the $3$-rd one at distance
$R_{13}\approx R_{23}\approx R\gg d$. Altogether we have $3\times2\times2=12$
terms in the sum. The terms with flipped $i\rightarrow j$ indices are
identical so we can reorder:\ %

\begin{align}
-\mathcal{\bar{F}}_{cVdW}^{(3)}/\beta &  =\left(  w_{1,22}+w_{1,33}%
+2w_{1,23}\right) \label{2b-1}\\
&  +\left(  w_{2,11}+w_{2,33}+2w_{2,13}\right) \nonumber\\
&  +\left(  w_{3,11}+w_{3,22}+2w_{3,21}\right) \nonumber
\end{align}
Whenever an index repeats (e.g. as in $w_{3,11},w_{3,22}$ etc) we have a
$2$-body force. Then the terms are symmetric and we have $w_{1,22}=w_{2,11}$ ,
$w_{2,33}=w_{3,22},$ $w_{1,33}=w_{3,11}:$%
\begin{align}
-\mathcal{\bar{F}}_{cVdW}^{(3)}/\beta &  =\left(  2w_{1,22}+2w_{1,33}%
+2w_{2,33}\right) \label{2b-2}\\
&  +\left(  2w_{2,13}+2w_{1,23}+2w_{3,21}\right) \nonumber
\end{align}
As $R_{13}\approx R_{23}\approx R$ and $R_{12}=d\ll R$ we have $w_{2,13}%
\approx w_{1,23}$ (as $\theta_{2,13}\approx\pi-\theta_{1,23}$ and so $\cos
^{2}\theta_{2,13}=\allowbreak\cos^{2}\theta_{1,23}$) , one has $w_{1,33}%
\approx w_{2,33}$ so that%
\begin{equation}
-\mathcal{\bar{F}}_{cVdW}^{(3)}/\beta\approx\left(  2w_{1,22}+4w_{1,33}%
\right)  +\left(  4w_{1,23}+2w_{3,21}\right)  . \label{2b-3}%
\end{equation}
Further we have $w_{1,22}\approx\frac{1}{d^{6}}\left(  3-1\right)  ,$
$w_{1,33}\approx w_{3,21}\approx\left(  3-1\right)  /R^{6}$ and $w_{1,23}%
\approx\left(  3\cos^{2}\theta_{1,23}-1\right)  /d^{3}R^{3}$. Keeping only the
lowest power in $R$ it simplifies to
\begin{equation}
-\frac{\mathcal{\bar{F}}_{cVdW}^{(3)}}{\beta}=\frac{4}{d^{6}}+\frac{4\left(
3\cos^{2}\theta_{1,23}-1\right)  }{d^{3}R^{3}}+O\left(  R^{-6}\right)  .
\label{2b-4}%
\end{equation}

\subsection{3-Body Free-Energy of Finite cVdW Chain}

For the finite chain we consider the limit of the chain being still long
enough that the two ends do not see each other (summations for each particle
are infinite in one direction). Then we have $3\cos^{2}\theta_{k,ij}-1=2$ and
we can split up the summation
\begin{equation}
\mathcal{\bar{F}}_{cVdW,ch}^{(3)}=-\beta\sum\nolimits_{k=1}^{N}f_{k}%
=-2\beta\sum\nolimits_{k=1}^{N/2}f_{k} \label{2c-1}%
\end{equation}
with%
\begin{align}
f_{k}  &  =\sum\nolimits_{j=1,j\neq k}^{N}\sum\nolimits_{i=1,i\neq k}^{N}%
\frac{2}{\left\vert \mathbf{R}_{ik}\right\vert ^{3}\left\vert \mathbf{R}%
_{kj}\right\vert ^{3}}\label{2c-2}\\
&  =2\left(  \sum\nolimits_{i=1,i\neq k}^{N}\frac{1}{\left\vert \mathbf{R}%
_{ik}\right\vert ^{3}}\right)  ^{2}\nonumber
\end{align}
The upper sum can be subdivided in two parts, one left and one right of the
particle $k$ with one of the sums approximated by an infinite boundary
$N=\infty$%
\begin{align}
\sum\nolimits_{i=1,i\neq k}^{N}\frac{1}{\left\vert \mathbf{R}_{ik}\right\vert
^{3}}  &  \approx\frac{1}{D^{3}}\left(  \sum\nolimits_{i=1}^{k-1}\frac
{1}{i^{3}}+\sum\nolimits_{l=1}^{\infty}\frac{1}{i^{3}}\right) \label{2c-3}\\
&  =\frac{2\zeta\left(  3\right)  -S_{k}}{D^{3}}\nonumber
\end{align}
with $S_{k}=\sum\nolimits_{i=k}^{\infty}(1/i^{3})$. Therefore $f_{k}=2\frac
{1}{D^{6}}\left(  2\zeta\left(  3\right)  -S_{k}\right)  ^{2}$ and the
free-energy per particle is given by
\begin{equation}
\frac{\mathcal{\bar{F}}_{cVdW,ch}^{(3)}}{N\beta}\approx-\frac{2}{D^{6}}%
\frac{1}{N}\sum\nolimits_{k=1}^{N}\left(  2\zeta\left(  3\right)
-S_{k}\right)  ^{2}. \label{2c-7}%
\end{equation}
For large $k$ we approximate $S_{k}$ by integral which gives $S_{k}%
\approx1/2k^{2}$ and this
\begin{equation}
\frac{\mathcal{\bar{F}}_{cVdW,ch}^{(3)}}{N\beta}\approx-\frac{8\zeta
^{2}\left(  3\right)  }{D^{6}}+\allowbreak\frac{3.4}{ND^{6}}. \label{A2c-Fch}%
\end{equation}
This first term is for the infinite chain, while the second is the leading
order correction as expected $O\left(  1/N\right)  $ .

\subsection{The 3-Body Free-Energy of Ring in cVdW Systems}

Here $\mathbf{R}_{i0}=R\left(  \cos(2\pi i/N),\sin(2\pi i/N)\right)  $ (note
$\mathbf{R}_{ij}=(R_{x,ij},R_{y,ij})$ \ , $\cos\theta_{k,ij}=\mathbf{R}%
_{ik}\cdot\mathbf{R}_{kj}/\left\vert \mathbf{R}_{ik}\right\vert \left\vert
\mathbf{R}_{kj}\right\vert $. All terms $k=0,1,..N-1$ give the same
contribution as the term $k=0$ due to symmetry. We can introduce the angle
$\phi_{1}=2\pi i/N$ and $\phi_{2}=2\pi j/N$ with $d\phi\approx2\pi/N$ (for
$N\rightarrow\infty)$ . Then $\mathbf{R}_{j0}\approx\mathbf{R}\left(
\phi\right)  $
\begin{align}
\left\vert \mathbf{R}\left(  \phi\right)  \right\vert ^{3}  &  =R^{3}\left(
2\left(  1-\cos\phi\right)  \right)  ^{\frac{3}{2}}\label{2d-1}\\
\text{ }\cos^{2}\theta\left(  \phi_{1},\phi_{2}\right)   &  =\left(
\frac{a_{1}a_{2}+b_{1}b_{2}}{2\sqrt{a_{1}a_{2}}}\right) \nonumber
\end{align}
where $a_{1,2}=(1-\cos\phi_{1,2})$ and $b_{1,2}=\sin\phi_{1,2}$. It also holds
$\cos^{2}\theta\left(  \phi_{1},\phi_{2}\right)  =\cos^{2}\left(  \frac
{\phi_{1}-\phi_{2}}{2}\right)  $. With a small distance cutoff angle
$c=2\pi/N$. Then we can write the free-energy:%
\begin{align}
\frac{\mathcal{\bar{F}}_{cVdW,ring}^{(3)}}{\beta N}  &  =-\sum\nolimits_{j=1}%
^{N-1}\sum\nolimits_{i=1}^{N-1}\frac{3\cos^{2}\theta_{0,ij}-1}{\left\vert
\mathbf{R}_{i0}\right\vert ^{3}\left\vert \mathbf{R}_{0j}\right\vert ^{3}%
}\label{2d-2}\\
&  =-\frac{1}{2R^{6}}\sum\nolimits_{j=1}^{N-1}\sum\nolimits_{i=1}^{N-1}%
\frac{1+3\cos\left(  \phi_{1}-\phi_{2}\right)  }{\left(  2a_{1}\right)
^{\frac{3}{2}}\left(  2a_{2}\right)  ^{\frac{3}{2}}}.\nonumber
\end{align}
The latter gives most contribution for $\phi_{1/2}$ small and can be expanded
around $\phi_{1/2}=0$. Note that we have $2$ identical contributions around
$i,j=1$ and around $N-1$ in each of the terms. Summing these expanded terms
(with $R=\frac{ND}{2\pi},$ $\phi=\frac{2\pi}{N}k$ ) we arrive at:%

\begin{equation}
\frac{\mathcal{\bar{F}}_{cVdW,ring}^{(3)}}{N\beta}\approx-\frac{8\zeta
^{2}\left(  3\right)  }{D^{6}}+\frac{16\zeta\left(  3\right)  \pi^{2}}{D^{6}%
}\frac{\ln N}{N^{2}} \label{2d-3}%
\end{equation}

\subsection{3-Body Free-Energy of The cVdW Spherical Shell}

For a spherical membrane (coloidosome) with classical elasticity or radius
$R$, we would have an energy density proportional to $1/R^{2}$
(=curvature$^{2}$). The total energy coming from the bending (i.e. without
self energy of beads) is then $\sim(1/R^{2})R^{2}\sim1$ constant. What happens
in the case of a coherent coloidosome? Starting again from:%
\begin{equation}
\frac{\mathcal{\bar{F}}_{cVdW,sph}^{(3),N}}{\beta}=-\sum\nolimits_{i,j,k}%
^{\prime}\frac{3\cos^{2}\theta_{k,ij}-1}{\left\vert \mathbf{R}_{ik}\right\vert
^{3}\left\vert \mathbf{R}_{kj}\right\vert ^{3}} \label{2e-1}%
\end{equation}
with fixed arbitrary $\mathbf{R}_{k}=R\left(  0,0,1\right)  $ (north pole of
the sphere, $\mathbf{R}=\left(  R_{x},R_{y},R_{z}\right)  $) and
$\mathbf{R}_{kj}=R\left(  -\sin\theta_{j}\cos\phi_{j},-\sin\theta_{j}\sin
\phi_{j},1-\cos\theta_{j}\right)  $, $|\mathbf{R}_{kj}|=R\sqrt{2\left(
1-\cos\theta_{j}\right)  }=2R\sin(\theta/2)$ and the apex angle is given by
$\cos\alpha=\mathbf{R}_{ik}\cdot\mathbf{R}_{kj}/\left\vert \mathbf{R}%
_{ik}\right\vert \left\vert \mathbf{R}_{kj}\right\vert $, where $\alpha
\equiv\alpha\left(  \theta_{1},\theta_{2},\phi_{1},\phi_{2}\right)  $. In
order to pick up the $N$-effects \ we replace the summation by integration
over two spheres where each of the two spheres contains $N=4\pi R^{2}%
/\rho_{sph}^{-1}\pi\left(  D/2\right)  ^{2}=\allowbreak16\rho_{sph}R^{2}%
/D^{2}$ beads giving the sphere radius: $(\sqrt{N}D/4\rho_{sph}^{1/2}%
)=\allowbreak R.$ The summation can be replaced by double integral over the
sphere with surface element $dA_{1,2}=R^{2}\sin\theta d\phi d\theta,$ and the
energy can be written as:
\begin{equation}
\frac{\mathcal{\bar{F}}_{cVdW,sph}^{(3)}}{N\cdot\Gamma}=-\int\left(
\frac{3\cos^{2}\alpha-1}{\left\vert \mathbf{R}\left(  \theta_{1}\right)
\right\vert ^{3}\left\vert \mathbf{R}\left(  \theta_{2}\right)  \right\vert
^{3}}\right)  dA_{1}dA_{2} \label{A2e-F-dA}%
\end{equation}
with $\Gamma=\beta N^{2}/\left(  4\pi R^{2}\right)  ^{2}$ and the latter is
calculated by an integration (in Mathematica)\ over $\phi_{1}$ and $\phi_{2}$
\begin{equation}
\frac{\mathcal{\bar{F}}_{cVdW,sph}^{(3)}}{N\beta}=-\frac{\rho_{pack}^{2}%
}{8R^{2}D^{4}}\left(  \int_{\theta_{\min}\left(  D\right)  }^{\pi}\frac
{3\cos\theta-1}{\sin^{3}\left(  \frac{\theta}{2}\right)  }\sin\theta
d\theta\right)  ^{2} \label{A2e-Fthe}%
\end{equation}
with $\theta_{\min}\left(  D\right)  $ being the angular cut-off resulting
from the spherical angle relation (surface area of bead / surface area of
whole sphere)\ :\ $\Omega\left(  D\right)  =\int_{0}^{2\pi}\int_{0}%
^{\theta_{\min}\left(  D\right)  }\sin\theta d\theta d\phi\approx
2\pi(1/2)\theta_{\min}^{2}$ but on the other hand $\Omega\left(  D\right)
=\pi\left(  D/2\right)  ^{2}/4\pi R^{2}=\rho_{sph}/N$ so that $\theta_{\min
}=\sqrt{\rho_{sph}/\pi N}.$ The integral can be done which gives the energy
per particle of the spherical shell:%

\begin{equation}
\frac{\mathcal{\bar{F}}_{cVdW,sph}^{(3)}}{N\beta}\approx-\frac{50\pi\rho
_{sph}^{2}}{D^{6}}\allowbreak\left(  1-\frac{4\sqrt{\rho_{sph}}}{\sqrt{\pi
}\sqrt{N}}\right)
\end{equation}

\subsection{The Microscopic Energy of the cVdW Tubular Membrane}

For an infinite cylinder of radius $R_{\perp}$ again we have a high symmetry
(all beads are the same) and we can choose the apex point anywhere, say at
$\left(  x,y,z\right)  =\left(  1,0,0\right)  .$ The other points along the
cylinder we parameterize cylindrically with coordinates $\left(
\phi,z\right)  $ , so that the difference vector becomes $\mathbf{R}%
_{kj}=\left(  R_{\perp}\left(  \cos\phi_{j}-1\right)  ,R_{\perp}\sin\phi
_{j},z\right)  $ and its length $R_{kj}=\sqrt{\allowbreak z_{j}^{2}+2R_{\perp
}^{2}\left(  1-\cos\phi_{j}\right)  }.$ The apex angle is $\cos\alpha
_{k,ij}=\mathbf{R}_{ki}\cdot\mathbf{R}_{kj}/R_{1}R_{2}$\ with $R_{1/2}%
=\sqrt{\allowbreak z_{1/2}^{2}+2R_{\perp}^{2}a_{1/2}}$.

The energy density consists of seven terms :%
\begin{equation}
\frac{\mathcal{\bar{F}}_{cVdW,tub}^{(3)}}{-\beta N}=\sum\nolimits_{i,j}%
^{\prime}\frac{3\cos^{2}\theta_{k,ij}-1}{\left\vert \mathbf{R}_{ik}\right\vert
^{3}\left\vert \mathbf{R}_{kj}\right\vert ^{3}} \label{2e-Ft}%
\end{equation}%
\begin{align}
\frac{\mathcal{\bar{F}}_{cVdW,tub}^{(3)}}{-\beta N}  &  =\frac{3\left(
C_{1}^{2}+C_{2}^{2}+C_{3}^{2}+2C_{4}^{2}\right)  }{R_{\perp}^{6}}\text{
\ }\label{A2f-F-C}\\
&  +\frac{3(2C_{5}^{2}+2C_{6}^{2})-C_{7}^{2}}{R_{\perp}^{6}}\nonumber
\end{align}
Some of the sums are trivially zero due to symmetry: $C_{4}=0$ \ ($R\left(
\phi,z\right)  $ even but $\left(  \cos\phi-1\right)  \sin\phi$ odd in$\phi$)
$\ C_{5}=0$ (odd in $z$ )\ , $C_{6}=0$ (odd in $z$ and $\phi$ ). Those which
are different from zero are%

\begin{equation}
C_{1}=\sum_{k=\pm1...,\pm\frac{1}{\delta}}\sum_{l=0,\pm1,..,\pm\infty}%
\frac{\left(  \cos\left(  \delta k\right)  -1\right)  ^{2}}{P^{5}(k,l)}
\label{C1}%
\end{equation}%
\begin{equation}
C_{2}=\sum_{k=\pm1...,\pm\frac{1}{\delta}}\sum_{l=0,\pm1,..,\pm\infty}%
\frac{\sin^{2}\left(  \delta k\right)  }{P^{5}(k,l)}, \label{C2}%
\end{equation}%
\begin{equation}
C_{3}=\sum_{k=\pm1..,\pm\frac{1}{\delta}}\sum_{l=0,\pm1,..,\pm\infty}%
\frac{\left(  \delta l\right)  ^{2}}{P^{5}(k,l)} \label{C3-4}%
\end{equation}%
\begin{equation}
C_{7}=\sum_{k=\pm1..,\pm\frac{1}{\delta}}\sum_{l=0,\pm1,..,\pm\infty}\frac
{1}{P^{3}(k,l)} \label{C5-7}%
\end{equation}
where $P(k,l)=\left(  l\allowbreak^{2}\delta^{2}+2\left(  1-\cos
k\delta\right)  \right)  ^{1/2}$. To obtain the scaling we introduced small
scale cutoffs $\delta=D/R_{\perp}$. Parameterizing the angle $\phi=\delta\cdot
k$ with $k=\pm1,\pm2,\pm3,...\pm\frac{1}{\delta}$ and the $z$ displacement as
$z=\delta\cdot l$ \ (with $l=0,\pm1,\pm2,...\pm\infty$) one can approximate
the summation over $l$ by integration. As the final result this gives for
$C_{1}\approx(2/3\delta^{2})$, $(C_{2}/2)\approx C_{3}$, $\approx
(4\zeta\left(  2\right)  /3\delta^{3})\approx C_{7}$. When inserting
$\delta=\frac{D}{R_{\perp}}$ we get then the final result for the cylinder
free-energy ($N=N_{1}N_{2},R_{\perp}\sim N_{1}$ - see \ Fig.11):%

\begin{equation}
\frac{\mathcal{\bar{F}}_{cVdW,tub}^{(3)}}{\beta N}\approx-\frac{8\pi^{4}}%
{27}\frac{1}{D^{6}}+\frac{40\pi^{2}}{27}\frac{1}{R_{\perp}D^{5}}.
\label{F-tub}%
\end{equation}%
\begin{equation}
\frac{\mathcal{\bar{F}}_{cVdW,sph}^{(3)}}{N\beta}\approx-\frac{50\pi\rho
_{sph}^{2}}{D^{6}}\allowbreak\left(  1-\frac{4\sqrt{\rho_{sph}}}{\sqrt{\pi
}\sqrt{N}}\right)  \label{A2e-Fsph-N}%
\end{equation}

\section{Appendix 3}

\subsection{Demagnetization Tensors of Spheroids and Cylinders}

In calculating magnetic fields of magnetized bodies and the corresponding
magnetostatic energy two kind of demagnetization tensors appear
\cite{1.Beleggia, 4.Beleggia, 2.Beleggia}. The first one is related to the
demagnetizing field of the uniformly magnetized body $\mathbf{H}%
_{D}(\mathbf{r})=-\hat{L}(r)\mathbf{M}(\mathbf{r})$ where $\mathbf{M}%
(\mathbf{r})=\mathbf{M}D(\mathbf{r})$. Here $\mathbf{M}=\mathbf{const}$ and
$D(\mathbf{r})$ is the dimensionless shape function which represents the
region of the space bounded by the body (sample) surface, i.e. $D(\mathbf{r}%
)=1$ inside the body and $D(\mathbf{r})=0$ outside it. Its Fourier transform
$D(\mathbf{k})$ - the shape amplitude, which is related to $\hat{L}(r)$
\cite{1.Beleggia, 4.Beleggia, 2.Beleggia}
\begin{align}
\hat{L}(\mathbf{r}) &  =\frac{1}{(2\pi)^{3}}\int d^{3}k\hat{L}(\mathbf{k}%
)e^{i\mathbf{k\cdot r}}\label{LrLk}\\
\hat{L}(\mathbf{k}) &  =\frac{D(\mathbf{k})}{k^{2}}\left\vert \mathbf{k}%
\right\rangle \left\langle \mathbf{k}\right\vert .\nonumber
\end{align}
Note, that $\left(  \hat{L}(\mathbf{k})\right)  _{ij}=D(\mathbf{k})k_{i}%
k_{j}/k^{2}$. The shape amplitude $D(\mathbf{k})$ and $\hat{L}(r)$ are
calculated for various bodies. For instance, for sphere of radius $R$ one has
$D(\mathbf{k})=(4\pi R^{2}/k)j_{1}(kR)$ where the spherical Bessel function of
first order $j_{1}(x)=(\sin x/x^{2})-\cos x/x$. For other body-shapes see more
in \cite{1.Beleggia, 4.Beleggia, 2.Beleggia} and references therein.

The second type of demagnetization tensor(factors) appears in the expression
for the magnetostatic (demagnetization) energy with the uniform magnetization
$\mathbf{M}(\mathbf{r})=\mathbf{M}$
\begin{align}
E_{D}  &  =-\frac{\mu_{0}}{2}\int_{V_{D}}d^{3}r\mathbf{M}(\mathbf{r}%
)\mathbf{H}_{D}(\mathbf{r})\label{A3-Ed}\\
&  =\frac{\mu_{0}}{2}V_{D}\mathbf{M}\left\langle \hat{L}(r)\right\rangle
\mathbf{M,}\nonumber
\end{align}
where $V_{D}$ is the volume of the body and $\left\langle \hat{L}%
(r)\right\rangle $ is the \textit{magnetometric} (volume averaged)
\textit{demagnetization tensor}, i.e.
\begin{align}
\left\langle \hat{L}(r)\right\rangle  &  =\frac{1}{V_{D}}\int_{V_{D}}%
d^{3}r\hat{L}(\mathbf{r})\label{A3-L-av}\\
&  =\frac{1}{(2\pi)^{3}V_{D}}\int d^{3}k\frac{D^{2}(\mathbf{k})}{k^{2}%
}\left\vert \mathbf{k}\right\rangle \left\langle \mathbf{k}\right\vert
\nonumber
\end{align}
It is easy to show that the trace of $\hat{L}(\mathbf{r})$ inside the body is
one, while outside is zero, i.e. $Tr\hat{L}(\mathbf{r})=D(\mathbf{r})$. The
magnetometric tensor fulfills $Tr\left\langle \hat{L}(r)\right\rangle =1$.
Note, that in Eq.(\ref{F-macro1}) enter the diagonal components of the
magnetometric demagnetization tensor $\left\langle \hat{L}(r)\right\rangle $.
Since we study magnetic bodies where $\hat{L}(r)=const=\hat{L}$ inside the
body, then $\left\langle \hat{L}(r)\right\rangle =\hat{L}$. Wi give the exact
and asymptotic expressions for the $L_{z}$ demagnetization factor for
ellipsoids and cylinders which are important for the studies in the main text.

(i) \textit{Demagnetization factors for ellipsoids} - If $a$, $b$, $c$ are the
semi-axis of ellipsoid with $\tau_{a}=(c/a)$, $\tau_{b}=(c/b)$, $k=\arcsin
\sqrt{1-\tau_{a}^{-2}}$, $m=(1-\tau_{b}^{-2})/(1-\tau_{a}^{-2})$ than one has
\cite{1.Beleggia, 4.Beleggia, 2.Beleggia}%
\begin{equation}
L_{z}(\tau_{a},\tau_{b})=\frac{1}{\tau_{a}\tau_{b}}\frac{F(k,m)-E(k,m)}%
{m\sin^{3}k}, \label{A3-Lz-ell}%
\end{equation}
where $E(k,m)$ and $F(k,m)$ are incomplete elliptic integrals
\cite{Abramowitz}. The symmetry implies $L_{x}(\tau_{a},\tau_{b})=L_{z}%
(\tau_{a}^{-1},\tau_{b}\tau_{a}^{-1})$ and $L_{y}(\tau_{a},\tau_{b}%
)=L_{z}(\tau_{a}\tau_{b}^{-1},\tau_{b}^{-1})$ - see \cite{1.Beleggia,
4.Beleggia, 2.Beleggia}. For oblate and prolate spheroids, where $\tau
_{a}=\tau_{b}=\tau_{s}$, one has
\begin{equation}
L_{z}(\tau_{s})=\frac{1}{1-\tau_{s}^{2}}\left[  1-\frac{\tau_{s}\arccos
(\tau_{s})}{\sqrt{1-\tau_{s}^{2}}}\right]  \label{A3-Lz-sp}%
\end{equation}
For $\tau_{e}\rightarrow0$ (extreme oblate, i.e. membrane-like spheroid) it
gives
\begin{equation}
L_{z}(\tau_{s})=1-\frac{\pi}{2}\tau_{s}+2\tau_{s}^{2}+O(\tau_{s}^{3}),
\label{A3-Lz-sp-0}%
\end{equation}
and for $\tau_{e}\rightarrow\infty$ (extreme prolate, i.e. chain-like
spheroid) one has
\begin{equation}
L_{z}(\tau_{s})=\frac{\ln(2\tau_{s}/e)}{\tau_{s}^{2}}+O(\tau_{s}^{-4}).
\label{chain-prol}%
\end{equation}

(ii) \textit{Demagnetization factors for cylinders} - For cylinders with
thickness (height) $t$ and radius $R$ with the aspect ratio $\tau=t/2R$ and
$\kappa=1/\sqrt{1+\tau^{2}}$ one has \cite{1.Beleggia, 4.Beleggia, 2.Beleggia}%
\begin{equation}
L_{z}^{cyl}(\tau)=1+\frac{4}{3\pi\tau}\left\{  1-\frac{1}{\kappa}\left[
(1-\tau^{2})E(\kappa^{2})+\tau^{2}K(\kappa^{2})\right]  \right\}
\label{A3-Lz-cyl}%
\end{equation}
where $E(\kappa^{2})$ and $K(\kappa^{2})$ are complete elliptic functions
\cite{Abramowitz}.

For \textit{very thin cylinder} where $\tau\rightarrow0$ one has%
\begin{equation}
L_{z}^{cyl}(\tau)=1+\frac{\tau}{\pi}\left(  1+2\ln\frac{\tau}{4}\right)
+O(\tau^{3}). \label{A3-Lz-cyl-t}%
\end{equation}
For \textit{very long cylinder} where $\tau\rightarrow\infty$ one has
\begin{equation}
L_{z}^{cyl}(\tau)=\frac{4}{3\pi\tau}-\frac{8}{\tau^{2}}+O(\tau^{-4}).
\label{A3-Lz-cyl-l}%
\end{equation}

\subsection{Surface and Demagnetization Factors For Deformed Sphere}

The aspect ratio of the spheroid is $\tau_{s}=c/a$. Close to the sphere one
has $\varepsilon=\tau_{s}-1\ll1$, where $\varepsilon<0$ is for oblate
ellipsoid while $\varepsilon>0$ is for the prolate one. By using
Eq.(\ref{A3-Lz-sp}) one has for small $\left\vert \varepsilon\right\vert \ll1$
one has%

\begin{align}
L_{z}^{prol/obl}  &  =\frac{1}{3}-\frac{4}{15}\varepsilon+\frac{6}%
{35}\varepsilon^{2}\label{Lz}\\
&  -\frac{32}{315}\varepsilon^{3}+\frac{40}{693}\varepsilon^{4}+O\left(
\varepsilon^{5}\right) \nonumber
\end{align}
Similarly for the surface of the prolate spheroid one has%

\begin{equation}
A_{prolate}=2\pi a^{2}\left(  1+\frac{\tau_{s}}{x}\sin^{-1}x\right)  \text{,}
\label{Apr}%
\end{equation}
with $\ x^{2}=1-1/\tau_{s}^{2}$ for $\tau_{s}>1$ and for the oblate one
\begin{equation}
A_{oblate}=2\pi a^{2}\left(  1+\frac{1-x^{2}}{x}\tanh^{-1}x\right)  ,
\label{Aob}%
\end{equation}
for $x^{2}=1-\tau_{s}^{2}$ \ for $\tau_{s}<1$. In the following analysis the
volume is fixed, i.e. $V=4\pi a^{2}c/3=4\pi a^{3}\tau_{s}/3$ , $a^{2}=\left(
3V/4\pi\tau_{s}\right)  ^{2/3}$ and in terms of the axial stretching one has%
\begin{equation}
\frac{A_{prolate}}{A_{0}}=\frac{1+\frac{\left(  \varepsilon+1\right)  ^{2}%
}{\sqrt{\varepsilon\left(  \varepsilon+2\right)  }}\arcsin\sqrt{1-\frac
{1}{\left(  \varepsilon+1\right)  ^{2}}}}{2\left(  \varepsilon+1\right)
^{2/3}} \label{Apr-eps}%
\end{equation}
for $\tau_{s}>1$ and%

\begin{equation}
\text{\ }\frac{A_{oblate}}{A_{0}}=\frac{1+\frac{\left(  \varepsilon+1\right)
^{2}}{\sqrt{-\varepsilon\left(  \varepsilon+2\right)  }}\operatorname{arctanh}%
\sqrt{-\varepsilon\left(  \varepsilon+2\right)  }}{2\left(  \varepsilon
+1\right)  ^{2/3}} \label{Aobl-eps}%
\end{equation}
for $\tau_{s}<1$,with $A_{0}=4\pi(3V/4\pi)^{2/3}$ the initial area of the
sphere. \ We can expand the surface area $A_{prol/obl}\left(  \varepsilon
\right)  $ \ (expansions coincide):%
\begin{align}
\frac{A_{prol/obl}\left(  \varepsilon\right)  \allowbreak}{A_{0}}  &
=1+\frac{8}{45}\varepsilon^{2}-\frac{584}{2835}\varepsilon^{3}\label{Apr-ob}\\
&  +\frac{118}{567}\varepsilon^{4}+O\left(  \varepsilon^{5}\right) \nonumber
\end{align}

\section{Appendix 4}

\subsection{Membrane-Bead Interaction in cVdW Systems}

In order to calculate the free-energy in Eq.(\ref{Wmb}) we need to know
$Tr\hat{\chi}_{m,eff}$ and $Tr\hat{\chi}_{b,eff}$ where%

\begin{equation}
\hat{\chi}_{m,eff}=(1-q_{m}q_{b}\varphi_{0}^{2}\hat{\chi}_{m}\hat{t}\hat{\chi
}_{b}\hat{t})^{-1}\hat{\chi}_{m}(1-q_{m}\varphi_{0}\hat{t}\hat{\chi}_{b}),
\label{A4-chi-m}%
\end{equation}
with $\hat{t}_{m-b}\equiv\hat{t}$ and $\hat{\chi}_{b,eff}$ is obtained by
replacing $m\leftrightarrow b$. For the assumed symmetry and geometry of the
problem we have $\left\vert \mathbf{n}_{m}\right\rangle =\left\langle
\mathbf{b}\right\vert $. Here, $q_{m,b}=V_{m,b}/V_{0}$, $V_{0}=V_{m}+V_{b}$,
and $\hat{\chi}_{m,b}$ is the membrane and bead susceptibility (with respect
to external field), respectively, $\varphi_{0}=V_{0}/4\pi R_{mb}^{3}$. By
defining $\hat{N}=\left\vert \mathbf{n}_{m}\right\rangle \left\langle
\mathbf{n}_{m}\right\vert $ and $\hat{P}=1-\hat{N}$ we have $\hat{N}^{2}%
=\hat{N}$, $\hat{P}^{2}=\hat{P}$, $\hat{N}\hat{P}=0$, $Tr\hat{N}=1$ and
$Tr\hat{P}=2$ and
\begin{equation}
\hat{t}=-2\hat{N}+\hat{P}. \label{A4-t}%
\end{equation}
Further we parameterize $\hat{\chi}_{m}=\chi_{\max}\hat{\chi}_{m}^{0}$,
$\hat{\chi}_{m}^{0}=p_{m}\hat{N}+\hat{P}$, $p_{m}=\chi_{\min}/\chi_{\max}$,
$\hat{\chi}_{b}=\chi_{b}\hat{1}$, $b=\alpha_{b}\alpha_{m}$, $\alpha_{m}%
=q_{m}\chi_{\max}\varphi_{0}$, $\alpha_{b}=q_{b}\chi_{b}\varphi_{0}$. In this
parametrization we have
\begin{align}
\hat{\chi}_{m,eff}  &  =\chi_{\max}\hat{A}^{-1}\hat{\chi}_{m}^{0}(1-\alpha
_{b}\hat{t})\label{A4-chi-mb}\\
\hat{\chi}_{b,eff}  &  =\chi_{b}\hat{A}^{-1}(1-\alpha_{m}\hat{\chi}_{m}%
^{0}\hat{t})\nonumber
\end{align}
and
\begin{equation}
\hat{A}=1-b\hat{\chi}_{m}^{0}\hat{t}^{2}. \label{A4-A}%
\end{equation}
By using the projecting properties of $\hat{N}$ and $\hat{P}$ one obtains the
inverse matrix $\hat{A}^{-1}$
\begin{equation}
\hat{A}^{-1}=\frac{1}{1-b}\left[  1-\frac{c}{1+c}\hat{N}\right]
\label{A4-A-1}%
\end{equation}
with $c=b(1-4p_{m})/(1-b)$. After some algebra one obtains the effective
membrane susceptibility%
\begin{equation}
\hat{\chi}_{m,eff}=\frac{\chi_{\max}}{1-b}\left\{  a\hat{1}+\left[
\frac{p_{m}(1+2\alpha_{b})}{1+c}-a\right]  \hat{N}\right\}
\label{A4-chi-meff}%
\end{equation}
where $a=(1-\alpha_{b})$. Then the trace is%
\begin{equation}
Tr\hat{\chi}_{m,eff}=\frac{\chi_{\max}}{1-b}\left\{  2(1-\alpha_{b}%
)+\frac{p_{m}(1+2\alpha_{b})}{1+c}\right\}  . \label{A4-Tr-chi-m}%
\end{equation}
Analogously one obtains the effective bead susceptibility $Tr\hat{\chi
}_{b,eff}$%
\begin{equation}
Tr\hat{\chi}_{b,eff}=\frac{\chi_{b}}{1-b}\left\{  \frac{3+2c}{1+c}+2\alpha
_{m}(\frac{p_{m}}{1+c}-1)\right\}  . \label{A4-Tr-chi-b}%
\end{equation}
By inserting Eq.(\ref{A4-Tr-chi-m}) and Eq.(\ref{A4-Tr-chi-b}) into
Eq.(\ref{Wmb}) one obtains $w_{m}$ and $w_{b\text{ }}$in Eq.(\ref{wm}) and
Eq.(\ref{wb}), respectively.

\subsection{Derivation of the 2-Membrane Interaction in cVdW Systems}

Having in mind two identical membranes (with volume $V_{m}$) we consider
susceptibilities $\hat{\chi}_{1}$ and $\hat{\chi}_{2}$ of two oblate
spheroids, which are differently oriented in space. In terms of their own
local coordinate systems (in Dirac bra-ket notation for tensors) they are
given by%
\begin{align}
\hat{\chi}_{1}  &  =\chi_{\min}\left\vert \mathbf{n}_{1}\right\rangle
\left\langle \mathbf{n}_{1}\right\vert +\chi_{\max}(\hat{1}-\left\vert
\mathbf{n}_{1}\right\rangle \left\langle \mathbf{n}_{1}\right\vert
)\label{M4}\\
\hat{\chi}_{2}  &  =\chi_{\min}\left\vert \mathbf{n}_{2}\right\rangle
\left\langle \mathbf{n}_{2}\right\vert +\chi_{\max}(\hat{1}-\left\vert
\mathbf{n}_{2}\right\rangle \left\langle \mathbf{n}_{2}\right\vert ),\nonumber
\end{align}
where the unit vectors $\left\vert \mathbf{n}_{1}\right\rangle ,$ $\left\vert
\mathbf{n}_{2}\right\rangle $ are the normals of the membranes $1$ and $2$
respectively. By using Eq.(\ref{M4}), and noting that $Tr\{\hat{\chi}_{1}%
\hat{\chi}_{2}\}=$ $Tr\{\hat{\chi}_{2}\hat{\chi}_{1}\}$, $Tr\{\hat{\chi}%
_{1}\hat{\chi}_{2}(\left\vert \mathbf{b}_{21}\right\rangle \left\langle
\mathbf{b}_{21}\right\vert )\}=Tr\{\hat{\chi}_{2}\hat{\chi}_{1}(\left\vert
\mathbf{b}_{12}\right\rangle \left\langle \mathbf{b}_{12}\right\vert )\}$ and
$Tr\{\left\vert \mathbf{n}_{i}\right\rangle \left\langle \mathbf{n}%
_{j}\right\vert \}=\mathbf{\langle n}_{i}\left\vert \mathbf{n}_{j}%
\right\rangle $ (with $\mathbf{a}\cdot\mathbf{b\equiv\langle a}\left\vert
\mathbf{b}\right\rangle $ the scalar product) it follows
\begin{equation}
Tr\{\hat{\chi}_{1}\hat{\chi}_{2}\}=\chi_{\max}^{2}\left[  1+2\gamma+c_{3}%
^{2}(1-\gamma)^{2}\right]  \label{M5}%
\end{equation}%
\begin{align}
\frac{Tr\{\hat{\chi}_{1}\hat{\chi}_{2}(\left\vert \mathbf{b}_{12}\right\rangle
\left\langle \mathbf{b}_{12}\right\vert )\}}{\chi_{\max}^{2}}  &
=[1-(1-\gamma)(c_{1}^{2}+c_{2}^{2})\label{M6}\\
&  +(1-\gamma)^{2}c_{1}c_{2}c_{3}],\nonumber
\end{align}
where $\gamma=(\chi_{\min}/\chi_{\max})$ and $c_{1}=\mathbf{n}_{1}%
\cdot\mathbf{b}_{12}$, $c_{2}=\mathbf{n}_{2}\cdot\mathbf{b}_{12}$,
$c_{3}=\mathbf{n}_{1}\cdot\mathbf{n}_{2}$ are factors describing the mutual
orientation of membranes. By replacing Eqs.(\ref{M5}-\ref{M6}) in
Eq.(\ref{M3}) (where $V_{b}$ in $\varphi_{12}$ is replaced by the membrane
volume $V_{m}$) one obtains Eq.(\ref{me-me}) in the manuscript.

\subsection{The Interaction Energy of The Cubic Shelf Structure}

The lattice sum $S(\equiv a_{M}^{3}\mathcal{\bar{F}}_{int}/\alpha)$,
$\alpha=9\eta\chi_{\max}^{2}\mu_{0}H_{0}^{2}V_{m}^{2}/16\pi$ (where
$\mathbf{r}_{i1}=\mathbf{R}_{i1}/a_{M}$ and $\eta=(1-\chi_{\min}/\chi_{\max
})/3$) for the cubic shelf structure is given by
\begin{equation}
S=\sum_{\mathbf{l},\kappa=x,y,z}\frac{\eta C_{3,\kappa}^{2}+C_{1,\kappa}%
^{2}+C_{2,\kappa}^{2}-3\eta C_{1,\kappa}C_{2,\kappa}C_{3,\kappa}-\frac{2}{3}%
}{r_{\mathbf{l},\kappa}^{3}}, \label{A4c-S}%
\end{equation}
where for compactness we define $C_{1/2/3,\kappa}=C_{1/2/3,\mathbf{l,}\kappa}%
$. The coefficients $C_{1}=\mathbf{n}_{1}\cdot\mathbf{b}_{12}$, $C_{2}%
=\mathbf{n}_{2}\cdot\mathbf{b}_{12}$, $C_{3}=\mathbf{n}_{1}\cdot\mathbf{n}%
_{2}$ where $\mathbf{n}_{1}$ and $\mathbf{n}_{2}$ are normals to membrane 1
and 2, respectively, while $\mathbf{b}_{12}$ $\mathbf{is}$ the unit bonding
vector. The summation over unit cells labeled by \textbf{$l$}$=(l_{x}%
,l_{y},l_{z})$ comprises interactions of membrane at the point $\mathbf{r}%
_{\mathbf{l},\kappa}=(0,0,0)$ and with the normal to the membrane plane
parallel to the z-axis, i.e. $\mathbf{n}_{z}^{0}=(0,0,1)$, with all others.
The summation over $\kappa=x,y,z$ means the interaction with membranes whose
normals $\mathbf{n}_{x}$, $\mathbf{n}_{y}$, $\mathbf{n}_{z}$ are along the
$x$, $y$, $z$-axis, respectively. For further calculations we parameterize
$\mathbf{n}_{1}=\left(  0,0,1\right)  $, $\mathbf{n}_{2}=\cos\phi_{2}%
\sin\theta_{2},\sin\phi_{2}\sin\theta_{2},\cos\theta_{2}$ and for $\mathbf{b}$
the same as for $\mathbf{n}_{2}$ but $\phi_{2}$, $\theta_{2}$ goes to
$\phi_{b}$, $\theta_{b}$. It is straightforward to show that $C_{3,\kappa
=x}=C_{3,\kappa=y}=0$. Similarly, $C_{3,\kappa=z}=1$, $C_{1,\kappa=z}%
=l_{z}/\sqrt{l_{x}^{2}+l_{y}^{2}+l_{z}^{2}}$ and that $C_{1,\kappa
=z}=C_{2,\kappa=z}$. It turns out that $C_{1,\kappa=x}=(l_{z}-1/2)/\sqrt
{(l_{x}-1/2)^{2}+l_{y}^{2}+(l_{z}-1/2)^{2}}$, and analogously for
$C_{2,\kappa=x}$, $C_{1,\kappa=y}$, $C_{2,\kappa=x}$. Based on these results
the sum in Eq.(\ref{A4c-S}) has the final form%

\begin{align}
S  &  =2\sum_{l_{x},l_{y}}\sum_{l_{z}=1}^{\infty}\frac{A(l_{x},l_{y},l_{z}%
)}{(l_{x}^{2}+l_{y}^{2}+l_{z}^{2})^{3/2}}\label{A4c-S-final}\\
&  +2\sum_{l_{x},l_{y}}\sum_{l_{z}=-\infty}^{\infty}\frac{B(l_{x},l_{y}%
,l_{z})}{(l_{x}^{2}+(l_{y}-\frac{1}{2})^{2}+(l_{z}-\frac{1}{2})^{2})^{3/2}%
}\nonumber
\end{align}
where%
\[
A(l_{x},l_{y},l_{z})=\alpha+\frac{l_{z}^{2}}{l_{x}^{2}+l_{y}^{2}+l_{z}^{2}%
}-\frac{2}{3}%
\]%
\[
B(l_{x},l_{y},l_{z})=\frac{(l_{y}-\frac{1}{2})^{2}+(l_{z}-\frac{1}{2})^{2}%
}{l_{x}^{2}+(l_{y}-\frac{1}{2})^{2}+(l_{z}-\frac{1}{2})^{2}}-\frac{2}{3}%
\]
$\alpha=(1-\chi_{\min}/\chi_{\max})/3$ and the sum over $l_{x},l_{y}$ goes
from $-\infty$ to $\infty$.


\begin{thebibliography}{99}
\bibitem {CommentNotation}We adopt the magnetic notation but all results apply
in the electric formalism (paramagnets become dielectrics)
\bibitem {NOTEAxilrodTeller}The $\chi^{2}$ scaling is valid for equilibrium
conditions, where a detailed ballance between the moments and (thermal or
quantum) bath hold. The scaling for icFI\ switches to $\propto\chi^{3}$ if the
fluctuating field is exogeneous (externally set) and non-equilibrium as in the
present cVdW case.
\bibitem {Note4Body}Note, when $i\neq j$ the Eq.(\ref{TrTSquare}) contains
both 2-body interactions as well as the non-local 3-body ones (for instance
$12,23$, etc.) - see discussion below. Similarly Eq.(\ref{TrTQube}) contains
besides the 3-body interactions also the non-local 4-body ones (for instance
$12,23,34$, etc), which property is contained also in all higher order terms.
\end{thebibliography}
\end{document}